\documentclass[english, aps,prd,twocolumn,twoside,superscriptaddress,floatfix, nofootinbib]{revtex4-1}

\usepackage[utf8]{inputenc}
\usepackage{amsmath}
\usepackage{geometry}
\usepackage{xcolor}
\usepackage{graphicx}
\usepackage{float}
\usepackage{lipsum, babel}
\usepackage{subcaption}
\usepackage{comment}

\usepackage{color}
\usepackage{hyperref}
\usepackage{ifthen,amsmath,amssymb, bm}

\newcommand{\beq} {\begin{equation}}
\newcommand{\eeq} {\end{equation}}
\newcommand{\bal} {\begin{aligned}}
\newcommand{\eal} {\end{aligned}}

\begin{document}

\title{Optimizing foreground mitigation for CMB lensing with combined multifrequency and geometric methods}

\author{Omar Darwish}
\affiliation{%
 Center for Theoretical Cosmology, DAMTP,\\
University of Cambridge, Cambridge CB3 0WA, UK
}%
 \email{od261@cam.ac.uk}
\author{Blake D.~Sherwin}%
\affiliation{%
 Center for Theoretical Cosmology, DAMTP,\\
University of Cambridge, Cambridge CB3 0WA, UK
}%

\author{Noah Sailer}
\affiliation{
Berkeley Center for Cosmological Physics, Department of Physics,\\
University of California, Berkeley, CA 94720, USA
}
\affiliation{Lawrence Berkeley National Laboratory, One Cyclotron Road, Berkeley, CA 94720, USA}
\author{Emmanuel Schaan}
\affiliation{Lawrence Berkeley National Laboratory, One Cyclotron Road, Berkeley, CA 94720, USA}
\affiliation{
Berkeley Center for Cosmological Physics, Department of Physics,\\
University of California, Berkeley, CA 94720, USA
}
\author{Simone Ferraro}
\affiliation{Lawrence Berkeley National Laboratory, One Cyclotron Road, Berkeley, CA 94720, USA}
\affiliation{
Berkeley Center for Cosmological Physics, Department of Physics,\\
University of California, Berkeley, CA 94720, USA
}

\begin{abstract}
    A key challenge for current and upcoming CMB lensing measurements is their sensitivity to biases from extragalactic foregrounds, such as Sunyaev-Zeldovich (SZ) signals or cosmic infrared background emission. Several methods have been developed to mitigate these lensing foreground biases, dividing broadly into multi-frequency cleaning approaches and modifications to the estimator geometry, but how to optimally combine these methods has not yet been explored in detail. In this paper, we examine which combination of lensing foreground mitigation strategies is best able to reduce the impact of foreground contamination for a Simons-Observatory-like experiment while preserving maximal signal-to-noise. Although the optimal combination obtained depends on whether bias- or variance-reduction are prioritized and on whether polarization data is used, generally, we find that combinations involving both geometric (profile hardening, source hardening or shear) and multifrequency (symmetric cleaning) methods perform best. For lensing power spectrum measurements from temperature (polarization and temperature), our combined estimator methods are able to reduce the bias to the lensing amplitude below $\sigma/4$ or $0.3\%$ ($0.1\%$), a factor of 16 (30) lower than the standard QE bias, at a modest signal-to-noise cost of only $18\%$ ($12\%$). In contrast, single-method foreground-mitigation approaches struggle to reduce the bias to a negligible level below $\sigma/2$ without incurring a large noise penalty. For upcoming and current experiments, our combined methods therefore represent a promising approach for making lensing measurements with negligible foreground bias.
\end{abstract}

\maketitle

\section{Introduction}
Along their paths to our telescopes, photons of the cosmic microwave background radiation (CMB) are deflected by the gravitational influence of matter in our Universe. This leads to a remapping of CMB photons that depends on a weighted integral of the matter perturbations along the line of sight. The ability to directly map the projected mass distribution out to high redshifts makes CMB lensing a powerful source of cosmological information (such as constraints on neutrino masses or dark energy properties) (e.g. \cite{Roland2009, Ade_2019}).

When reconstructing the projected matter fluctuations from observations, CMB foreground contamination in temperature is expected to induce $\sim 5\%$-level biases in the CMB lensing power spectrum and lensing cross-correlations with tracers of the matter field~\cite{Osborne, Engelen_2014, kSZBias, Baxter2018, SmithForegrounds, DasForegrounds, Madhavacheril:2018, 2019SchaanFerraro, sailer_paper1}. These biases are especially concerning for high-resolution ground-based CMB experiments, such as AdvACT, SPT-3G and Simons Observatory, since these experiments still rely heavily on lensing reconstruction from temperature and derive more information from small angular scales which have higher levels of foreground contamination.

Several mitigation strategies have been proposed to address this key foreground challenge for CMB lensing. Broadly, these strategies divide into geometric methods and frequency-based methods. Geometric methods \cite{Namikawa_2013, 2019SchaanFerraro, sailer_paper1}, which include bias hardening, profile hardening, and shear reconstruction, aim to modify the lensing estimators' weight functions in order to null or reduce the biases induced by foregrounds. Frequency-based methods \cite{Delabrouille_2008, Remazeilles_2010, Madhavacheril:2018, Abylkairov_2021, Madhavacheril_2020, PlanckComponent2020, bleem2021cmbksz}, which include foreground deprojection in a constrained or partially constrained ILC, as well as gradient cleaning and symmetric cleaning, use the departure of the foregrounds' spectral properties from a blackbody to null or reduce the foreground bias levels.

While many of these methods can reduce foreground biases quite effectively, with increasing measurement precision the requirements on bias mitigation are becoming increasingly stringent, and further improvements are becoming well motivated. A key question for upcoming experiments therefore is: which combination of bias mitigation strategies minimizes the biases most effectively while preserving as much signal-to-noise as possible?

To address this question, we consider estimators composing -- i.e., simultaneously applying -- geometric and frequency-based mitigation methods; we also consider linear combinations of different estimators at the lensing reconstruction level and optimise this linear combination to best mitigate foreground-induced lensing biases while maximising the signal-to-noise.


The remainder of this paper is structured as follows. In \S\ref{sec:combination}, we introduce our method for combining different estimators and mitigation approaches. In \S\ref{sec:biasnoise}, we define bias and noise measures for the combined estimator. In \S\ref{sec:optimality} we present the optimization formalism, followed by our results and their discussion in \S\ref{sec:optinpractice}. We conclude in \S\ref{sec:conclusion}.

\section{CMB Lensing foreground biases for estimators\label{sec:combination}}

\subsection{Quadratic CMB Lensing Estimators}

Weak gravitational lensing of the CMB induces correlations between different modes of the CMB, because a fixed lensing field breaks the statistical isotropy of the primordial CMB. One can use these off-diagonal correlations to write a quadratic estimator for the CMB lensing convergence field from the observed beam-deconvolved lensed CMB temperature field $T$. This estimator has the form \cite{Hu:2002}\footnote{We focus on temperature-only reconstruction for now, and later we will discuss including polarization.}

\begin{equation}
    \hat{\kappa}(\vec{L}) = \int_{\vec{\ell}} T(\vec{\ell})T(\vec{L}-\vec{\ell})g(\vec{\ell}, \vec{L}-\vec{\ell}),
\end{equation}
where $\int_{\vec{\ell}} \equiv \int \frac{d^2\vec{\ell}}{(2\pi)^2}$ and $g$ is a weight-function that satisfies the unit response condition to the true CMB lensing convergence field, $\langle\hat{\kappa}\rangle_\text{CMB}=\kappa$, where $\langle\cdots\rangle_\text{CMB}$ denotes averaging over the primordial CMB while fixing the lensing mode to be reconstructed. This, in turn implies that
\begin{equation}
    \int_{\vec{\ell}} f(\vec{\ell}, \vec{L}-\vec{\ell}) g(\vec{\ell}, \vec{L}-\vec{\ell}) = 1,
\end{equation}
 where $f$ is a response function that encodes the response of the off-diagonal CMB temperature two-point correlation function to lensing, $\langle T({\vec{\ell}})T(\vec{L}-\vec{\ell}) \rangle = f(\vec{\ell}, \vec{L}-\vec{\ell}) \kappa(\vec{L})$.

\subsection{Origin of foreground biases}

The observed CMB temperature field $T$ contains contributions from both the lensed CMB $T_\text{CMB}$ and foregrounds $T_{f}$, so that $T = T_\text{CMB}+T_{f}$, where here we ignore any noise contribution.
The foregrounds, which are non-Gaussian and also correlated with the lensing convergence, give rise to biases in both the auto-correlation of a CMB lensing quadratic estimator $\hat{Q}[T,T]$, and to the cross-correlation of $\hat{Q}$ with an external matter tracer $g_m$.
As discussed in \cite{vanEngelenForegrounds, Osborne, 2019SchaanFerraro}, and explained in detail in Appendix \ref{app:cmblensingbiases},
foregrounds induce three bias terms to the reconstructed auto-spectrum:
a trispectrum term of the form $T_f^4$, and two bispectrum terms that involve $T_f^2\times \kappa$. For a CMB lensing cross-spectrum with an LSS tracer, the bias appears as a bispectrum of the form $\sim T_f^2 \times g_m$
. These biases can easily lead to an incorrect inference in parameters, e.g.,~the amplitude of matter fluctuations, if not treated properly.

\subsection{Current foreground mitigation methods and estimators}

Several modifications to CMB lensing estimators have been proposed to extract an unbiased CMB lensing signal; they differ both in the degree to which they mitigate foreground biases and to which they lose signal-to-noise.
In this paper we will consider the following foreground-mitigating CMB lensing estimators, some of which rely on geometric mitigation of the estimator kernels, and others on multi-frequency methods:

\begin{itemize}
    \item The standard Quadratic Estimator \cite{Hu:2002} (QE): this estimator has the smallest possible variance. On the other hand, it is not immune to the biases induced by foregrounds due to the contamination of the temperature maps.
    \item Bias hardened estimators \cite{Namikawa_2013,sailer_paper1}, such as point source-hardening (PSH) and profile hardening (PH), which modify the estimator weights to null a component of known response at the cost of increasing the variance. 
    \item The shear estimator \cite{2019SchaanFerraro} (SH), which isolates the quadrupole (shear) component of the lensing response $f$, and is generally insensitive to foregrounds with symmetric profiles about the line of sight.
    \item The symmetric multi-frequency cleaned estimator \cite{Madhavacheril:2018,Darwish2020} (Symm), which allows for the nulling of some foreground effects using multi-frequency cleaned data, but has been constructed to reduce the noise cost incurred.
\end{itemize}

In this paper we use a tSZ-like profile for the PH estimator, identical to the one considered in \cite{sailer_paper1}, i.e. the square root of a fiducial tSZ power spectrum on small scales.

\begin{figure}
    \centering
    \includegraphics[width=\columnwidth]{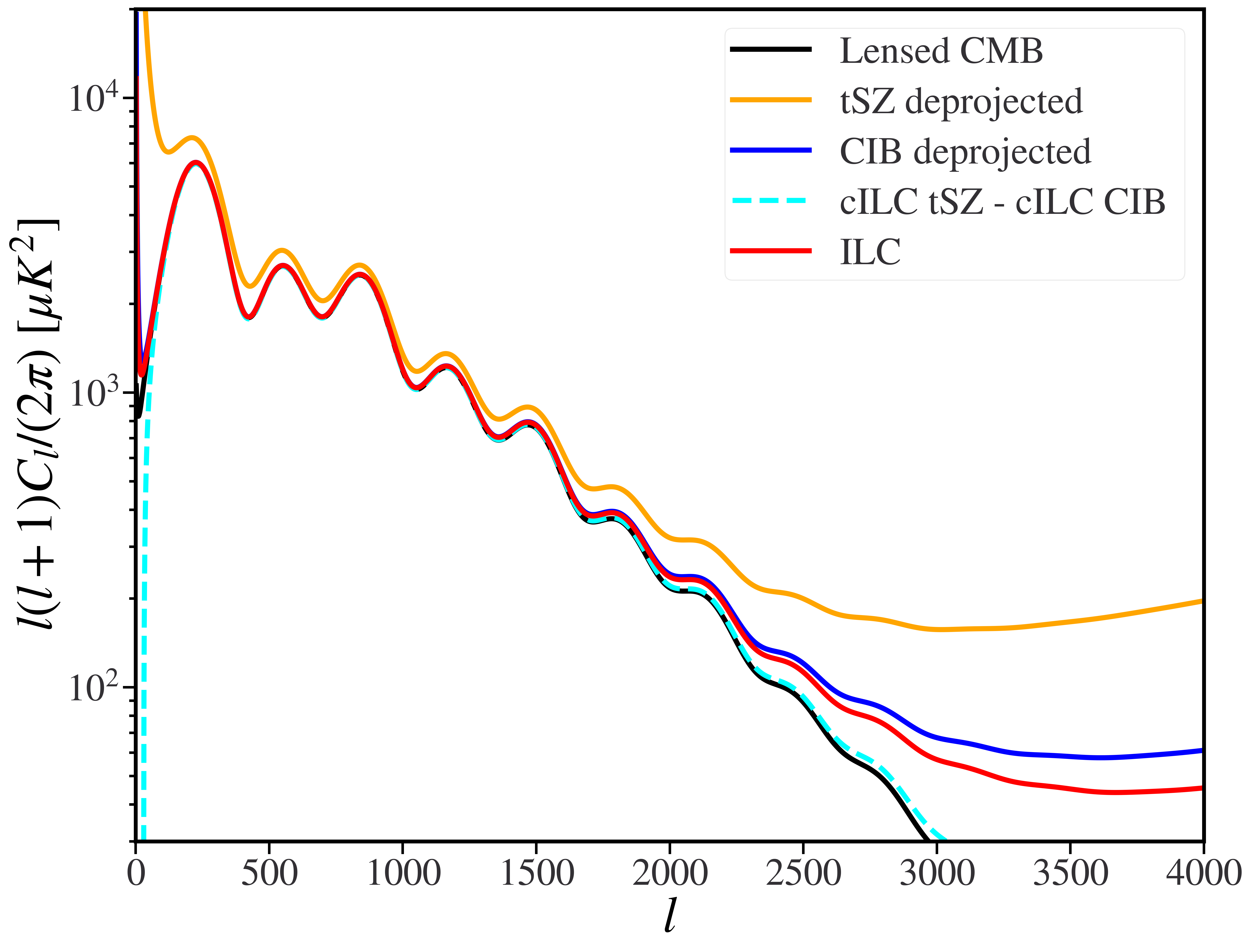}%
    \caption{The temperature power spectra of different multi-frequency combinations of frequency maps for a Simons Observatory-like experiment: in red is the standard internal linear combination (ILC), in blue is the ILC with CIB deprojection, in orange is the ILC with tSZ deprojection. In cyan we show the cross correlation between the tSZ-deprojected and CIB-deprojected combinations. In black is the lensed CMB theory curve. For the tSZ-deprojected curve in orange, the large power on large scales is due to atmospheric noise.
    } \label{fig:cmb}
\end{figure}

\begin{figure}
    \includegraphics[width=\columnwidth]{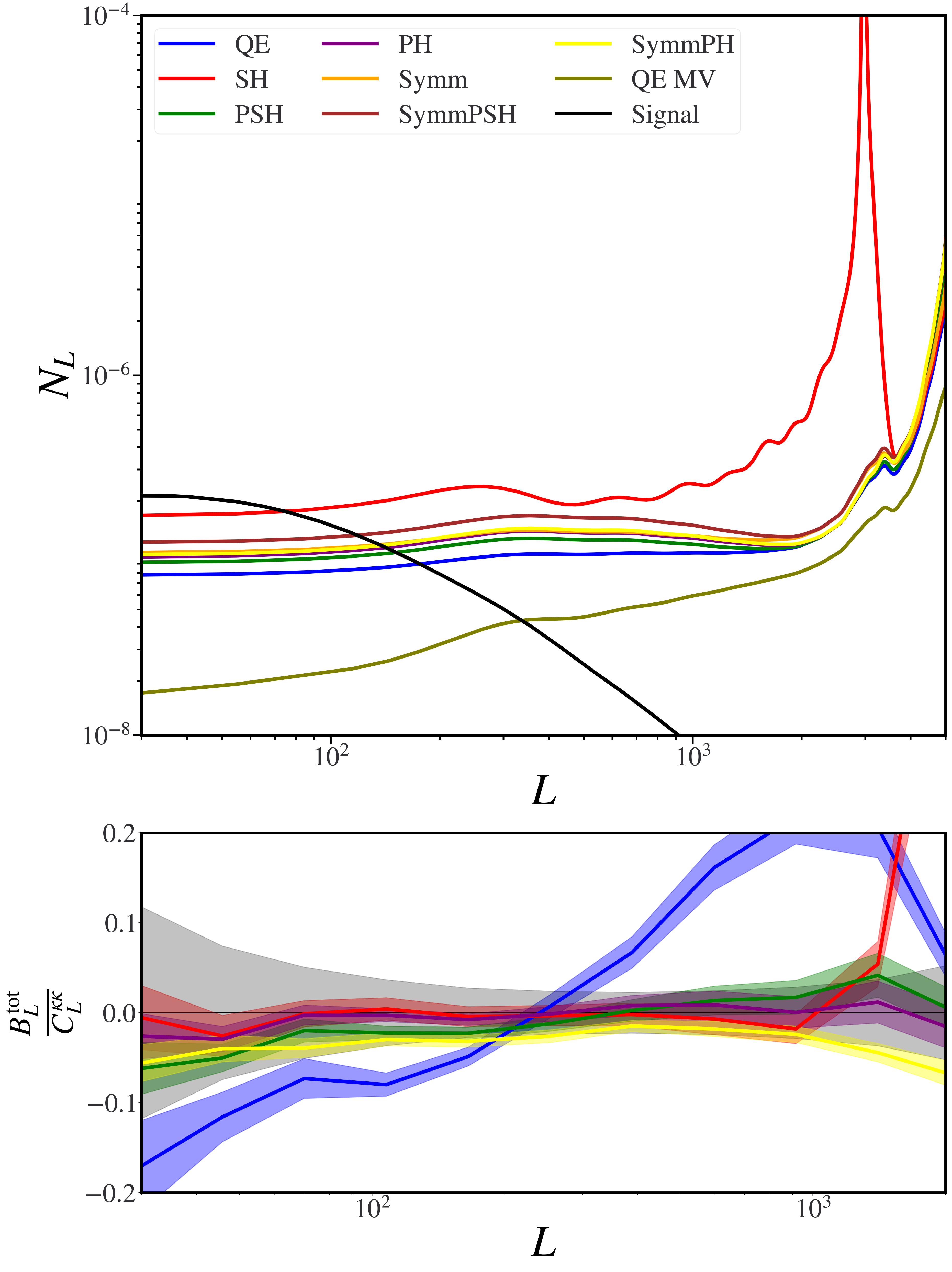}%
    \caption{CMB lensing reconstruction noise curves for the standard quadratic estimator (QE), shear (SH), point source hardened estimator (PSH), and the profile hardened estimator with a tSZ profile (PH), all applied to a minimum variance temperature-only ILC map, as well as noise curves for the standard symmetric estimator with tSZ-deprojection (Symm) and two new estimators that we introduce later in the text: point source hardening on the symmetric estimator with tSZ-deprojection (SymmPSH), profile hardening on the symmetric estimator with CIB-deprojection (SymmPH), and finally the minimum variance estimator with both temperature and polarization (QE MV). For the temperature maps we take $\ell_{\mathrm{min}}, \ell_{\mathrm{max}}=30, 3500$ in this figure, while for polarization we assume $\ell_{\mathrm{min}}=30, \ell_{\mathrm{max,pol}}=5000$. In black we show the CMB lensing signal $C_L^{\kappa\kappa}$. In the bottom panel we show the total bias to the CMB lensing autospectrum (as a fraction of the signal power) for some of the estimators. The relative statistical error for the standard QE is shown in grey. Note that the standard QE is biased at most scales.} \label{fig:noisegaussianlensing}
\end{figure}

In Figure \ref{fig:cmb} we plot the temperature power spectra of several multi-frequency combinations for an SO-like experiment, while in Figure \ref{fig:noisegaussianlensing} we show the lensing reconstruction noise for different estimators, assuming $\ell_\text{max}=3500$ is used in the reconstruction, as well as the total foreground biases for a few estimators (see Appendix \ref{app:biaseslensing}, Figure \ref{fig:biases_auto_per_L} for the biases of the remaining estimators).

\section{Optimal combination of CMB lensing estimators: formalism\label{sec:biasnoise}}

\subsection{Amplitude shift on the CMB lensing power \label{sec:amplshiftsingle}}

For our investigations we must be able to appropriately describe the impact of foregrounds on the measured CMB lensing power spectrum.
For simplicity we will focus on constraints on the amplitude $A$ of the lensing power spectrum or the cross-correlation of the lensing convergence with a matter tracer -- we expect this parameter to be a good proxy for the most relevant applications of the lensing power spectrum, such as measuring $\sigma_8$ or the neutrino mass. Effectively we have $\hat{A}^\alpha_L \propto \hat{C}_L^{\kappa \alpha} $, with $\alpha = \kappa$ or $g$, where $g$ is a tracer of the underlying matter distribution (e.g. galaxies). Therefore, to lowest order, a bias in the measured power spectrum leads to a bias to the inferred amplitude, i.e. $\delta A^\alpha_L \propto \delta C_L^{\kappa \alpha} $. Let us define the measured lensing amplitude per mode $\vec{L}$ with respect to a fiducial cosmology:
\begin{equation}
    \hat{A}(\vec{L}) =  \hat{C}_{L}^{\kappa\alpha}/C_L^{\kappa\alpha,\mathrm{fid}},
\end{equation}
where we've suppressed the explicit dependence of $\hat{A}$ on $\alpha$. Assuming that $\kappa$ and $g_m$ are Gaussian fields, for on single mode the variance of this estimator is related to the variance of the power spectrum estimator $\hat{C}_{L}^{\kappa\alpha}$

\begin{equation}
    \sigma^2_L = (C_L^{\kappa \alpha}+N_L^{\kappa \alpha})^2+(C_L^{\kappa \kappa}+N_L^{\kappa \kappa})(C_L^{\alpha \alpha}+N_L^{\alpha \alpha}),
\end{equation}
where $N_L^{\alpha \beta}$ a noise component, equal to the lensing reconstruction noise if $\alpha=\beta=\kappa$, the Poisson shot noise if $\alpha=\beta=g_m$, and zero if $\alpha\neq\beta$. The bias to the estimator $\hat{A}(\vec{L})$ is simply
\begin{equation}
    \mathcal{B}(\hat{A}(\vec{L})) = \delta C_{L}^{\kappa\alpha}/C_L^{\kappa\alpha,\mathrm{fid}}.
\end{equation}

Given a range of modes, we can combine the measurements of $A$ at each $\vec{L}$ to obtain a global estimator:
\begin{equation}
    \hat{A}=  \int_{\vec{{L}}}w(\vec{L})\hat{A}(\vec{L})\ ,
\end{equation}
where the weights $w$ satisfy $\int_{\vec{L}}w(\vec{L})=1$ to ensure that the estimator $\hat{A}$ is unbiased to the fiducial cosmology ($\langle \hat{A}\rangle = A_{\mathrm{fid}} = 1$). When foregrounds are present, the estimator $\hat{A}$ acquires a bias:
\begin{equation}
   b(\hat{A}) =  \int_{\vec{{L}}}w(\vec{L})\mathcal{B}(\hat{A}(\vec{L})).
\end{equation}
The variance of the estimator $\hat{A}$ is given by
\beq
\bal
    \sigma^2(\hat{A}) &=\frac{1}{4\pi f_\text{sky}} \int_{\Vec{L}}w^2_L\frac{\sigma^2_L}{(C_L^{\kappa\alpha,\rm{fid}})^2},
\eal
\eeq
where $f_\text{sky}$ is the covered sky fraction. Of particular relevance is the minimum variance estimator, which has weights
\begin{equation}
\label{eq:mvweight}
    w^{\mathrm{MV}}_L= \frac{(C_L^{\kappa\alpha, \rm{fid}})^2}{\sigma^2_L}\left[ \int_{\Vec{L}} \frac{(C_L^{\kappa\alpha, \rm{fid}})^2}{\sigma^2_L} \right]^{-1}.
\end{equation}

\begin{figure}
    \centering
    \includegraphics[width=\columnwidth]{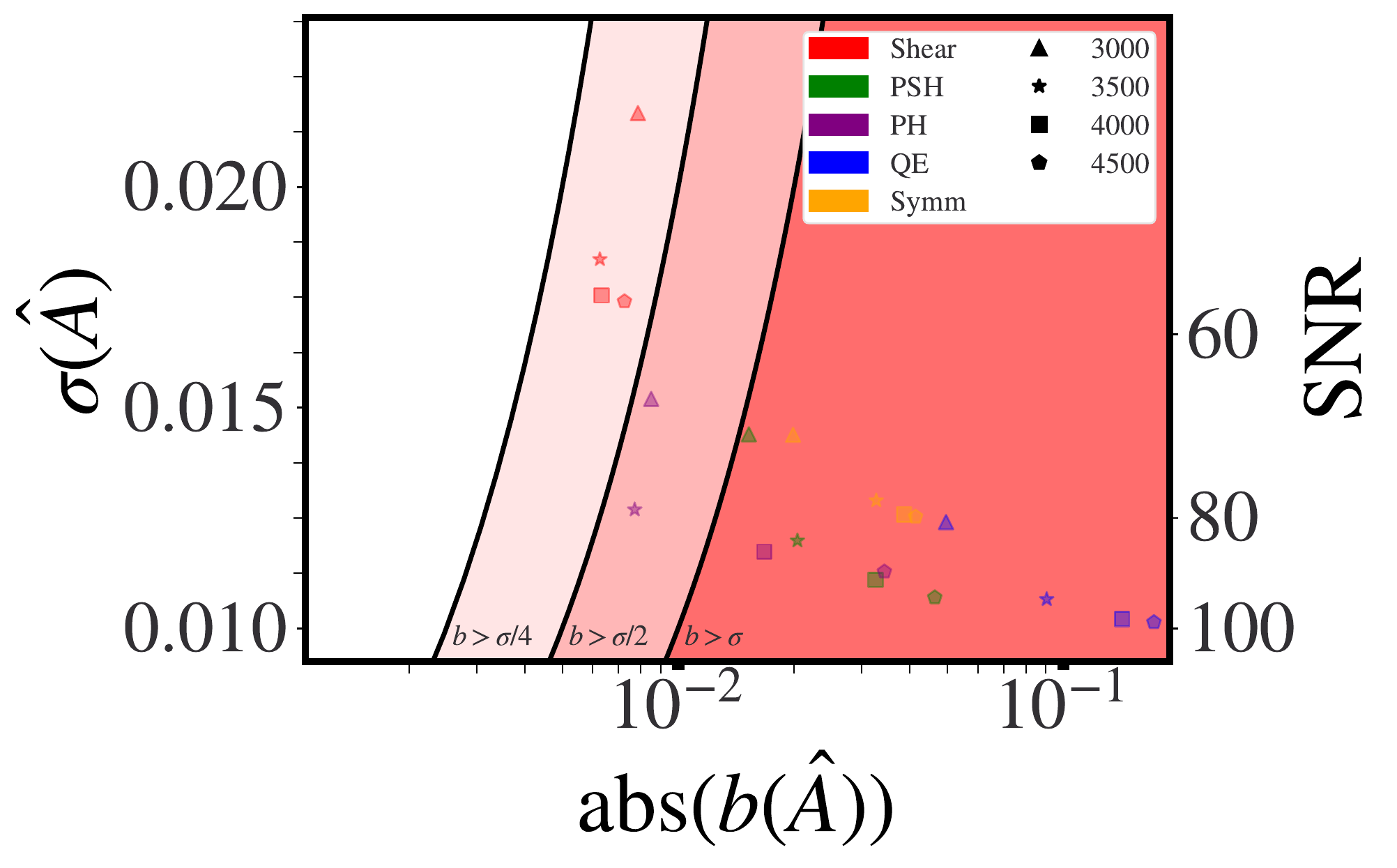}%
    \caption{Values of measurement standard deviation $\sigma$ and bias $b$ for the CMB lensing power spectrum amplitude $A$, assuming only temperature ($TT$) reconstruction is used in the lensing power spectrum measurement. Results are shown for the different estimators with different types of foreground mitigation that have been proposed so far; several values are assumed and plotted for the maximum CMB multipole used in the lensing reconstruction. In this and the following plots, [QE, SH, PSH, PH, Symm] indicate the standard quadratic estimator, shear estimator, point source hardened estimator, profile hardened estimator, and symmetric tSZ-deprojected estimator respectively. The colored bands represent regimes where the bias is greater than a certain fraction of the statistical noise. We can see clearly that a trade-off between bias and noise exists for the different estimators proposed thus far: estimators that aim to reduce biases unfortunately incur a significant noise penalty. We would like to be able to reduce both foreground bias and variance of lensing estimators simultaneously (obtaining new estimators that are somewhat closer to the lower left corner of this plot).} \label{fig:alensperestimatororiginal}
\end{figure}

In Figure \ref{fig:alensperestimatororiginal} we plot the CMB lensing error and bias for the foreground-mitigating estimators proposed thus far, assuming several different values for the reconstruction $l_{\mathrm{max,TT}}$ and summing over CMB lensing modes $30 \leq |\vec{L}|\leq 1200$.

In calculating the bias, we choose to replace the ``true'' biases $B^{i}_L$ for the estimator $i$, as measured from a set of simulations described in section \ref{sec:optinpractice}, with a smoothed version of the absolute value of the biases. This choice will be justified later in section \ref{sec:optimality}.

It is clear that in most cases we are in a regime where bias is not negligible. For the estimators which achieve $b<\sigma$ we are in general paying a substantial noise penalty, even for profile hardening (PH) at $l_{\mathrm{max,TT}}=3500$. Furthermore, to ensure that bias is truly subdominant and negligible (despite uncertainties in simulations and modeling), we would like to be in a regime where bias is well below the statistical uncertainty, such as $b \leq \sigma/4$ -- and this cannot be realistically achieved with the currently-proposed estimators.

\subsection{Amplitude shift on CMB lensing power from composing geometrical and multi-frequency methods \label{sec:composingestimators}}

To improve upon the current situation, we first propose two new CMB lensing estimators; these are constructed by simultaneously applying (or ``composing'') both multi-frequency cleaning and the bias hardening operation in an estimator. In particular, first we apply the symmetric estimator (which involves multifrequency cleaning), then we apply bias hardening on the symmetric estimator as explained in Appendix \ref{app:cmblensingcomposition}. For the first new estimator, we consider a symmetric estimator with a tSZ-deprojected map, and then apply point source hardening. For the second new estimator, we consider using a symmetric estimator with a CIB-deprojected map (see \cite{Madhavacheril_2020} for the first implementation of CIB deprojection on real CMB data), and then apply profile hardening. The logic behind these choices is that we employ one mitigation method that targets the tSZ (tSZ deprojection or profile hardening) alongside another that focuses on the CIB (CIB deprojection or point source hardening). We refer to these two new estimators as SymmPSH (using tSZ deprojection with point source hardening), and SymmPH (using CIB deprojection with profile hardening) respectively.

\begin{figure}
    \centering
    \includegraphics[width=\columnwidth]{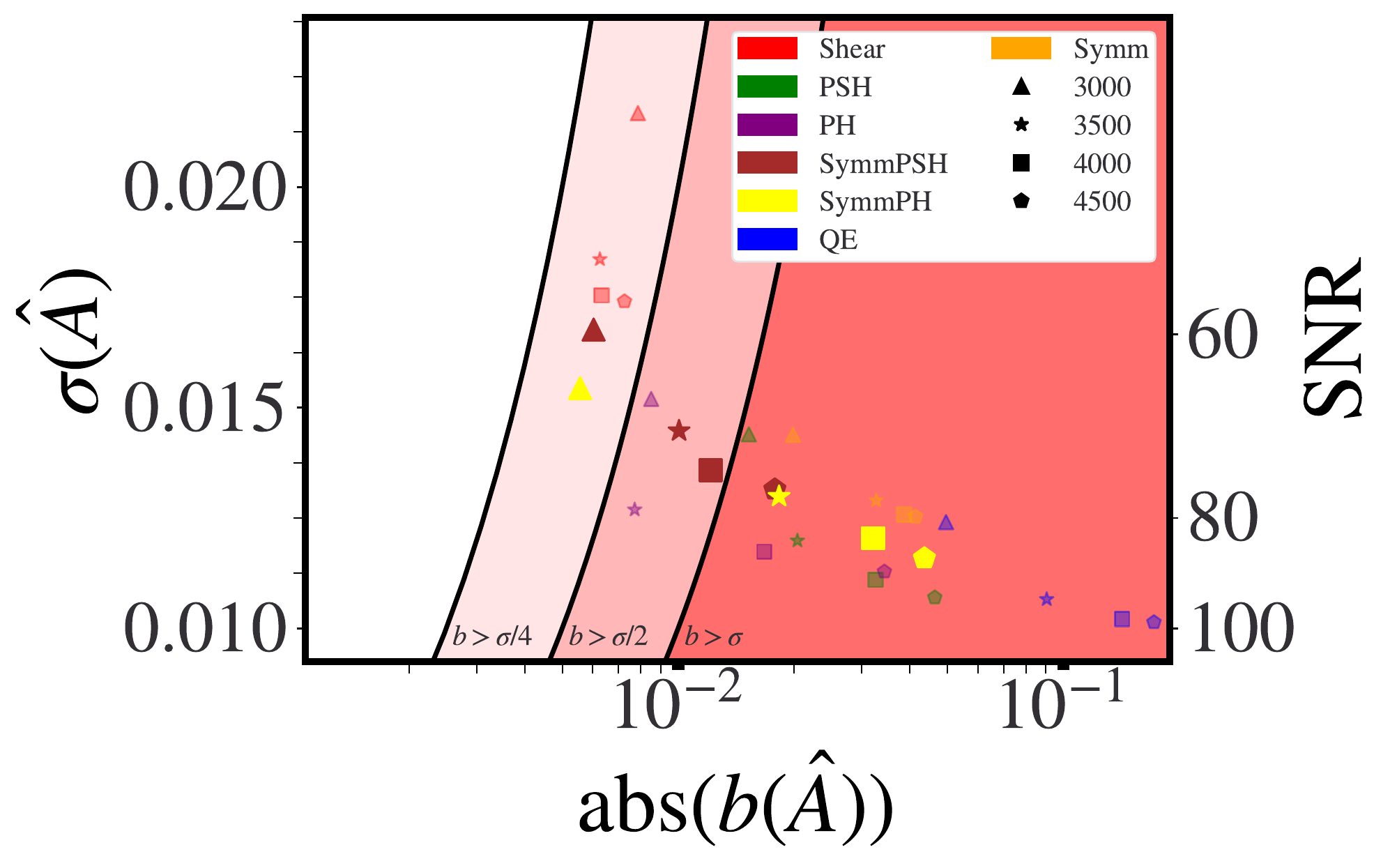}%
    \caption{As for Figure \ref{fig:alensperestimatororiginal}, but with the addition of two new combined estimators: i) SymmPSH, which stands for point source hardening applied to a symmetric estimator with tSZ deprojected; ii) SymmPH, given by profile hardening applied to a symmetric estimator with CIB deprojected. We can see that, for some configurations, these new estimators (which are highlighted with black circles) allow us to obtain somewhat lower bias at the same noise in comparison with the existing estimators shown in Figure \ref{fig:alensperestimatororiginal}, although potential remains for further improvements.} \label{fig:alensperestimatornew}
\end{figure}

We can see from Figure \ref{fig:alensperestimatornew} that these estimators perform well: they generally have lower biases given the same variance as the previously proposed estimators and have $b<\sigma$. However, further improvements are still motivated, since we have not yet found that biases are completely negligible such that $b \ll \sigma$. For this purpose, we will now investigate linear combinations of different estimators $\hat Q_i$.

\subsection{Amplitude shift on CMB lensing power for combined estimators \label{sec:amplshiftmultiple}}

Using the same definitions as above, we can straightforwardly derive the bias and variance of the lensing amplitude estimator $\hat{A}$ when computed from a linear combination of different individual quadratic estimators (with different foreground mitigation methods) $\hat{Q}_i$. We can define this linear combination as follows:
\begin{equation}
    \hat{Q}(\vec{L}) \equiv \sum_i a_i(L) \hat{Q}_i(\vec{L}),
\end{equation}
where $\sum_i a_i(L) = 1$ to ensure unit response to the CMB: $\langle  \hat{Q} \rangle_\text{CMB} = \kappa_\text{CMB}$. For the estimator $\hat{Q}$, the measured power spectrum $\hat{C}^{\kappa\alpha}_L$ takes the form:
\begin{equation}
    \hat{C}^{\kappa\alpha}_L = \sum_{i,j}a_i(L) b_j(L) \hat{C}_L^{\kappa\alpha,ij},
\end{equation}
where $\hat{C}_L^{\kappa\alpha,ij}$ is the reconstructed power spectrum from the correlation of the individual $i,j$ estimators, and $b_j(L) = a_j(L)$ when $\alpha=\kappa$, or $b_j(L) = \delta^K_{j1}$ when $\alpha=g$.

The estimator for the amplitude now takes the form
\begin{equation}
     \hat{A} = \int_{\vec{L}} w(\vec{L}) \sum_{i,j} a_ib_j \hat{C}_L^{\kappa\alpha,ij} /C_L^{\kappa \alpha,\rm{fid}}.
     \label{eq:combinedAlens}
\end{equation}
In the presence of foregrounds, this estimator obtains a bias, which can be expressed as
\begin{equation}
\label{eq:bias}
b(\hat{A})  =\int_{\vec{L}} w(\vec{L})\sum_{i,j}a_ib_j\frac{B^{ij}_L}{C_L^{\kappa\alpha,\rm{fid}}}
\end{equation}
where $B^{ij}_L = \delta C^{\kappa\alpha,ij}_L$ is the foreground-induced lensing bias from the cross-correlation of the individual $i,j$ estimators. The variance (assuming the lensing reconstruction noise can be approximated as Gaussian) is given by
\begin{equation}
\label{eq:noise}
    \sigma^2(\hat{A})= \frac{1}{4\pi f_\text{sky}}\int_{\vec{L}} w_{L}^2  \sum_{ijmn}  \frac{\Theta^{ijmn}_L a_ia_jb_mb_n}{(C_L^{\kappa \alpha,\rm{fid}})^2}\ ,
\end{equation}
where we have supressed the $L$ dependence of $a_i$ and $b_i$, and we have defined
\beq
\bal
  \Theta^{ijmn}_L &\equiv
  (C^{\kappa\alpha}_L + N^{\kappa\alpha,in}_L)(C^{\kappa\alpha}_L + N^{\kappa\alpha,jm})\\
  &+(C^{\kappa\kappa}_L + N^{\kappa\kappa,im}_L)(C^{\alpha\alpha}_L+N^{\alpha\alpha,jn}_L).
\eal
\eeq
Again, when $\alpha\neq\beta$, the noise term $N^{\alpha\beta,ij}_L=0$. When $\alpha=\beta=g$, $N^{gg,ij}_L \equiv N^{gg}$ is just the Poisson shot noise. When $\alpha=\beta=\kappa$, $N^{\alpha\beta,ij}_L $ is the lensing noise associated with the cross-correlation of the $i,j$ estimators.

We note that while the variance $\sigma^2(\hat{A})$ is guaranteed to decrease as more $L$'s are summed over, the same is not true for the bias. Thus, even in the case where the bias $B_L$ is subdominant to the noise $\sqrt{\Theta_L}$ for each $L$, one can still have $b^2(\hat{A})>\sigma^2(\hat{A})$.

\section{Finding the optimal combination\label{sec:optimality}}

Now that we have introduced all the relevant definitions, we can begin to address the key question of this paper: what is the optimal linear combination for minimizing both foreground bias and measurement noise? 
To answer this question, we define a loss function, inspired by the mean squared error $\langle (\hat{A}-A)^2 \rangle$, which allows us to parameterize our goals in this optimization: i.e. to what extent we prioritize bias or variance. To this end we consider the following loss function
\beq
\label{eq:totalfunction}
   \mathcal{L}[a_i,w;f_b] =  \sigma^2(\hat{A})+f_b^2 b^2(\hat{A})
\eeq
where $\sigma^2(\hat{A})$ and $b(\hat{A})$ are defined in Eqs.~\eqref{eq:bias} and \eqref{eq:noise} respectively, and the $f_b$ parameter regulates the importance of the bias in the total functional, with a higher $f_b$ assigning a higher importance to bias reduction.\footnote{One can introduce an extra function $r$ too,  monotonic in the weights $a$ and biases matrix $\mathbf{B}$: this should be chosen to act as a prior on the weights, or as a 'regularizer'. The main reason for such term is to not overfit the specific set of simulations or theoretical models used for calculating the biases, as there is an inherent modeling uncertainty in these. A regularizer will be also useful as a 'smoother' of the optimal solution, to have more stable, non-oscillating solutions.} Eq.~\eqref{eq:totalfunction} represents a general quartic optimization problem, which we numerically minimize by varying $a_i(L)$ and $w(\vec{L})$ subject to the constraints
\begin{equation}
    \sum_i a_i(L) = \int_{\vec{L}} w(\vec{L}) = 1.
\end{equation}
We additionally impose the constraint $a_i(L)\geq 0$ on the coefficients, which make our optimized solutions less finely-tuned to the simulated biases.\footnote{For example, if the biases of two estimators were known perfectly well, and if the coefficients weren't restricted to be non-negative, one could easily solve for coefficients which cancelled these biases. This cancellation is simulation dependent, and could result in significant biases on real data if the biases to the individual estimators are non-negligible.}

In our optimization we choose to replace the ``true'' biases $B^{ij}_L$, as measured from simulations, with a smoothed version of the absolute value of the biases. That is, when minimizing Eq.~\eqref{eq:totalfunction}, we replace $B^{ij}_L$ in Eq.~\eqref{eq:bias} with $|B^{ij}_L|_s>0$, where the subscript $s$ denotes a smoothing operation, which is described in Appendix \ref{app:optdetails}. Doing so avoids potential exact cancellations in bias among different $L$'s in order to be conservative (and less simulation-dependent), as seen for example in Fig.~\ref{fig:noisegaussianlensing}. We discuss how this choice impacts our results in Appendix \ref{app:optdetails}.

We note that the case $f_b=0$ corresponds to minimizing the variance $\sigma^2(\hat{A})$. Since the integrand in Eq.~\eqref{eq:noise} is positive, minimizing the full variance corresponds to minimizing the integrand for each $\vec{L}$. In other words, the ideal solution corresponds to the minimum variance combination of the individual estimators, and setting $w(\vec{L})=w^\text{MV}_L$ (i.e. Eq.~\eqref{eq:mvweight} with $\sum_{ijmn}\Theta^{ijmn}_L a_ia_jb_mb_n$ replacing $\sigma^2_L$).

\section{\label{sec:optinpractice}Optimisation in practice}

\subsection{Calculating foreground CMB lensing biases}

In this paper we consider a Simons Observatory-like experiment, with six observational frequency channels ($[27, 39, 93, 145, 255, 280]\ \mathrm{GHz}$). We assume the goal noise levels \cite{Ade_2019} for the detector and atmospheric noise contributions.\footnote{The noise curves are calculated using the noise calculator $\mathrm{V3\_calc}$ available at a private repository on the SO Github.} We follow \cite{Dunkley_2013} when modeling the foreground power spectra and their SEDs, as implemented in the code \rm{LensQuEst}.\footnote{https://github.com/EmmanuelSchaan/LensQuEst}

The foreground-induced CMB lensing biases that we consider here come from the total sum of extragalactic foregrounds, namely tSZ + CIB + kSZ + radio PS, as given by the non-Gaussian foreground simulations of \cite{Sehgal_2010} at 150 GHz. To create ILC combinations from these simulations we do the following: first, we mask the sum of extragalactic foregrounds with a mask obtained by inpainting \cite{Bucher_2012} disks with a 3 arcmin radius around individual detected point sources with flux density higher than 5 mJy at 150 GHz, picking up also clusters. After masking, we Fourier transform the simulations, rescaling them from 150 GHz to the corresponding SO frequencies using the SEDs of \cite{Dunkley_2013},\footnote{Note that we are implicitly assuming that CIB can be simply rescaled, but in reality there is a dechoerence among CIB observed at different frequencies, i.e. the correlation among CIB maps at different frequencies is not one, although quite high in the simulations.} and create the ILC/cILC  combinations, with a total theory power spectrum given by lensed CMB added to experimental noise, galactic dust, and extragalactic foreground contributions from theory spectra, as given by \cite{Dunkley_2013}.
Then we take the inverse Fourier transform and mask again with the same mask (equal to one outside the inpainted disks).

We then follow the method of \cite{2019SchaanFerraro, sailer_paper1} to estimate the foreground-induced non-Gaussian CMB lensing biases and hence obtain the matrix $B^{ij}$ for the bias arising from the cross of estimator $i$ with estimator $j$. We describe the specific calculations in Appendix \ref{app:cmblensingbiases}.

\subsection{Results}

In the following subsections we present our optimization results. Aside from $f_b=0$, for all cases we optimize over the [SH, PSH, PH, SymmPSH, SymmPH] estimators as explained in Appendix \ref{app:optdetails} (recall that these abbreviations indicate shear, point source hardening, profile hardening, symmetric tSZ deprojection with point source hardening, symmetric CIB deprojection with profile hardening); we also allow the maximum multipole to vary separately for each estimator in our optimization, with the options $l_{\mathrm{max, TT}}=[3000, 3500, 4000, 4500]$.

\subsubsection{Lensing auto-spectrum}
\paragraph*{Temperature only lensing reconstruction}

\begin{figure}
    \centering
    \includegraphics[width=\columnwidth]{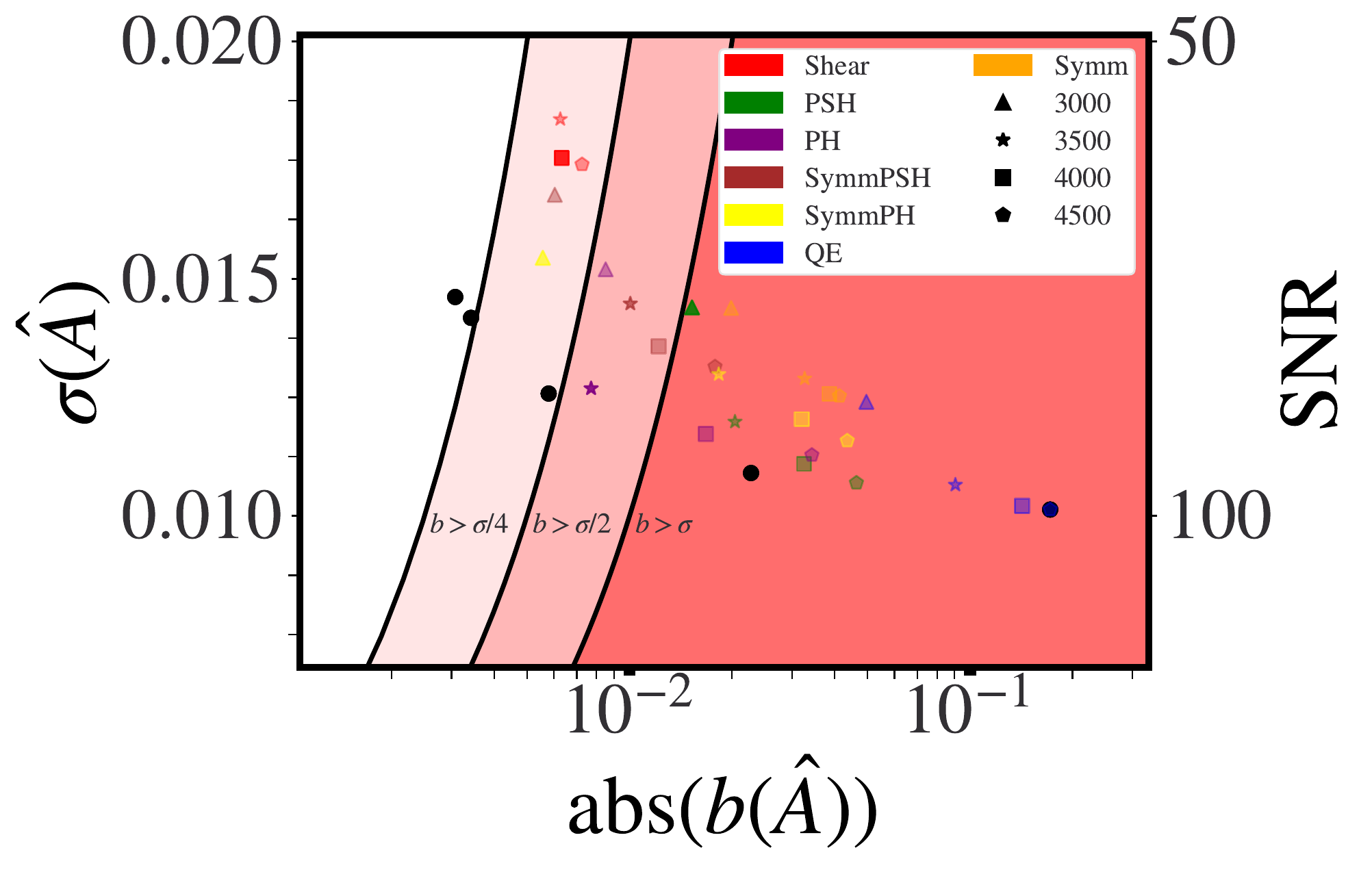}%
  \caption{As for Figures \ref{fig:alensperestimatororiginal} and \ref{fig:alensperestimatornew}, but now including results from performing an optimized linear combination of individual estimators (in this plot we only consider lensing measurement from temperature). The solid black dots show the optimized points, with results shown for optimization with $f_b=0, 0.1, 1, 2, 4$ going from dots on the right to the left (recall that $f_b$ parametrizes the importance of bias-squared relative to variance in the optimization). We can see that the black dots representing the linear combinations of estimators perform significantly better than single estimators; using these optimized linear combinations we are able to reach a regime with a negligible bias with respect to the noise at only a modest noise cost.} \label{fig:autoTTresultsblackdots}
\end{figure}

\begin{figure}
    \centering
    \includegraphics[width=\columnwidth]{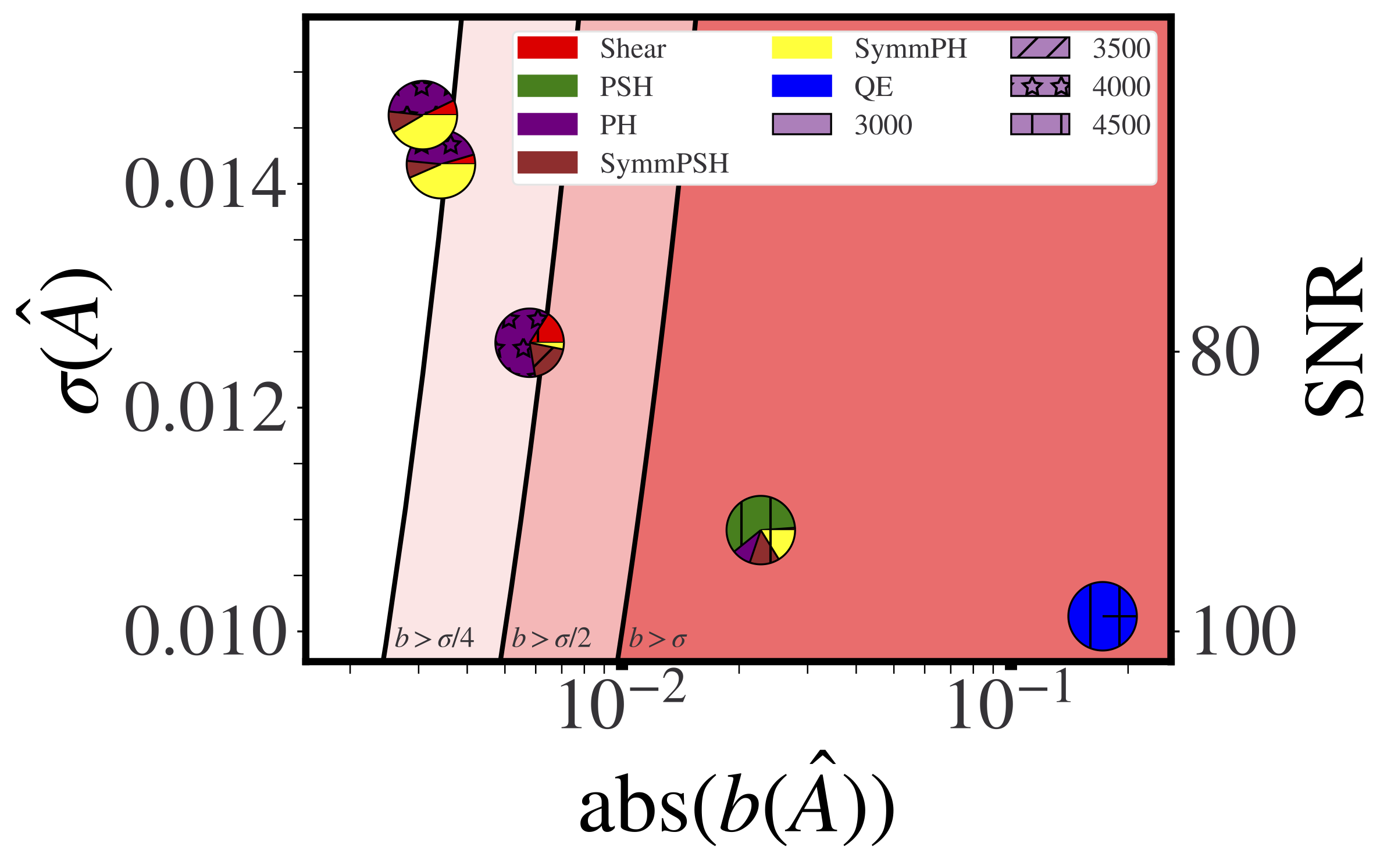}%
  \caption{Composition of the optimized linear combination estimators shown in the solid black points in the previous figure. The pie charts show the fractional contribution made by each individual estimator in the linear combination, calculated as $\int_{\vec{L}} w(L) a_{i}(L)$. In the pie charts, colors indicate the type of estimator and shading indicates the maximum CMB multipole used in the reconstruction. When assuming $f_b=0$, i.e. optimizing for minimum variance alone, as expected the standard QE alone is dominant. For an optimization which heavily penalizes bias using $f_b=4$, a combination of PH and SymmPH is the best combination; this combination only has a $0.5\%$ bias, although the signal-to-noise falls below 70. We may notice that for high $f_b$ PH with a high $l_{\mathrm{max}}$ is chosen, and not with a low one that we might expect to give lower bias. The reason is that at the important scales for the calculation of the CMB lensing amplitude, the PH with higher $l_{\mathrm{max}}$ is lower in bias compared to the ones with lower $l_{\mathrm{max}}$. This is because there are internal cancellations in the calculation of the bias, depending on the specific $l_{\mathrm{max}}$ (and set of simulations).} \label{fig:autoTTresults}
\end{figure}

In Figure \ref{fig:autoTTresultsblackdots} we show our results for combining estimators using only temperature data; here solid black dots show the performance of our new optimal combinations (on the x-axis we again plot the absolute value of the induced foreground bias on the amplitude of the CMB lensing power spectrum, as described in Appendix \ref{app:optdetails}. and on the y-axis the noise on the lensing amplitude). For black dots from right to left, we vary $f_b$ from $f_b=0, 0.1, 1, 2, 4$, progressively increasing the importance of the bias in the noise-bias trade-off. We can see that all the combined estimators perform significantly better than the single estimators, giving a lower bias for the same noise. Figure  \ref{fig:autoTTresults}  shows in more detail the composition of the these optimal combinations (each piechart in this figure describes the corresponding solid black dot in the previous Figure \ref{fig:autoTTresultsblackdots}).

Our reference case for $TT$ will be QE at $l_{\mathrm{max}}=3000$, for which we have a bias of $5 \%$; this estimator gives a total SNR of $80$.

When $f_b=0$, we have the minimum variance solution, which is just the standard QE. The signal-to-noise of the temperature-only reconstruction in this case is close to 100.

For $f_b=0.1$, PSH with a high $l_{\mathrm{max}}=4500$ gains importance, although there is also some contribution from other estimators. In this case bias is nearly irrelevant in our optimization, so that the non-QE estimator with the highest signal-to-noise dominates. This combination decreases the bias by a factor of $2.2$ with respect to QE at $l_{\mathrm{max}}=3000$ (although the bias is still quite large, at the level of a few percent); this reduction in bias still has a lower noise compared to the reference case of QE, with a SNR above 90.

\begin{figure}
    \centering
  \includegraphics[width=1.\columnwidth]{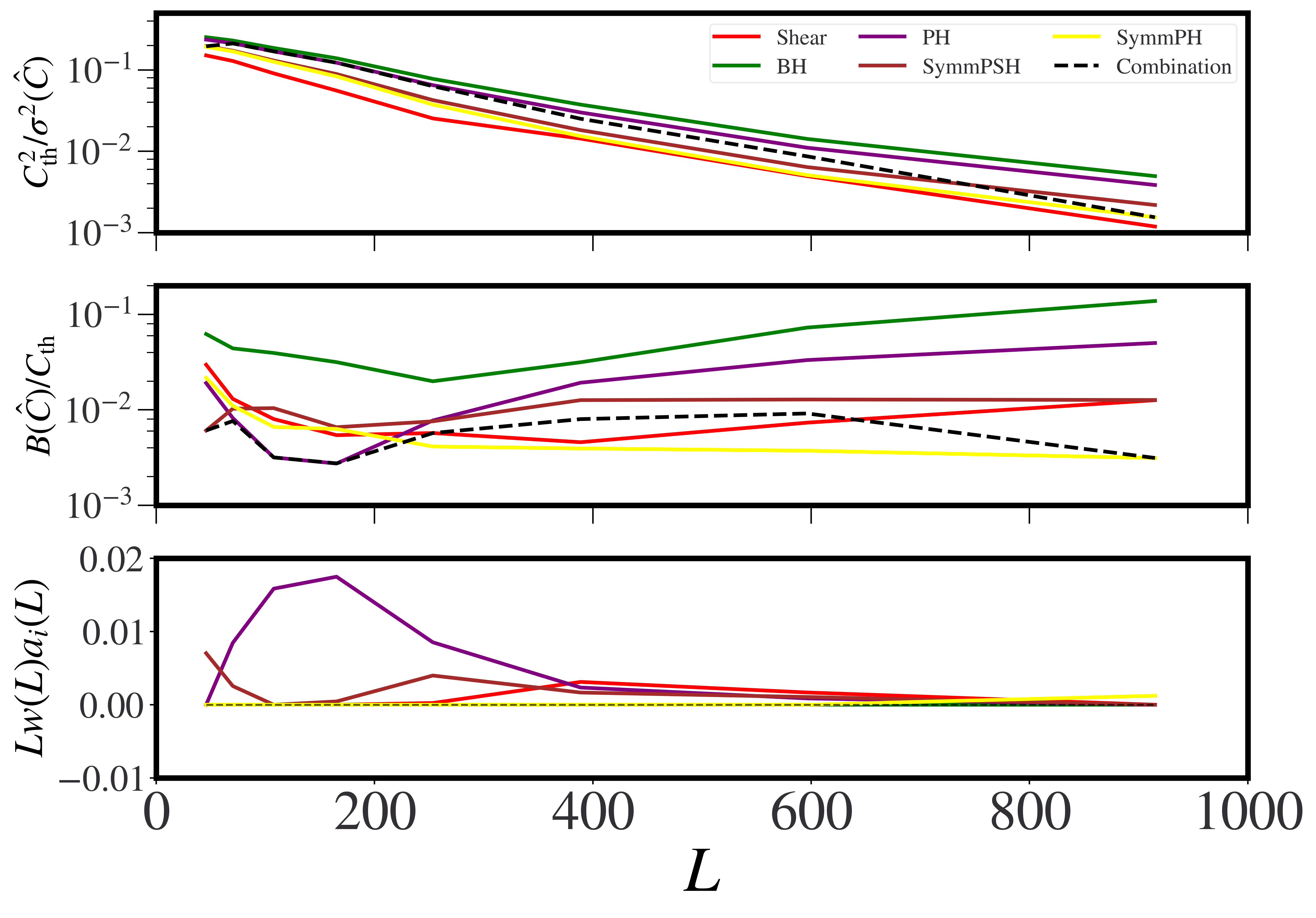}%
  \caption{More detailed illustration of the estimator linear combination for the case of temperature-lensing-only power spectrum optimization with $f_b=1$. The top panel shows the signal-to-noise ratio squared per mode for the lensing reconstruction, the middle panel shows the Gaussian-smoothed the absolute value of the total foreground bias for each estimator. On the bottom, the weights per mode are shown for each estimator; this illustrates the relative contribution of each estimator to the linear combination. It can be seen that, for $f_b=1$ where bias and variance are assigned equal importance, the optimization tries to compromise between bias and noise per mode. Most of the constraining power for the CMB lensing amplitude derives from $L \leq$ 500.} \label{fig:autoTTresults_per_L}
\end{figure}

For $f_b=1$, a case where we assign minimizing bias and variance similar priority, our solution decreases the bias by $8.6$ times compared to QE at $l_{\mathrm{max}}=3000$. The solution gives a bias of $\approx 0.6\%$, corresponding to $1/2 \sigma$, at a moderate noise cost similar to the simple QE, corresponding to an SNR of 80. In Figure \ref{fig:autoTTresults_per_L} we illustrate more clearly the contribution from each estimator for this case ($f_b=1$); in particular, we plot the signal-to-noise per mode, the total foreground-induced bias, and the contribution of each of the estimators in the combination. The solution is dominated by profile hardening, although it can be seen that there are contributions from SymmPSH on large scales and shear on small scales as well.

\begin{figure}
    \centering
    \includegraphics[width=\columnwidth]{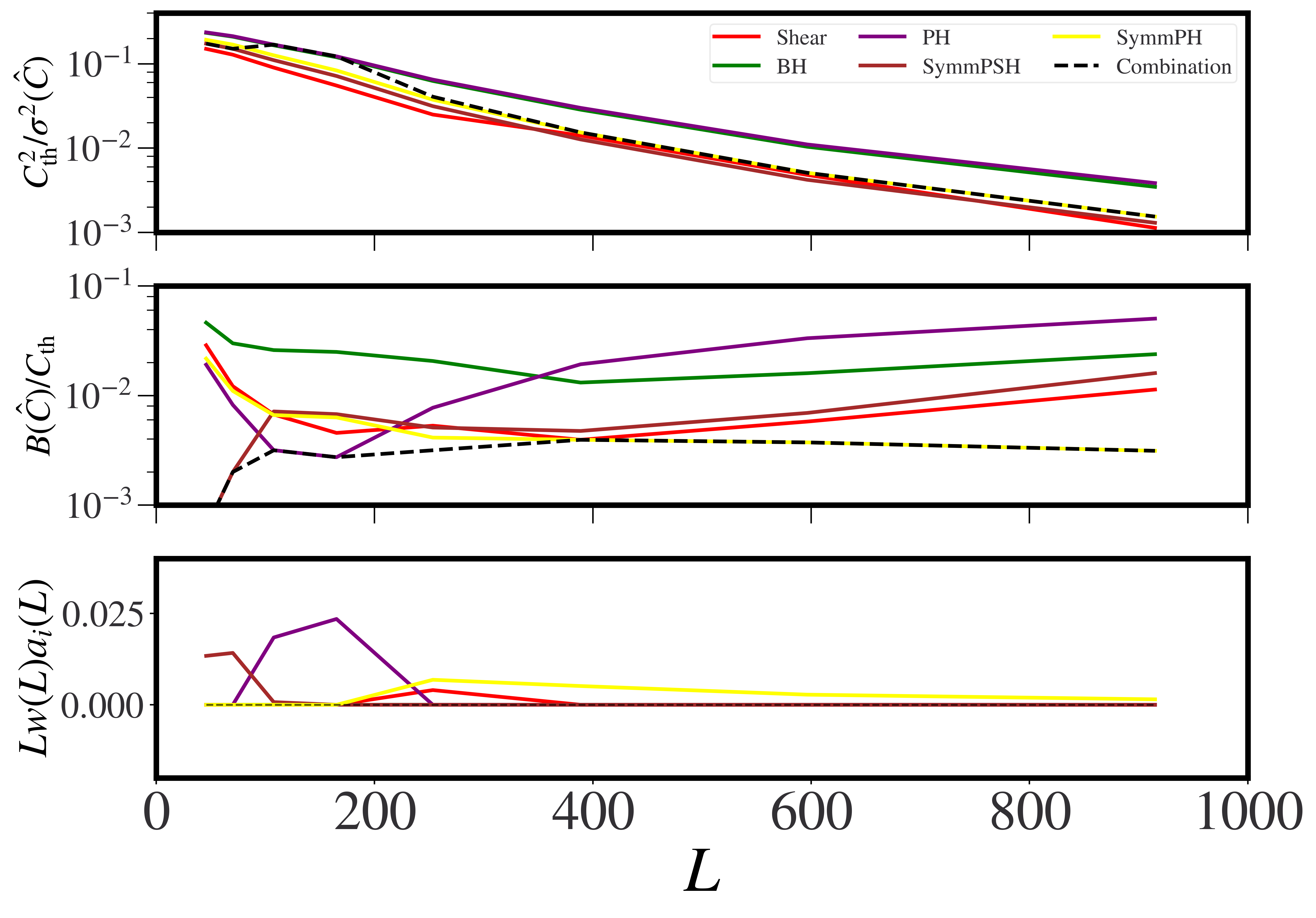}%
  \caption{As for Figure \ref{fig:autoTTresults_per_L}, but here showing results for an optimization run with $f_b=4$, i.e. the bias is assigned approximately four times more importance in the optimization than the noise. It can be seen that, as expected, this optimization effectively selects the estimator at each $L$ that has the lowest bias per mode.} \label{fig:autoTTresults_per_L_fb4}
\end{figure}

$f_b=4$ is a case where minimizing bias takes on higher priority in the optimization. In this case, again, profile hardening is one of the dominant estimators; we obtain a solution where we mix PH at a moderate-high $l_{\mathrm{max}}=4000$ with SymmPH with CIB deprojection at low $l_{\mathrm{max}}=3000$. Here we reduce the bias by a large factor $\mathcal{O}(16)$, to nearly $1/4 \sigma$ or $\approx 0.3\%$; however, this comes at a noise cost of around $18\%$, resulting in an SNR of below 70. We show the per-mode contribution from each estimator in figure \ref{fig:autoTTresults_per_L_fb4}. As indicated previously, profile hardening is the dominant estimator over the most relevant range of scales; however, on small scales there is a significant contribution from SymmPH. It can be seen that the optimizer selects the estimator with the lowest bias at each scale, which is as expected since $f_b=4$ prioritizes bias minimization.

We note that for the each choice of $f_b$ we could have multiple near-optimal configurations, in the sense that some configurations with different $l_{\mathrm{max}}$-es differ by a only few percent in the loss function. Since our per-mode plots are only for the single optimal configuration, we check that the three best-performing configurations are composed similarly. In Appendix \ref{app:cmblensingbiases} we discuss how these optimized results depend also to some extent on the masking choice.

\paragraph*{Minimum variance lensing reconstruction (temperature and polarization)}

The situation changes significantly when including polarization data alongside temperature in the lensing estimator. Our reference case for reconstruction from both temperature and polarization will be QE plus polarization, at $l_{TT, \mathrm{max}}=3000$, $l_{pol, \mathrm{max}}=5000$, for which we have a bias of $3 \%$, with an SNR of $143$. Note that in our analysis in this paper, we assume that no foregrounds are present in the polarization maps themselves. In this case, the total foreground bias is given by
\begin{equation}
     B = \vec{\alpha}\cdot\mathbf{B}\vec{\alpha} = \alpha_{TT}^2B_{TT}+2\alpha_{TT}\sum_{XY\in\mathrm{pol}}\alpha_{XY}B^{TT,XY} \label{eq:biaspolspecific}
\end{equation}
where $B_{TT}$ is the bias per mode for the $TT$ estimator lensing power spectrum, which can in turn be written as $\vec{a}\cdot B^{TT,TT} \vec{a}$. Here $B^{TT,TT}$ is the Gaussian-smoothed absolute value of the bias matrix for each temperature estimator, where we define the bias matrix as a matrix with elements $ij$ given by the bias arising from the correlation of estimator $i$ with estimator $j$; similarly, $B^{TT,XY}$ is the smoothed absolute value of the bias matrix for each temperature estimator crossed with polarization, $XY \neq TT$. Finally $\vec{\alpha}$ are minimum variance $TT$ and polarization weights, defined in Appendix \ref{app:cmblensingbiases}. We note that we do not allow for cancellations between the $TT$ contribution and the polarization contribution when calculating the total bias when using both temperature and polarization data: this is done by taking the absolute value of the bias in temperature, and the absolute value of the bias with polarization data.


We find the following results for the minimum variance lensing power spectrum.

\begin{figure}
    \centering
    \includegraphics[width=\columnwidth]{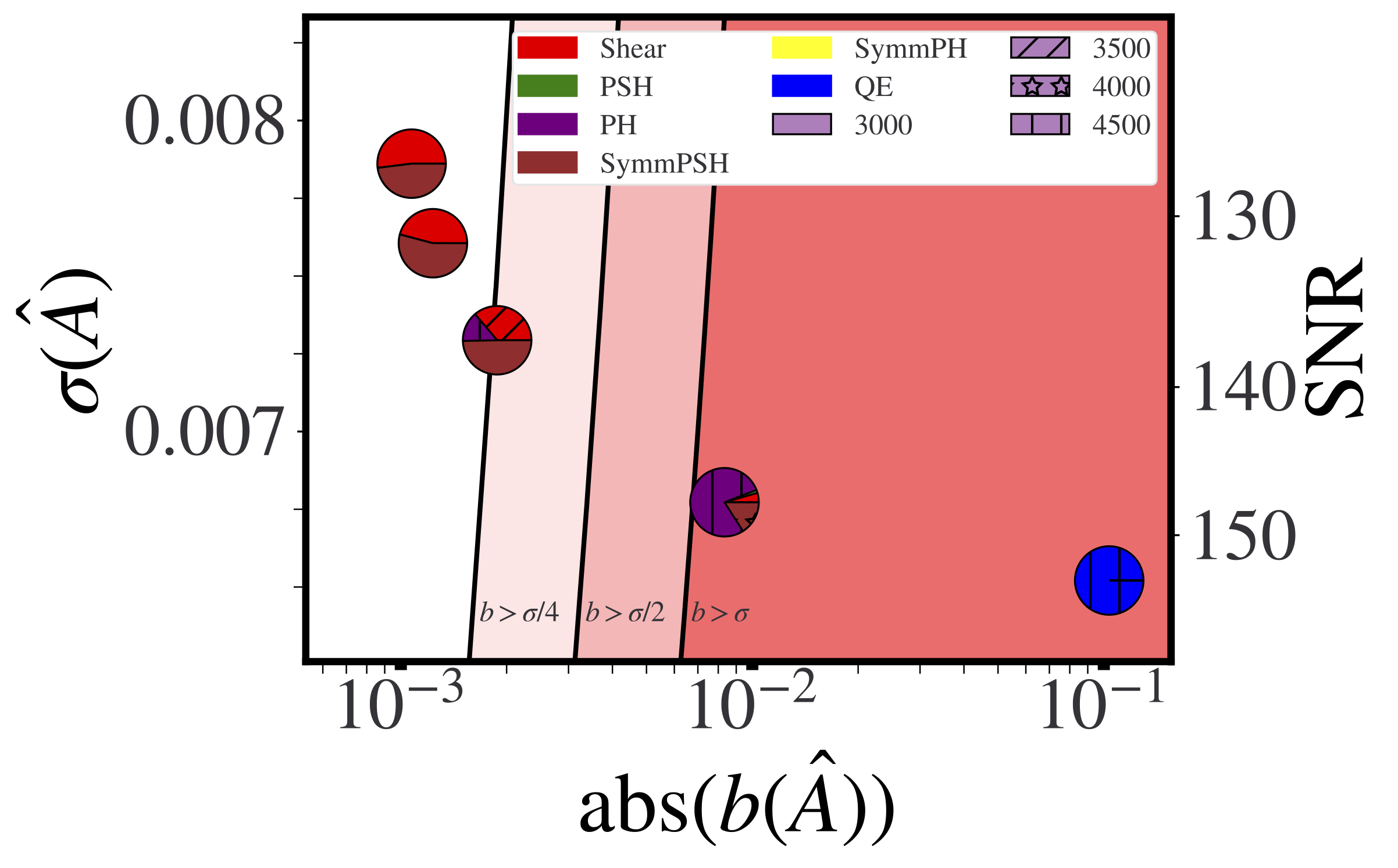}%
  \caption{Linear-combination optimization results for the lensing power spectrum as in Figure 6; however, in this plot, we consider not just temperature data but are also including polarization data. As before, the pie charts represent the contribution from each estimator, calculated as $\int_{\vec{L}} w(L) a_{i}(L)$. We see that when bias-removal is prioritized, at left, the optimization selects a combination of geometric and multi-frequency cleaning, namely Shear and SymmPSH. This is different than the result for the optimal temperature ($TT$) only combination, due to the higher importance of the primary bias term when also including polarization data as explained in the text.} \label{fig:auto_pol_results}
\end{figure}

We immediately see a change in which estimator dominates. For higher $f_b$, we see that the dominant estimator is no longer PH but instead SymmPSH (with also large contributions from shear).

\begin{figure}
    \centering
    \includegraphics[width=\columnwidth]{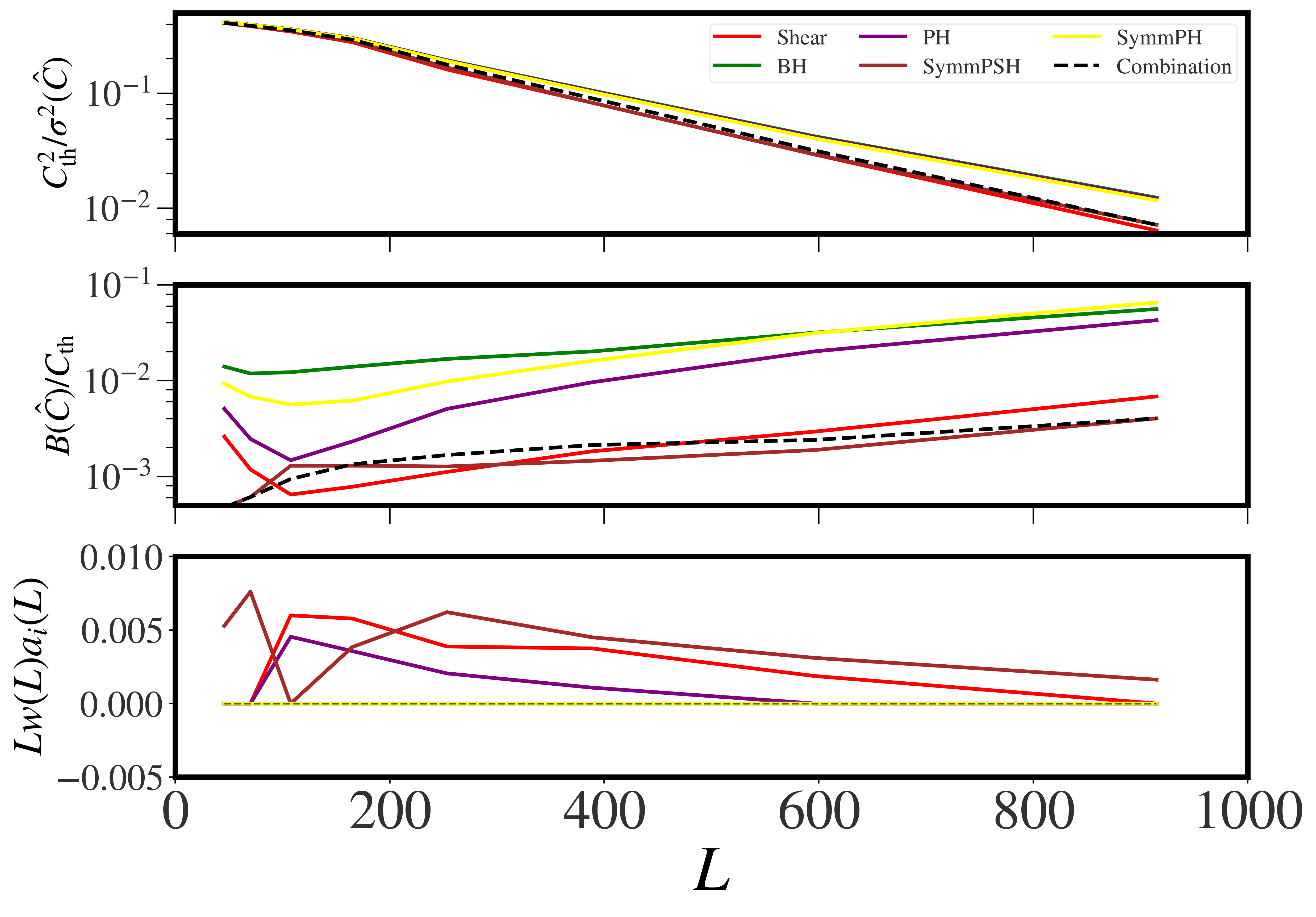}%
  \caption{Illustration of the estimator linear combination for power spectrum optimization with $f_b=1$ -- now assuming both polarization and temperature data are used in the reconstruction. The top panel shows the signal-to-noise ratio squared per mode for the lensing reconstruction, the middle panel shows the Gaussian-smoothed absolute value of the total foreground bias (equation \ref{eq:biaspolspecific}) for each estimator and the combination. In the bottom panel, the weights per mode are shown for each estimator. It can be seen that, for $f_b=1$ where bias and variance are assigned equal importance, the optimization tries to compromise between bias and noise per mode.} \label{fig:autopolresults_per_L_fb1}
\end{figure}

In particular, for $f_b=1$ we see that the SymmPSH-dominated, $l_{\mathrm{max}}=3000$, combination performs very well. We find that the bias is reduced by a factor $\mathcal{O}(17)$ with respect to QE MV with $TT$ at $l_{\mathrm{max}}=3000$, giving a bias that is well below percent level at around $0.18\%$, corresponding to just above $\sigma/4$. This bias reduction comes at a modest noise cost of only $4\%$, with the total SNR still reaching nearly 137. The contributions of the different estimators are shown in Figure \ref{fig:autopolresults_per_L_fb1} for $f_b=1$. We can see that SymmPSH dominates on smaller scales, although on large scales a mix of estimators, SymmPSH, SH and PH, is chosen.

\begin{figure}
    \centering
    \includegraphics[width=\columnwidth]{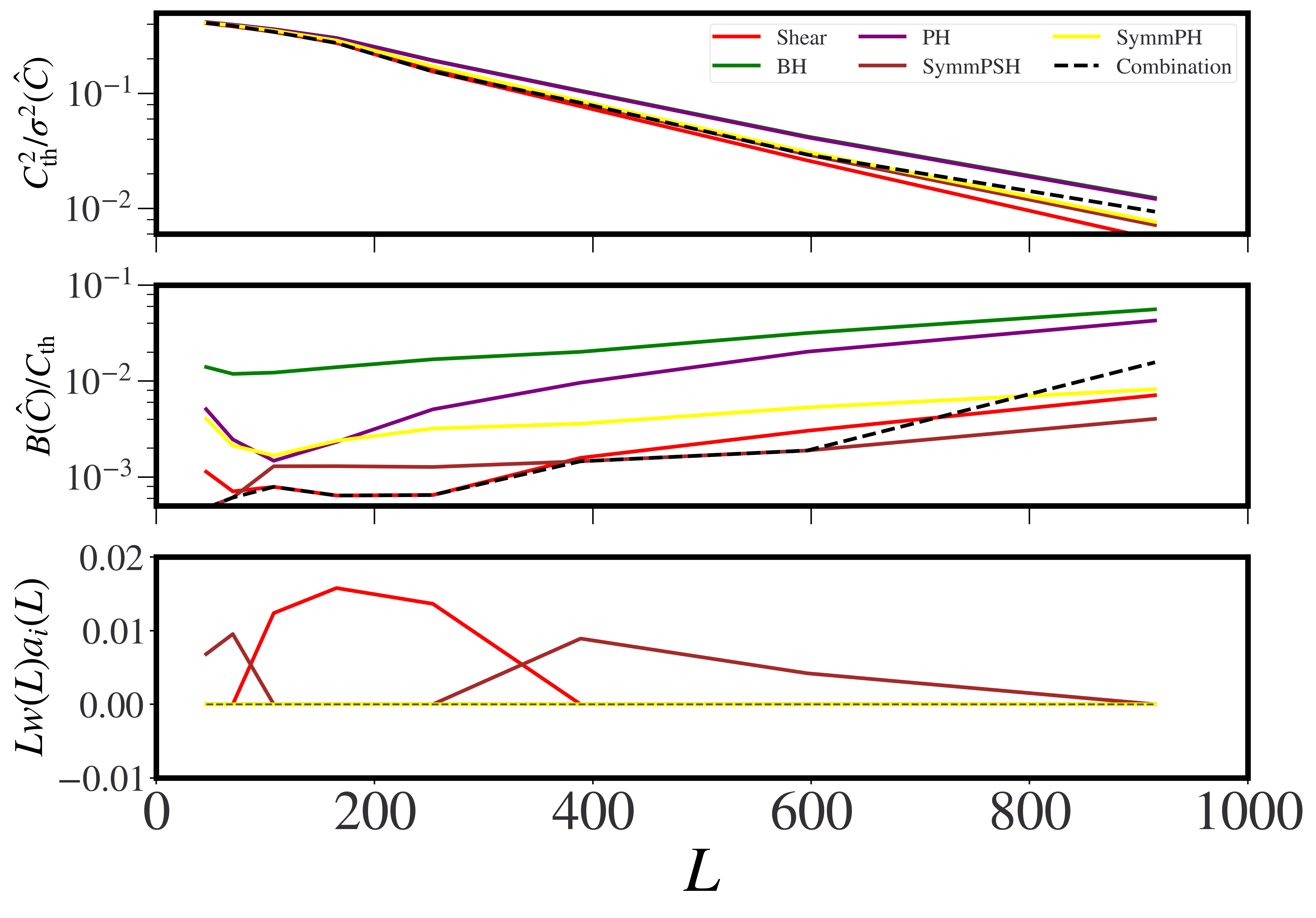}%
  \caption{As for Figure \ref{fig:autopolresults_per_L_fb1}, but now showing results for an $f_b=4$ optimisation where bias reduction is prioritized. As expected, generally, the estimator with lowest bias gets selected in the combination.} \label{fig:autopolresults_per_L_fb4}
\end{figure}

For $f_b=4$ the bias is reduced by nearly $\mathcal{O}(30)$ with respect to QE MV with $TT$ at $l_{\mathrm{max}}=3000$, giving a bias $0.1\%$ level and below $\sigma/4$. The noise cost of roughly $12\%$ is still only moderate. The contributions from different estimators are shown in Figure \ref{fig:autopolresults_per_L_fb4}; for $f_b=4$, generally the estimator with the lowest bias, dominates, although there are some scales to which this does not apply.\footnote{We note that the combined bias and the combined signal over noise squared in Figures \ref{fig:autopolresults_per_L_fb1}, \ref{fig:autopolresults_per_L_fb4} for the $TT$ plus polarization case, are not just given by combining the contributions plotted for each single estimator. The reason is that for example the combined bias in this case is $\vec{\alpha}\cdot\mathbf{B}\vec{\alpha}$, as described in equation \ref{eq:biaspolspecific}, but for each estimator, the bias plotted is $\vec{\alpha}'\cdot\mathbf{B}'\vec{\alpha}'$, where now we specify that the MV weights for the single estimator are different than those for the combined estimator. For this reason, in Figure \ref{fig:autopolresults_per_L_fb4} the bias is not always the smallest one among estimators.}

\begin{figure}
    \centering
    \includegraphics[width=\columnwidth]{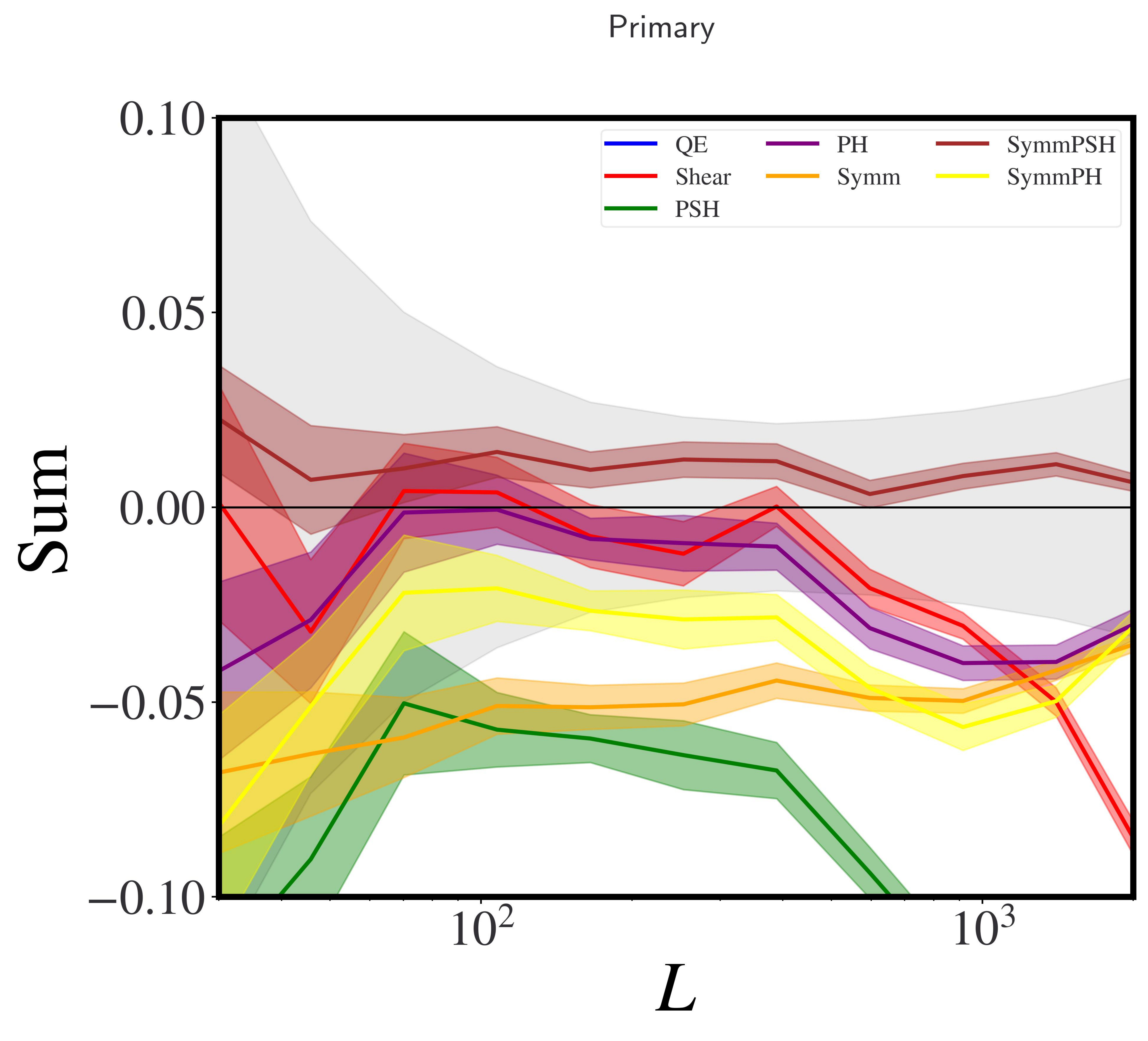}%
  \caption{The primary contribution $\langle \kappa Q_{TT+pol}[T_f, T_f] \rangle$ relative to the CMB lensing autospectrum signal for $l_{\mathrm{max}}$ for a few estimators. This type of plot becomes important when adding polarization data, as the estimators with the lowest primary bias contribution tend to become favoured in this case, in comparison to the $TT$ only case.} \label{fig:auto_pol_biases_primary}
\end{figure}

Why does the optimizer give such different results when including polarization data? The reason for this change when considering the minimum variance estimator can be understood as follows: when adding polarization data (assumed to be foreground-free), the importance of different bias contributions from each estimator changes. As explained in Appendix \ref{app:cmblensingbiases}, in this case the primary bias $\langle \kappa~ Q_{TT+pol}[T_f, T_f] \rangle$ contribution, which arises also from the temperature-polarization estimator cross-correlation, becomes more important. This can be seen in a simple computation of the bias, assuming that polarization-only spectra do not contribute:\footnote{Actually, as we explain in Appendix \ref{app:cmblensingbiases}, $B^{TT,TE}$, a secondary contraction with the $TE$ estimator, usually small on large scales, and more important on smaller scales, but where the modes are downweighted more due to the noise. For now we ignore it, although we include it in the numerical calculations. And we will completely ignore $B^{TE,TE}$ as it enters with $\alpha_{TE}^2$, and $\alpha_{TE}$ is already small.}

\begin{equation}
\begin{aligned}
 B &= \vec{\alpha}\cdot\mathbf{B}\cdot\vec{\alpha} \\
 &= \alpha_{TT}^2B_{TT}+2\alpha_{TT}\sum_{XY\in\mathrm{pol}}\alpha_{XY}B^{TT,XY} \\
 &\approx \alpha_{TT}^2B_{ TT}+2\alpha_{TT}(\alpha_{EB}B^{TT,TE}+\alpha_{EE}B^{TT,EE}
 \\ &+ \alpha_{EB}B^{TT,EB}) \\
 &= \alpha_{TT}^2 B_{TT} + \alpha_{TT}(\alpha_{EE}+\alpha_{EB}+\alpha_{TE})P_{TT} \\
 &\approx \alpha_{TT}^2 B_{TT}+\frac{\alpha_{TT}(1-\alpha_{TT})}{2} P_{TT}
\end{aligned}
\end{equation}
 where $\vec{\alpha}$ are minimum variance estimator (MV) weights, $P_{TT}$ contains only a ``primary'' bias contribution (arising from one of the bispectrum terms), and where we neglect the $TB$ estimator on all scales. Since the inclusion of polarization further up-weights the importance of the primary bias (which is dominant on large scales anyway), the estimators with the lowest primary bias contribution become the most important ones in the combination when polarization data is included. An example can be seen in Figure \ref{fig:auto_pol_biases_primary} for $l_{\mathrm{max}}=3500$; this represents a different ordering compared to the total foreground bias shown in Figure \ref{fig:biases_auto_per_L}.

\subsubsection{Cross-correlations}

\paragraph*{Temperature-only lensing reconstruction}

\begin{figure}
    \centering
    \includegraphics[width=\columnwidth]{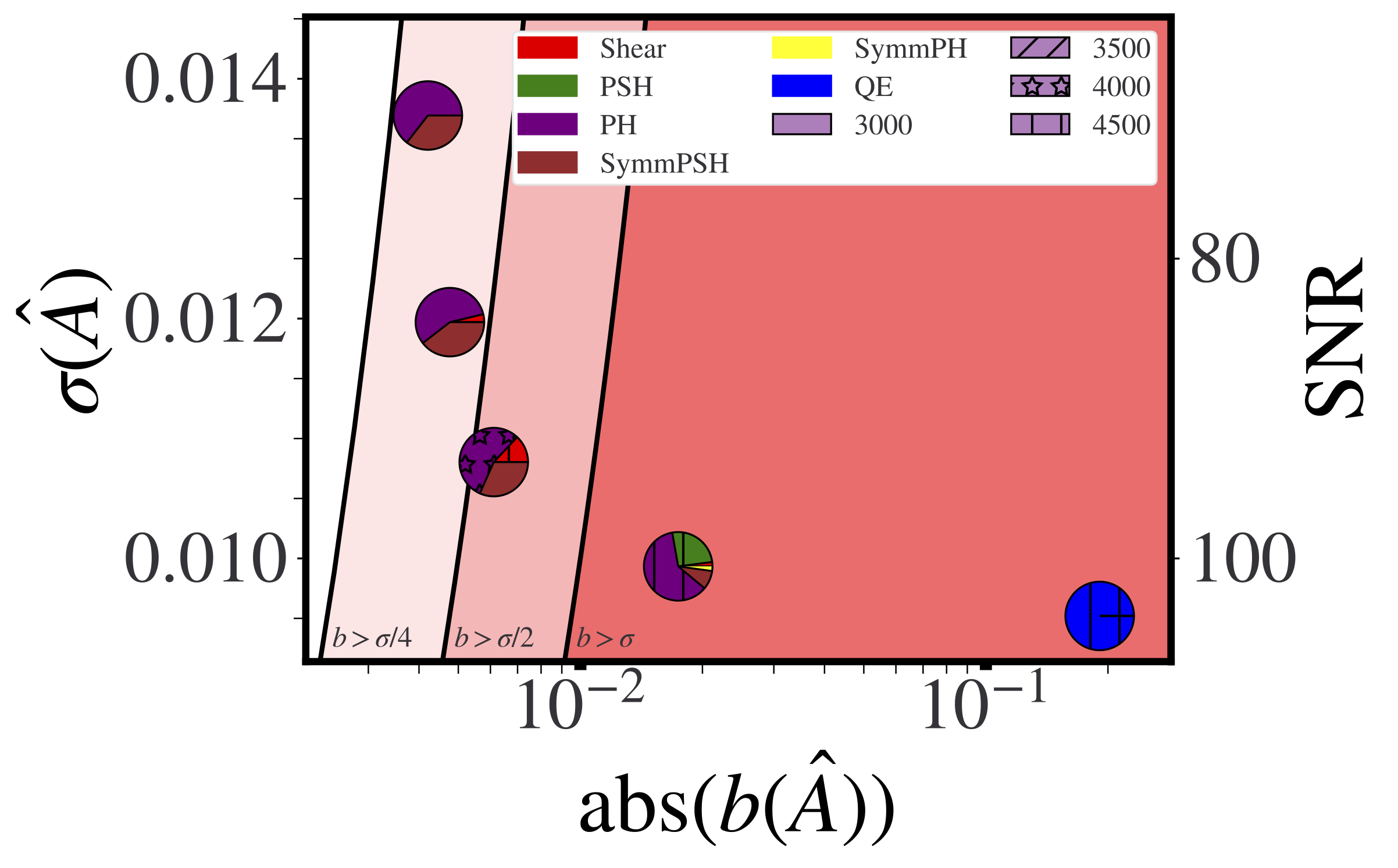}%
  \caption{Optimization results for the cross-correlation of lensing from temperature-only ($TT$) data with an LSST-like sample. The pie charts represent the contribution from each estimator, calculated as $\int_{\vec{L}} w(L) a_{i}(L)$. We see that for a high deprojection of the bias, at far left, the optimization selects a combination of geometric and multi-frequency cleaning, namely PH and SymmPSH.} \label{fig:crosscorrTTresults}
\end{figure}

We now turn to optimization results for a cross-correlation with an LSST-like sample. Our reference case for $TT$ will again be the QE at $l_{\mathrm{max}}=3000$, for which we have a bias of around $8 \%$, with an SNR of roughly $94$. Note that since the LSST galaxies do not have the same redshift distribution as the CMB lensing redshift kernel, different biases may take on a different importance; in particular, we expect that the SZ effect will become somewhat more important than CIB emission since the lower redshifts are somewhat more emphasized in this cross-correlation. We expect that this will lead to different optimization results. In Figure \ref{fig:crosscorrTTresults} we show the results for $TT$ only data.

For $f_b=1$, the bias is reduced by a factor of 12.6 (giving a sub-percent bias of around $0.6\%$), at a noise cost of around $2\%$ (still giving an SNR of more than 90), with respect to QE at $l_{\mathrm{max}}=3000$. We see that PH dominates the combination; in part, this may be because tSZ mitigation is of the highest importance in this cross-correlation. Going to higher $f_b$, PH still dominates, with some contribution from SymmPSH. For $f_b=4$ we are able to reduce the bias by a factor of $\mathcal{O}(18)$ compared to QE at $l_{\mathrm{max}}=3000$, although this comes at a price of a roughly $29\%$ increase in noise.
In the Appendix we show an example of the per mode solution for $f_b=1$.

\paragraph*{Minimum variance estimator}

Our reference case for $TT$ plus polarization will be QE plus polarization, at $l_{TT, \mathrm{max}}=3000$, $l_{pol, \mathrm{max}}=5000$, for which we have a bias of $1 \%$, with an SNR of $124$.

\begin{figure}
    \centering
    \includegraphics[width=\columnwidth]{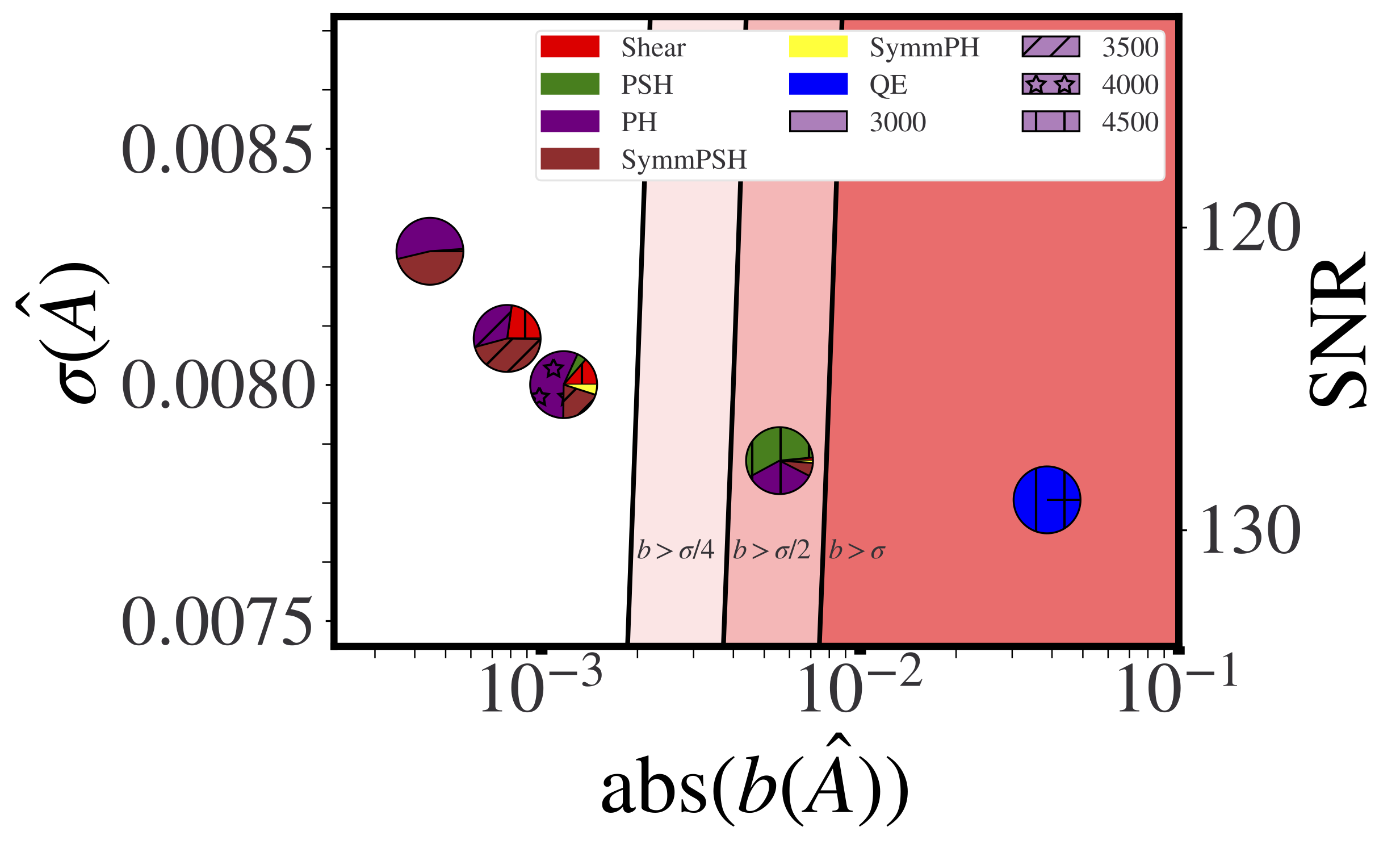}%
    \caption{Optimization results for the cross correlation of lensing from polarization and temperature data with an LSST-like sample. The pie charts represent the contribution from each estimator, calculated as $\int_{\vec{L}} w(L) a_{i}(L)$. As the primary cross-bias has no contributions from (foreground-free) polarization, we can see that for high deprojection of the bias, at far left, the results are very similar to the $TT$ only cross-correlation case.} \label{fig:crosscorrPOLresults}
\end{figure}

When adding polarization data to temperature data, we see from Figure \ref{fig:crosscorrPOLresults} that the optimization results are fairly similar, unlike for the autospectrum. The reason is that for cross-correlations, assuming no foregrounds are contained in the polarization maps, there is no contribution from the polarization estimators to the foreground-induced bias. However, due to the resulting lower total bias, is possible to increase the $l_{\mathrm{max}}$ of the temperature estimators, and still obtain small biases. For $f_b=4$, we can see that the bias is reduced by a factor of $\mathcal{O}(22)$ to a negligible level of less than $0.1\%$. This comes at only a modest, $3\%$ noise cost compared to QE MV at $l_{\mathrm{max}}=3000$, reaching an SNR of around $120$, while remaining in a regime where the bias is negligible.

\subsubsection{Simplifying the estimator combinations}

A question that arises is to what extent the complex estimator combinations considered previously can be simplified without degrading their performance.
Indeed, some of the optimal estimator combinations we derived above may be burdensome to implement in practice, as they require implementing multiple lensing quadratic estimators, applied to temperature maps from several multi-frequency linear combinations. We therefore investigate how much these complex estimators improve over simple two-estimator combinations.

In particular, we omit all but the two most-contributing estimators in an optimized linear combination, and we recalculate the weights for the two most-contributing estimators by maintaining their relative proportion in the $a(\vec{L})$ weight, and then recalculating $w(\vec{L})$ (note this is not equivalent to an optimization over two estimators).

\begin{figure}
     \centering
     \includegraphics[width=\columnwidth]{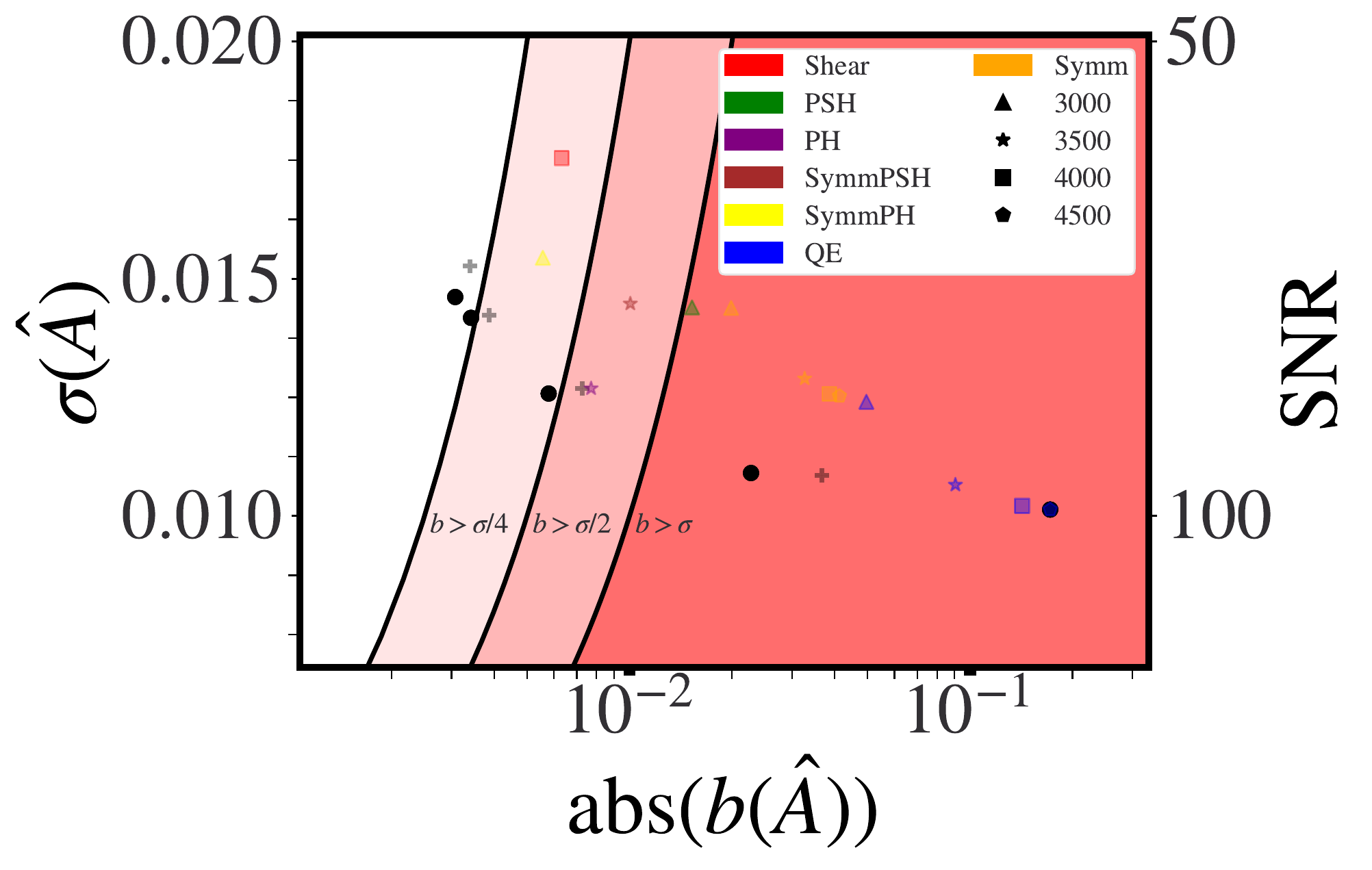}
     \subcaption{$TT$ only case.}
     \label{fig:dotsautoTT}
     \includegraphics[width=\columnwidth]{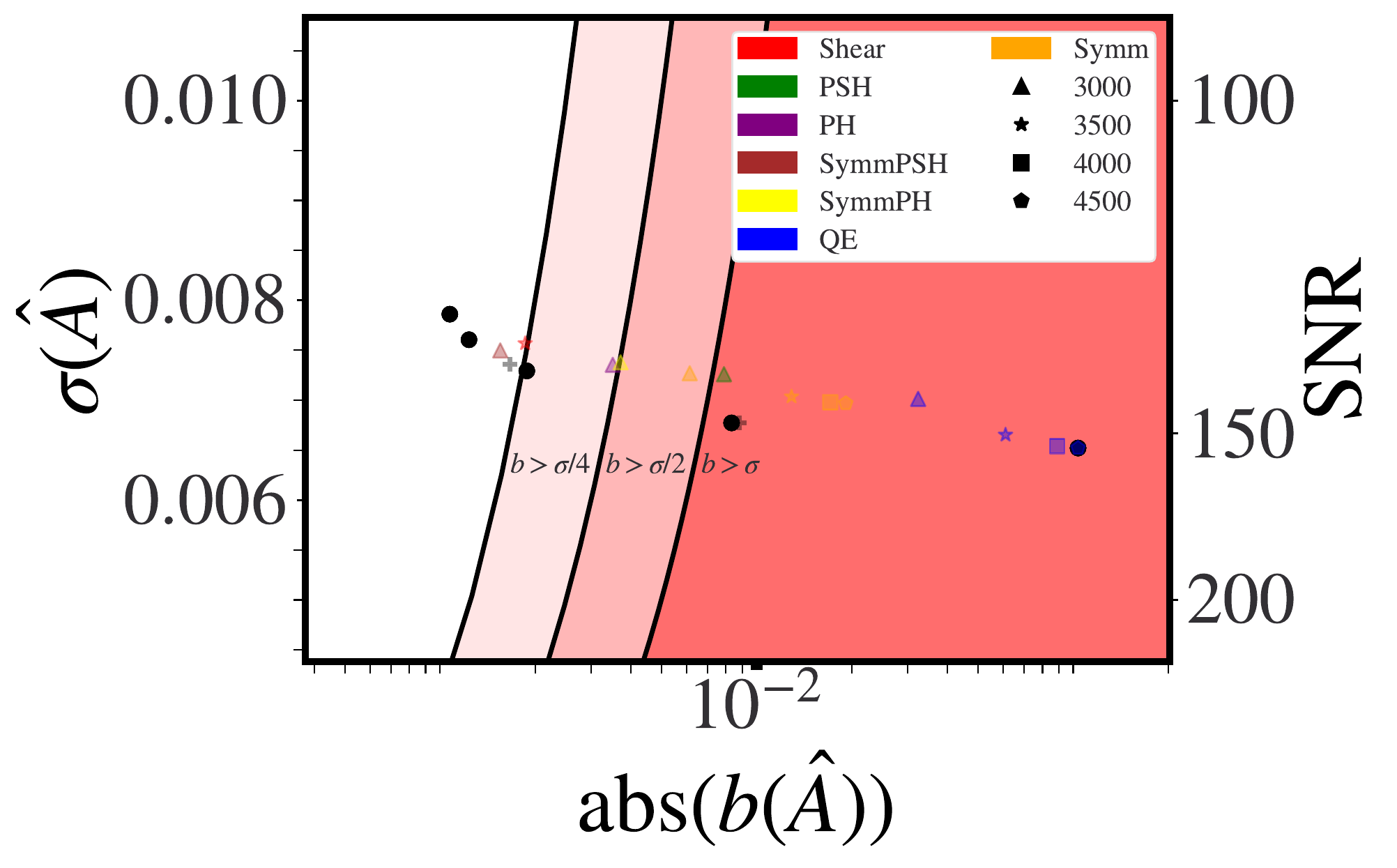}
     \subcaption{$TT$+polarization case.}
     \label{fig:dotsautopol}
     \caption*{Can the (sometimes complex) optimal linear combinations of estimators be easily simplified? Black dots show the optimal points for the CMB lensing auto-spectrum shown in previous plots. The grey cross is the shift in this optimal point, if we choose to simplify the combination by using only the two estimators which contribute most. In the top panel show the $TT$ only case, and in the bottom panel we also include polarization data. We can see that in general we have at most percent level shifts, going from black dot to grey cross, so that simplifying the estimator combinations works well.}
     \label{fig:dotsauto}
\end{figure}

\begin{figure}
     \centering
     \includegraphics[width=\columnwidth]{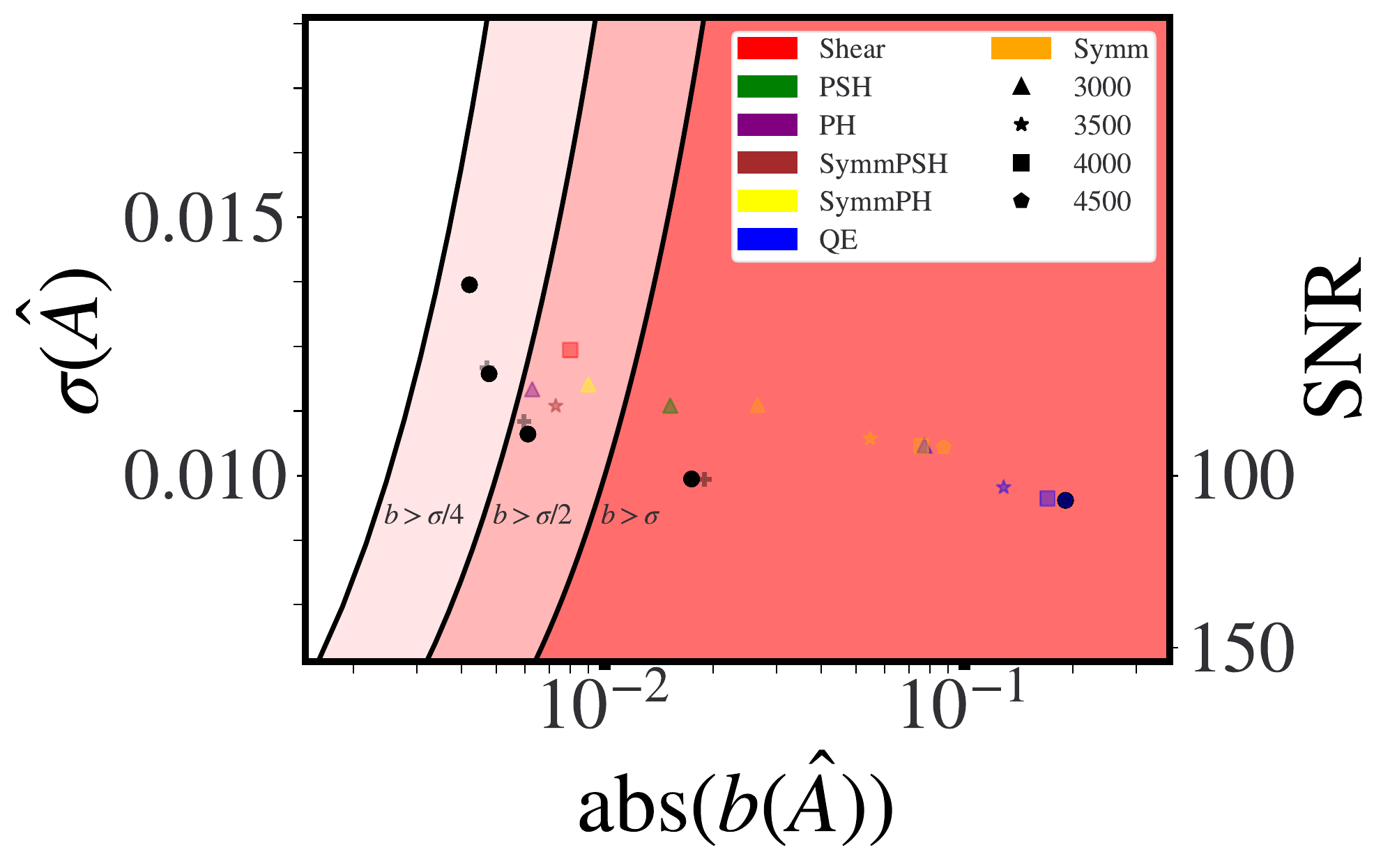}
     \subcaption{$TT$ only case.}
     \label{fig:dotscrossTT}
     \includegraphics[width=\columnwidth]{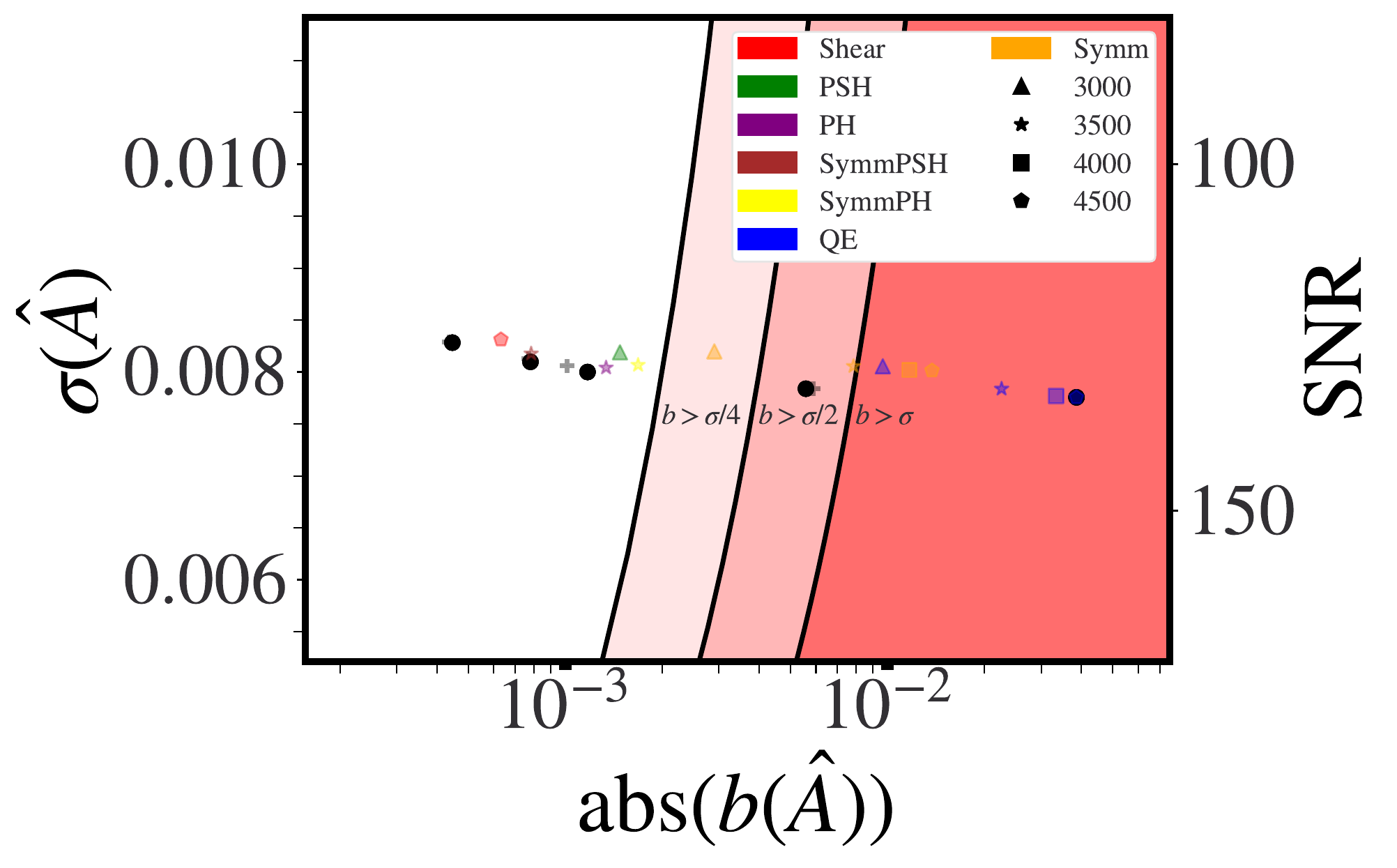}
     \subcaption{$TT$+polarization case.}
     \label{fig:dotscrosspol}
     \caption*{As for the previous plot, but for cross-correlations with LSST. We again conclude that the combined estimators can be simplified without significantly degrading their performance.}
     \label{fig:dotscross}
\end{figure}

The results are summarized in Figure \ref{fig:dotsauto} for the auto-spectrum, and Figure \ref{fig:dotscross} for the cross-spectrum. We find that in most cases, this simplification has very little impact on the estimator performance: we typically find only small, percent-level shifts in bias and noise on the amplitude (we also plot the performance of some individual estimators for comparison).

For high-$f_b$ estimators, we also find that as a general rule of thumb, these are composed from the two single estimator configurations with lowest bias. In particular, SymmPSH and PH are favoured for cross-correlations with low-z tracers, which can be understood as follows: PH is ideal for eliminating tSZ effects and SymmPSH, with tSZ deprojected, performs well at eliminating both tSZ and CIB biases; both together are sufficient to reach a regime of $b<\sigma/2$. For the auto-spectrum these two estimators similarly form a robust combination.

\section{Conclusions}
\label{sec:conclusion}
In this paper we have investigated which combination of different lensing foreground mitigation methods minimizes the foreground biases while maximizing the signal-to-noise ratio of CMB lensing power spectrum and cross-spectrum measurements.

We explore two different approaches to combine mitigation methods for this purpose.

The first one is composing (i.e., simultaneously using) geometric and multi-frequency deprojection methods, namely bias hardening applied to a symmetric estimator in which a particular foreground has been removed via multifrequency deprojection. In this way, we propose two new estimators: i) SymmPSH, given by point source hardening applied to a symmetric estimator with tSZ deprojected; ii) SymmPH, given by profile hardening applied to a symmetric estimator with CIB deprojected.

The second approach, used to further mitigate foregrounds, is to linearly combine different types of foreground-reducing lensing estimators, and minimize a loss function given by the sum of the squares of the noise and bias on the lensing amplitude. We find that when using both temperature and polarization data, assuming no intrinsic foregrounds in the polarization, the optimal combination for a high reduction in bias is given by a mixture containing mainly point source hardening applied to the tSZ-deprojecting symmetric estimator; this gives a total bias that is below a half-percent (more than an order of magnitude smaller than the QE bias), at only a modest $11\%$ increase in noise beyond the QE. For cross-correlating with an LSST-like sample, we find that the best combination for a high reduction in bias is given by a mixture of point source hardening applied to the CIB-deprojecting symmetric estimator combined with the profile hardened estimator on its own; this results in a bias to the cross-correlation amplitude that is less than $0.1\%$ (significantly smaller than the QE bias) at a cost of only a $3\%$ increase in noise.

While the exact combination depends on the observable considered, generally we find that the the most robust estimators that can be used for auto and cross-spectrum analyses consist mainly of combinations of: the profile hardened estimator, point source hardening applied to the tSZ-deprojecting symmetric estimator, and, although it has higher noise, the shear estimator (SH).

We caution that the results of this paper may have some dependence on the specific set of simulations used, although we have made efforts to perform our analysis without sensitivity to precise cancellations or other fine details. In this paper, we have focused on the simulations of \cite{Sehgal_2010}, but exploring our results with other sets of simulations, such as as Websky \cite{Stein_2020}, is well motivated. Future work could also include both extragalactic and galactic foregrounds in the polarization map simulations or -- departing from simulations altogether -- a theoretical modeling of foreground biasees. We defer such investigations to future work.

\begin{acknowledgments}
We thank Colin Hill for useful feedback on an earlier draft.
This research made use of standard python packages (numpy, scipy, matplotlib), the LensQuest code,\footnote{\url{https://github.com/EmmanuelSchaan/LensQuEst}.} symlens,\footnote{\url{https://github.com/simonsobs/symlens}}, pixell,\footnote{\url{https://github.com/simonsobs/pixell}} and mystic.\footnote{\url{https://github.com/uqfoundation/mystic}}

OD is funded by the STFC CDT in Data Intensive Science.
OD, BDS acknowledge support from the European Research Council (ERC) under the European Unions Horizon 2020 research and innovation programme (Grant agreement No. 851274). BDS further acknowledges support from an STFC Ernest Rutherford Fellowship.
N.S. is supported by the NSF. E.S. is supported by the Chamberlain fellowship at Lawrence Berkeley National Laboratory. S.F. is supported by the Physics Division of Lawrence Berkeley National Laboratory. This work used resources of the National Energy Research Scientific Computing Center, a DOE Office of Science User Facility supported by the Office of Science of the U.S. Department of Energy under Contract No. DE-AC02-05CH11231.

\end{acknowledgments}

\newpage
\onecolumngrid
\appendix

\section{CMB Lensing biases for a combination of estimators\label{app:cmblensingbiases}}

\subsection{Combination of estimators}

Suppose we observe the lensed CMB at several frequencies $\nu_i, i \in \{1,..., N_f\}$. We can exploit the frequency and scale dependence of foregrounds to build a quadratic estimator $Q(\vec{L})= \hat{\kappa}(\vec{L})$ for extracting the CMB lensing signal by minimizing the impact of such foregrounds
\begin{equation}
    Q(\vec{L})= \sum_{\nu \nu'} \int_{\vec{\ell}_1}\int_{\vec{\ell}_2} g_{\nu\nu'}(\vec{\ell}_1,\vec{\ell}_2 ) T_{\nu}(\vec{\ell}_1)T_{\nu'}(\vec{\ell}_2)|_{\vec{\ell}_1+\vec{\ell}_2=\vec{L}} \label{eq:generalproblem}
\end{equation}
where we sum over pair of frequencies, as we perform the quadratic reconstruction on a pair of CMB temperature maps observed at such frequencies, and we weight with some weighting function $g$ (that in principle might not be symmetric in $\vec{\ell}_1, \vec{\ell}_2$ and $\nu, \nu'$). This weight function depends not only on the observables, but also on choices like the maximum mode used in reconstruction $\ell_\text{max}$.

We can then impose the usual requirement of obtaining an unbiased estimate of the CMB lensing signal $\langle Q(\vec{L}) \rangle_{\mathrm{CMB}} = \kappa(\vec{L})$ (with the quadratic estimator applied only on CMB maps) and the minimization of a function $\mathcal{L}$ that depends monotonically on the variance and on the induced foreground biases. What is then the optimal form of $g$ in this case?

This is a difficult constrained optimization problem that could be simplified as follows. First, we further decompose the sum over frequencies and filters as
\begin{equation}
    Q(\vec{L}) =  \sum_{i \nu \nu'} \int_{\vec{\ell}_1}\int_{\vec{\ell}_2} g_{i}(\vec{\ell}_1,\vec{\ell}_2 ) a_{i\nu\nu'}(\vec{\ell}_1, \vec{\ell}_2) T_{\nu}(\vec{\ell}_1)T_{\nu'}(\vec{\ell}_2)|_{\vec{\ell}_1+\vec{\ell}_2=\vec{L}}
\end{equation}
where now we are summing over weighting functions $g_i$, independent from frequencies, and we absorb the frequency dependence in coefficients $a_{i\nu\nu'}$ that mix the frequency space and the CMB temperature multipoles with the CMB lensing space. Then, we make $a_{i\nu\nu'}$ seperable to obtain

\begin{equation}
    Q(\vec{L}) = \sum_{i \nu \nu'} \int_{\vec{\ell}_1}\int_{\vec{\ell}_2} a^{i}(\vec{L}) g_{i}(\vec{\ell}_1,\vec{\ell}_2 ) a^{i}_{\nu}(\vec{\ell}_1)a^{i}_{\nu'}(\vec{\ell}_2) T_{\nu}(\vec{\ell}_1)T_{\nu'}(\vec{\ell}_2)|_{\vec{\ell}_1+\vec{\ell}_2=\vec{L}}\ .
\end{equation}
In this paper we will focus on the case where we fix the CMB temperature multipole weights $a^i_\nu(\vec{\ell})$ and focus only on varying the lensing weights. Therefore, our optimization problem is to look for optimal coefficients that give an unbiased lensing estimator $Q$ that minimizes the impact of foreground biases without strong degradation in the signal over noise with the following form
\begin{equation}
    Q(\vec{L}) = \sum_{i} \int_{\vec{\ell}_1}\int_{\vec{\ell}_2} a^{i}(\vec{L}) g_{i}(\vec{\ell}_1,\vec{\ell}_2 ) T_{A}(\vec{\ell}_1)T_{B}(\vec{\ell}_2)|_{\vec{\ell}_1+\vec{\ell}_2=\vec{L}}
\end{equation}
where $T_{A}, T_{B}$ are two (possibly different) linear combinations of the individual temperature maps at each frequency. The problem of optimizing the estimator in Eq.~$\eqref{eq:generalproblem}$ has therefore been decomposed into two steps:
\begin{itemize}
    \item Choose best combination of frequency CMB temperature data, with respect to some optimisation request.
    \item Choose best combination of geometric CMB lensing methods, with respect to a similar optimisation request.
\end{itemize}
In this paper we will focus on the second problem, while \cite{sailer2021optimal} focuses on the first.

\subsection{Temperature data}

We will refer to a generic quadratic estimator $i$ that takes two temperature maps $T_{\mathrm{i1}},T_{\mathrm{i2}}$, with $\mathcal{Q}_{\mathrm{i}}[T_{\mathrm{i1}}, T_{\mathrm{i2}}](\Vec{L})$. Note, a priori that the estimator is \emph{not symmetric} with respect to its arguments. We now expand each $T_{\mathrm{ik}}$ ($\mathrm{k} \in \{1, 2\}$) as $T_{\mathrm{ik}}=T_{\mathrm{CMB}}+T_{\mathrm{f},\mathrm{ik}}+T_{\mathrm{n,ik}}$, where the first term is the lensed CMB (equal for each frequency and method of combination), the second is the foreground $\mathrm{f}$ for the map $\mathrm{i}$ in the leg $\mathrm{k}$, and the last one is the noise map for the map $\mathrm{i}$ in the leg $\mathrm{k}$. From now on, we will ignore this last term in the next calculations, keeping in mind that the experimental noise contribution (and large-scale galactic dust too) will only enter the CMB lensing filters used for reconstruction.

We now consider a set $\mathcal{S}$ of estimators. They could represent even the same estimator with different input maps (e.g. QE appearing once with maps at one frequency, and then at another frequency). From now on, any sum over the index $i$ will implicitly assume that $i\in \mathcal{S}$.

Our goal is to determine the different foreground-bias-induced contributions to a total estimator given by the combination of other estimators
\begin{equation}
    \mathcal{Q}_T(\Vec{L}) = \sum_i a_i(\vec{L}) \mathcal{Q}_i[T_{\mathrm{i1}}, T_{\mathrm{i2}}](\Vec{L})
\end{equation}
Using the bi-linearity of the quadratic estimators in their arguments
\begin{equation}
    \mathcal{Q}_T(\Vec{L})
    = \sum_i a_i(\vec{L}) \Big[  \mathcal{Q}_i[T_{\mathrm{CMB}},T_{\mathrm{CMB}}]+ \mathcal{Q}_i[T_{\mathrm{f,i1}},T_{\mathrm{CMB}}]+\mathcal{Q}_i[T_{\mathrm{CMB}},T_{\mathrm{f,i2}}]+\mathcal{Q}_i[T_{\mathrm{f,i1}},T_{\mathrm{f,i2}}] \Big]
\end{equation}
Let's now calculate the autospectrum of $\mathcal{Q}_T$, given by:
\begin{equation}
    \langle \mathcal{Q}_T(\Vec{L}) \mathcal{Q}^*_T(\Vec{L})\rangle -\langle \mathcal{Q}_T(\Vec{L}) \rangle \langle \mathcal{Q}^*_T(\Vec{L}) \rangle
    =
    \sum_{ij} a_i(\vec{L}) a^*_j(\vec{L}) \langle \mathcal{Q}_i(\Vec{L}) \mathcal{Q}_j^*(\Vec{L})\rangle\ - \langle \mathcal{Q}_T(\Vec{L}) \rangle \langle \mathcal{Q}^*_T(\Vec{L}) \rangle.
\end{equation}
To obtain this, we have to consider the spectrum obtained from crossing the CMB lensing estimator $i$ with $j$:\footnote{Note that we have dropped all terms of the form $ \langle  \mathcal{Q}_i[T_{\mathrm{f,i1}},T_{\mathrm{CMB}}] \mathcal{Q}_j[T_{\mathrm{f,j1}}, T_{\mathrm{f,j2}}] \rangle$, which indeed vanish. To see this, consider taylor expanding the lensed CMB field in powers of the convergence, schematically we have $T_\text{CMB}\sim T_\text{CMB}^0 + \kappa_\text{CMB} T_\text{CMB}^0 + \cdots$, where $T_\text{CMB}^0$ is the unlensed CMB. Since the unlensed CMB is uncorrelated with the convergence and the foregrounds, $ \langle  \mathcal{Q}_i[T_{\mathrm{f,i1}},T_{\mathrm{CMB}}] \mathcal{Q}_j[T_{\mathrm{f,j1}}, T_{\mathrm{f,j2}}] \rangle \propto \langle T_\text{CMB}^0\rangle = 0$. Likewise, all terms with three $T_\text{CMB}$'s will vanish, since the unlensed CMB has a null bispectrum.}
\begin{equation}
\begin{aligned}
    \langle \mathcal{Q}_i[T_{\mathrm{i1}}, T_{\mathrm{i2}}] \mathcal{Q}_j[T_{\mathrm{j1}}, T_{\mathrm{j2}}]\rangle &=
    \langle \mathcal{Q}_i[T_{\mathrm{CMB}},T_{\mathrm{CMB}}] \mathcal{Q}_j[T_{\mathrm{CMB}},T_{\mathrm{CMB}}]\rangle +\langle \mathcal{Q}_i[T_{\mathrm{CMB}},T_{\mathrm{CMB}}] (\mathcal{Q}_j[T_{\mathrm{f,j1}},T_{\mathrm{CMB}}] +\mathcal{Q}_j[T_{\mathrm{CMB}}, T_{\mathrm{f,j2}}]) \rangle\\ &+\langle \mathcal{Q}_j[T_{\mathrm{CMB}},T_{\mathrm{CMB}}] (\mathcal{Q}_i[T_{\mathrm{f,i1}},T_{\mathrm{CMB}}] +\mathcal{Q}_i[T_{\mathrm{CMB}}, T_{\mathrm{f,i2}}]) \rangle\\
    &+\langle  \mathcal{Q}_j[T_{\mathrm{CMB}},T_{\mathrm{CMB}}] \mathcal{Q}_i[T_{\mathrm{f,i1}}, T_{\mathrm{f,i2}}] \rangle
    +\langle  \mathcal{Q}_i[T_{\mathrm{CMB}},T_{\mathrm{CMB}}] \mathcal{Q}_j[T_{\mathrm{f,j1}}, T_{\mathrm{f,j2}}] \rangle\\
    &+ \langle \big( \mathcal{Q}_i[T_{\mathrm{f,i1}}, T_{\mathrm{CMB}}]+\mathcal{Q}_i[ T_{\mathrm{CMB}}, T_{\mathrm{f,i2}}] \big)
    \big( \mathcal{Q}_j[T_{\mathrm{f,j1}}, T_{\mathrm{CMB}}] +\mathcal{Q}_j[ T_{\mathrm{CMB}}, T_{\mathrm{f,j2}}]\big) \rangle \\
    &+ \langle \mathcal{Q}_i[T_{\mathrm{f,i1}}, T_{\mathrm{f,i2}}] \mathcal{Q}_j[T_{\mathrm{f,j1}}, T_{\mathrm{f,j2}}] \rangle.
\end{aligned}
\end{equation}
Recalling that $\mathcal{Q}_i[T_\text{CMB},T_\text{CMB}] = \kappa_\text{CMB} + \text{ noise}$, the above simplifies to:
\begin{equation}
\begin{aligned}
  \langle \mathcal{Q}_i[T_{\mathrm{i1}}, T_{\mathrm{i2}}] \mathcal{Q}_j[T_{\mathrm{j1}}, T_{\mathrm{j2}}]\rangle &=\langle \kappa_{\mathrm{CMB}}\kappa_{\mathrm{CMB}} \rangle \\ &+ \langle \kappa_{\mathrm{CMB}} (\mathcal{Q}_i[T_{\mathrm{f,i1}}, T_{\mathrm{f,i2}}]+\mathcal{Q}_j[T_{\mathrm{f,j1}}, T_{\mathrm{f,j2}}]) \rangle\\
  &+\langle \big( \mathcal{Q}_i[T_{\mathrm{f,i1}}, T_{\mathrm{CMB}}]+\mathcal{Q}_i[ T_{\mathrm{CMB}}, T_{\mathrm{f,i2}}] \big)
  \big( \mathcal{Q}_j[T_{\mathrm{f,j1}}, T_{\mathrm{CMB}}] +\mathcal{Q}_j[ T_{\mathrm{CMB}}, T_{\mathrm{f,j2}}]\big) \rangle\\
  &+ \langle \mathcal{Q}_i[T_{\mathrm{f,i1}}, T_{\mathrm{f,i2}}] \mathcal{Q}_j[T_{\mathrm{f,j1}}, T_{\mathrm{f,j2}}] \rangle,
\end{aligned}
\end{equation}
where we have neglected the Gaussian contribution from the noise, or the $N^0$ component. On the first row we have the CMB lensing signal, and on the others the foreground biases. These can be deconstructed in the following way, useful for calculation:
\begin{itemize}
    \item the second row can be viewed as a cross correlation between the CMB lensing convergence field and the map coming from applying the estimators to the foreground maps. Following \cite{Osborne, Engelen_2014, 2019SchaanFerraro}, we call this \textbf{Primary Bispectrum} term $ P_{ij}$.
    \item the third row constitutes a secondary contraction (i.e. an integral) over the primary term,
    also known as \textbf{Secondary Bispectrum}.
    Following \cite{2019SchaanFerraro} we will calculate the secondary bias as a cross correlation, by expanding $T_{\mathrm{CMB}}=T_{\mathrm{CMB},0}+T_{\mathrm{CMB},1}+\mathcal{O}(\kappa^2)$, and calculate it as
    \begin{equation}
    \begin{aligned}
    S_{ij} &= \langle
    (\mathcal{Q}_i[T_{\mathrm{f,i1}}, T_{\mathrm{CMB,0}}]+\mathcal{Q}_i[T_{\mathrm{CMB,0}}, T_{\mathrm{f,i2}} ]) (\mathcal{Q}_j[T_{\mathrm{f,j1}}, T_{\mathrm{CMB,1}}]+\mathcal{Q}_j[T_{\mathrm{CMB,1}}, T_{\mathrm{f,j2}} ])
    \rangle\\&+\langle  (\mathcal{Q}_i[T_{\mathrm{f,i1}}, T_{\mathrm{CMB,1}}]+\mathcal{Q}_i[T_{\mathrm{CMB,1}}, T_{\mathrm{f,i2}} ]) (\mathcal{Q}_j[T_{\mathrm{f,j1}}, T_{\mathrm{CMB,0}}]+\mathcal{Q}_j[T_{\mathrm{CMB,0}}, T_{\mathrm{f,j2}} ])
    \rangle
    \end{aligned}
    \end{equation}
    \item the last row comes from the autospectrum of the foregrounds only. This includes a signal part plus a Gaussian one. We subtract the latter by creating randomizing the phases in the foreground map,\footnote{Note, this is not equivalent to "Gaussianize" a field. Homogeneity and gaussianity of a field imply that real and imaginary parts of the field are independent, and the phases are randomly drawn from a uniform distribution on $[0, 2\pi[$. The converse is in general not true \cite{Leclercq:2014jda}.} with the same power spectrum as the non Gaussian ones. The difference is the \textbf{Trispectrum} term $ T_{ij} $.
\end{itemize}
In the end, for a combination of estimators the total foreground induced CMB lensing bias is
\begin{equation}
    B_{TT} = \vec{a}\cdot (\bm{T}+\bm{P}+\bm{S}) \cdot \vec{a}
\end{equation}

\subsection{Including polarization data}

A practical lensing analysis will ideally extract all the signal-to-noise available from observations, and this will include the use of polarization data to extract the lensing signal \cite{Hu:2002}
\begin{equation}
    \hat{\kappa} \equiv \mathcal{Q}_{\mathrm{MV}} = \sum_{XY \in \mathcal{E}} \alpha_{XY} \mathcal{Q}_{XY}[X, Y]
\end{equation}
such that $\sum_{XY \in \mathcal{E}} \alpha_{XY}=1$ to obtain an unbiased estimator, and $XY\in \mathcal{E}=\{TT,TE,TB,EE,EB\}$.
The combined power spectrum using temperature and polarization data is
\begin{equation}
\begin{aligned}
    C &= \vec{\alpha}^T\mathbf{C}\vec{\alpha} =  \sum_{XY, WZ} \alpha_{XY}\alpha_{WZ} C_L^{XY, WZ}\\& = \sum_{XY, WZ} \alpha_{XY}\alpha_{WZ} C_L^{\kappa\kappa} + \sum_{XY, WZ} \alpha_{XY}\alpha_{WZ} N_L^{XY,WZ}+\cdots\\& =C_L^{\kappa\kappa} + \sum_{XY, WZ} \alpha_{XY}\alpha_{WZ} N_L^{XY,WZ}+\cdots \label{eq:spectrumgeneral}
\end{aligned}
\end{equation}
If we want to minimize the Gaussian disconnected noise $N_L=\sum_{XY, WZ} \alpha_{XY}\alpha_{WZ} N_L^{XY,WZ}$  subject to the constraint of having an unbiased estimator, then we have that
\begin{equation}
    \vec{\alpha} = \frac{N^{-1}\vec{e}}{\vec{e}^TN^{-1}\vec{e}}
\end{equation}
where $\vec{e}$ is a vector of only ones, and $N^{-1}$ is the inverse, per mode, of the spectrum matrix $N^{XY,WZ}$ for $XY,WZ\in\mathcal{E}$. Correlations among estimators are negligible (e.g. for QE at most at the $10\%$ percent level, see \cite{Hu:2002}), so we neglect them when calculating the weights, and set the matrix $N$ to be diagonal. As in the main text, we have that the variance on the estimated amplitude per mode
\begin{equation}
    \sigma^2(\hat{A}(\vec{L}))=\frac{\sigma^2_{MV}}{(C_L^{\kappa\kappa})^2}\ ,
\end{equation}
and the bias on the estimated amplitude per mode becomes
\begin{equation}
    b(\hat{A}(\vec{L})) = \sum_{XY,WZ\in\mathcal{S}}\frac{\alpha_{XY}\alpha_{WZ}B_L^{XY, WZ}}{C_L^{\kappa\kappa}}= \frac{\vec{\alpha}^TB\vec{\alpha}}{C_L^{\kappa\kappa}}
\end{equation}
Our goal is to again calculate the total function
\begin{equation}
    \mathcal{L} = \sigma^2(\hat{A})+f_b^2b(\hat{A})^2
\end{equation}
The calculations are very similar to the temperature only case. The results are
\begin{multline}
     \mathcal{L} = \int_{\vec{L}} w^2(\vec{L}) \times \sum_{XY,WZ, AB,CD} \left( \frac{\Theta^{XY,WZ,AB,CD}(\vec{L})}{(C_L^{\kappa\kappa})^2} \alpha_{XY}(\vec{L})\alpha_{WZ}(\vec{L})\alpha_{AB}(\vec{L})\alpha_{CD}(\vec{L}) +\right. \\ \left. {} + f^2_{b}  w(\vec{L})\alpha_{XY}(\vec{L})\alpha_{WZ}(\vec{L}) \frac{B^{XY,WZ}(\vec{L})}{C_L^{\kappa\kappa}} \times  \int_{\vec{L'}} w(\vec{L'}) \alpha_{AB}(\vec{L})\alpha_{CD}(\vec{L}) \frac{B^{AB,CD}(\vec{L'})}{C_{L'}^{\kappa\kappa}} \right)
\end{multline}
Any $TT$ part will be given by the combination of temperature estimators (e.g. $B_{\mathrm{comb},TT}= \vec{a}\cdot B_{TT} \vec{a}$). The polarization contributions will be from a single estimator, and we will use the standard quadratic estimator on any polarization pair $XY \in \{ TE,EE,EB,TB \}$.\footnote{It should be possible to generalize the other non QE estimators, for example bias hardening, to the polarization. For simplicity we will just suppose to have QE to retain maximum signal over noise, as we assume no foregrounds for CMB polarization maps.} \footnote{We ignore $ET$.}

We now calculate the total bias per mode, as given from equation \ref{eq:spectrumgeneral}

\begin{equation}
    B = \vec{\alpha}\cdot\mathbf{B}\cdot\vec{\alpha} = \alpha_{TT}^2B_{TT}+2\alpha_{TT}\sum_{XY\in\mathrm{pol}}\alpha_{XY}B^{TT,XY}\ +\cdots,
\end{equation}
as $B^{XY, WZ} = B^{WZ, XY}$, and we therefore need $B^{XY, WZ}(\vec{L})$. To this end, we assume each map to be composed of a CMB contribution and an astrophysical foreground contribution, $X=\mathrm{CMB}+\mathrm{fg}$ (ignoring the experimental noise component, or any large-scale galactic foreground contribution). We can then write the contributions to the combined temperature and polarization power spectrum as (ignoring correlations of the type $\langle X_{CMB}ff \rangle$).
\begin{equation}
\begin{aligned}
  C^{XY, WZ}  \rightarrow   &\langle \mathcal{Q}_I[X_{\mathrm{I1}}, Y_{\mathrm{I2}}] \mathcal{Q}_J[W_{\mathrm{J1}}, Z_{\mathrm{J2}}]\rangle \\&= \langle \mathcal{Q}_I[X_{\mathrm{CMB}},Y_{\mathrm{CMB}}] \mathcal{Q}_J[W_{\mathrm{CMB}},Z_{\mathrm{CMB}}]\rangle \\&+\langle \mathcal{Q}_I[X_{\mathrm{CMB}},Y_{\mathrm{CMB}}] (\mathcal{Q}_J[W_{\mathrm{f,J1}},Z_{\mathrm{CMB}}] +\mathcal{Q}_J[W_{\mathrm{CMB}}, Z_{\mathrm{f,J2}}]) \rangle
  \\&+\langle \mathcal{Q}_J[W_{\mathrm{CMB}},Z_{\mathrm{CMB}}] (\mathcal{Q}_I[X_{\mathrm{f,I1}},Y_{\mathrm{CMB}}] +\mathcal{Q}_I[X_{\mathrm{CMB}}, Y_{\mathrm{f,I2}}]) \rangle
  \\
  &+\langle  \mathcal{Q}_J[W_{\mathrm{CMB}},Z_{\mathrm{CMB}}] \mathcal{Q}_i[X_{\mathrm{f,I1}}, Y_{\mathrm{f,I2}}] \rangle
  \\
  &+ \langle  \mathcal{Q}_I[X_{\mathrm{CMB}},Y_{\mathrm{CMB}}] \mathcal{Q}_J[W_{\mathrm{f,J1}}, Z_{\mathrm{f,J2}}] \rangle \\
  &+ \langle \big( \mathcal{Q}_I[X_{\mathrm{f,I1}}, Y_{\mathrm{CMB}}]+\mathcal{Q}_I[ X_{\mathrm{CMB}}, Y_{\mathrm{f,I2}}] \big)\big( \mathcal{Q}_J[W_{\mathrm{f,J1}}, Z_{\mathrm{CMB}}] +\mathcal{Q}_J[ W_{\mathrm{CMB}}, Z_{\mathrm{f,J2}}]\big) \rangle
  \\ &+ \langle \mathcal{Q}_I[X_{\mathrm{f,I1}}, Y_{\mathrm{f,I2}}] \mathcal{Q}_J[W_{\mathrm{f,J1}}, Z_{\mathrm{f,J2}}] \rangle
\end{aligned}
\end{equation}
where $I, J$ are shorthands to indicate an estimator for the combinations $XY, WZ$ respectively.
Furthermore, we assume that the polarization data is immune from foregrounds, under the assumption that a point source mask is enough to remove effects from polarized point sources. 
We want to focus on the part of the CMB lensing estimator that includes polarization data and gives foregrounds bia contributions, under the previous assumption. Let's set $X \in \{ T \}$, $Y\in\{T,E,B\}$ $W \in \{T, E, B \}$, $Z \in \{E, B\}$
\begin{equation}
\begin{aligned}
   C^{TY, WZ} =  C^{WZ, TY}   \rightarrow   & \langle \mathcal{Q}_I[T_{\mathrm{I1}}, Y_{\mathrm{I2}}] \mathcal{Q}_J[W_{\mathrm{J1}}, Z_{\mathrm{J2}}]\rangle\\
   &= \langle \mathcal{Q}_I[T_{\mathrm{CMB}},Y_{\mathrm{CMB}}] \mathcal{Q}_J[W_{\mathrm{CMB}},Z_{\mathrm{CMB}}]\rangle+ \\
   &+\langle \mathcal{Q}_I[T_{\mathrm{CMB}},Y_{\mathrm{CMB}}] (\mathcal{Q}_J[W_{\mathrm{f,J1}},Z_{\mathrm{CMB}}]) \rangle
   \\
   &+\langle \mathcal{Q}_J[W_{\mathrm{CMB}},Z_{\mathrm{CMB}}] (\mathcal{Q}_I[T_{\mathrm{f,I1}},Y_{\mathrm{CMB}}] +\mathcal{Q}_I[T_{\mathrm{CMB}}, Y_{\mathrm{f,I2}}]) \rangle
   \\
   &+\langle  \mathcal{Q}_J[W_{\mathrm{CMB}},Z_{\mathrm{CMB}}] \mathcal{Q}_I[T_{\mathrm{f,I1}}, Y_{\mathrm{f,I2}}] \rangle
   \\
   &+ \langle \big( \mathcal{Q}_I[T_{\mathrm{f,I1}}, Y_{\mathrm{CMB}}]+\mathcal{Q}_I[ T_{\mathrm{CMB}}, Y_{\mathrm{f,I2}}] \big) \times \mathcal{Q}_J[W_{\mathrm{f,J1}}, Z_{\mathrm{CMB}}] \rangle
\end{aligned}
\end{equation}
These correlations give the following
\begin{equation}
    \langle  \kappa_{\mathrm{CMB}}\kappa_{\mathrm{CMB}} \rangle + \langle \kappa_{\mathrm{CMB}}  \mathcal{Q}_I[T_{\mathrm{f,I1}},  Y_{\mathrm{f,I2}}] \rangle
    +  \langle \big( \mathcal{Q}_I[T_{\mathrm{f,I1}}, Y_{\mathrm{CMB}}]+\mathcal{Q}_I[ T_{\mathrm{CMB}}, Y_{\mathrm{f,I2}}] \big) \times \mathcal{Q}_J[W_{\mathrm{f,J1}}, Z_{\mathrm{CMB}}] \rangle
\end{equation}
Now, the only way to have a primary bispectrum term is if $Y\in\{T\}$, given our assumptions, and for our case this is simply
\begin{equation}
     \frac{1}{2}P_{TT, I}=\frac{1}{2}\sum_i a_i  \mathcal{Q}_i[T_{\mathrm{f,i1}},  T_{\mathrm{f,i2}}] = \frac{1}{2}\sum_i a_i  P_{ii}
\end{equation}
where $P_{ii}$ is the primary that we have already calculated from $TT$ data for the temperature $i$ estimator.

\begin{figure}
    \centering
    \includegraphics[width=0.7\columnwidth]{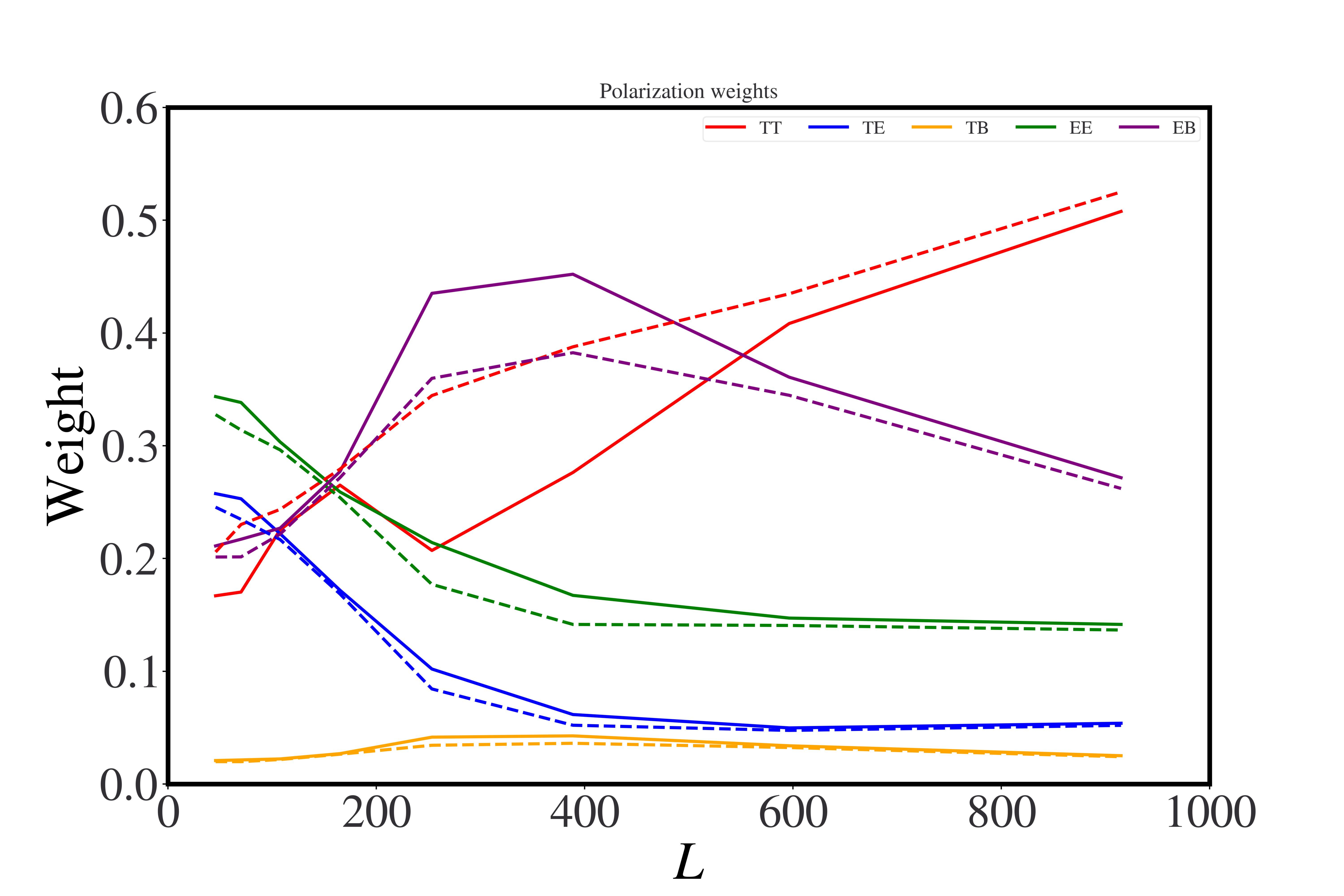}%
  \caption{Polarization weights change when accounting for the presence of biases. The solid line represents the minimum variance weights for each polarization estimator when including the presence of biases in the optimization, with $f_b=4$. The dashed line is the minimum variance weight when we do not consider the presence of biases. We can see that the combined $TT$ estimator becomes less important when including biases, and the $EB$ one is favored.} \label{fig:polweights}
\end{figure}

For the secondary bispectrum part, $W\in\{T\}$, as we assume no foregrounds in polarization maps. If $Y\in\{T\}$, then we have $S^{TT,TE},S^{TT,TB}$, and this is (we consider only $TE$)\footnote{If $Y\in\{E,B\}$, $S^{TE,TE},S^{TB,TB}, S^{TE,TB}, S^{TB,TE}$ we still have contributions, that we here ignore, as the polarization weight for $TB$ is practically negligible, and for $TE$ becomes not too important when squared.}
\begin{equation}
  S^{TT,TE} = \langle \big( \mathcal{Q}_I[T_{\mathrm{f,I1}}, T_{\mathrm{CMB}}]+\mathcal{Q}_I[ T_{\mathrm{CMB}}, T_{\mathrm{f,I2}}] \big) \times \mathcal{Q}_J[T_{\mathrm{f,J1}}, E_{\mathrm{CMB}}] \rangle
\end{equation}
and we use again the trick to calculate the secondary by decomposing $X_{\mathrm{CMB}}=X_{\mathrm{CMB}}^0+X_{\mathrm{CMB}}^1,\ X \in \{T,E\}$ in powers of $\kappa.$ To recap, we obtain a symmetric bias matrix $B^{XY,WZ}$ whose components are\footnote{
We ignore the secondary contraction for $TETE$ as it enters with a small weight of $\alpha_{TE}^2$. We do not consider a secondary for $TB$. Even setting $B^{TT,TB}=0$ should be fine, given the negligible weight for $TB$ in the $MV$ combination.}

\begin{equation}
    B^{TT,TT} = \vec{a}\cdot B_{TT, TT-\mathrm{data}}\cdot \vec{a}
\end{equation}
\begin{equation}
    B^{TT,EB} = B^{TT,EE} =  \frac{1}{2}\sum_i a_i  P_{ii}
\end{equation}
\begin{equation}
    B^{TT,TE} =  \frac{1}{2}\sum_i a_i  P_{ii}  + S^{TT,TE}
\end{equation}
\begin{equation}
    B^{TT,TB} =  \frac{1}{2}\sum_i a_i  P_{ii}
\end{equation}
For the optimization, we then take its absolute value and smooth it with a Gaussian as with the $TT$ case, and calculate
\begin{equation}
    B_{TT+\mathrm{pol}} = \vec{\alpha}\cdot \bm{B} \cdot\vec{\alpha}
\end{equation}
where $B$ has components $B^{XY,WZ}$. In Figure \ref{fig:polweights} we show how an example of how minimum variance weights when combining temperature and polarization data change when including bias contributions.\footnote{To calculate the Gaussian lensing noise from polarization data, we use a minimum variance noise from the SO polarization noise frequencies.}
Finally, given our assumptions, for the cross-correlation we do not expect to have a significant foreground bias arising from the polarization estimator contributions.

\section{Composition of CMB lensing estimators\label{app:cmblensingcomposition}}

In this section we discuss how we compose, i.e. simultaneously apply, the symmetric estimator and the bias hardening operation. First, we will review the symmetric estimator. Then, we review the bias hardening operation.

\subsection{The symmetric estimator}

The symmetric estimator is built on a foundation of the gradient cleaning method \cite{Madhavacheril:2018}. The idea behind the gradient cleaned estimator is the following. The standard quadratic
estimator for CMB lensing estimation can be obtained from the divergence of $T_{\mathrm{filt,1}}\vec{\nabla}T_{\mathrm{filt,2}}$. $T_{\mathrm{filt,1}}$ is
an inverse variance weighted CMB map. On the other hand, $T_{\mathrm{filt,2}}$ is a Wiener
filtered CMB map useful for the estimation of the gradient of the unlensed CMB map. Because
of diffusion damping, the information coming from the latter saturates after CMB scales at $l \sim 2000$ \cite{Hu2007}. The intuition behind the gradient cleaned estimator is then to use a noisier foreground-free cleaned
CMB map $T_{\mathrm{depr,2}}$ in the gradient leg, $T_{\mathrm{filt,1}}\vec{\nabla}T_{\mathrm{depr,2}}$. This results in cleaned cross-correlations with matter tracers, or auto-correlations without trispectrum or primary bispectrum terms \cite{Madhavacheril:2018}.

Now, let's write the temperature maps as the superposition of two sets of modes $T = T_{\mathrm{low}} + T_{\mathrm{high}} $, where $\mathrm{low}$ indicates modes below some scale $l_{c}$, and $\mathrm{high}$ otherwise. The problem of the gradient cleaned estimator is that it misses the
$\mathrm{high}-\mathrm{high}$ CMB lensed multipole correlations. This results in a high noise on large scales in the reconstructed CMB lensing map. The symmetric estimator in \cite{Darwish2020} recovers these modes, by creating a linear combination of gradient cleaned estimators, by taking the divergence of $aT_{\mathrm{filt,1}}\vec{\nabla}T_{\mathrm{depr,2}}+bT_{\mathrm{depr,2}}\vec{\nabla}T_{\mathrm{filt,1}}$. This can also be written in Fourier space as $\hat{\kappa}_{\mathrm{symm}}=\int_{\vec{l}}g_{\mathrm{symm}}T_1({\vec{l}})T_{\mathrm{depr,2}}(\vec{L}-\vec{l})$, with formulae that can be found in \cite{Darwish2020}. The result is a much lower noise-cost in the symmetrised CMB lensing map gaussian noise, compared to gradient cleaning.

\subsection{Bias Hardening}

Bias hardening \cite{Namikawa_2013} methods have already been explored in detail; see for example \cite{sailer_paper1}. Here, we  briefly summarize the idea.

If the total temperature map is $T=T_{\mathrm{CMB}}+s$, where $s$ is a foreground source, we have that

\begin{equation}
    \langle T(\vec{l})T(\vec{L}-\vec{l}) \rangle = f^{\kappa}(\vec{l}, \vec{L}-\vec{l})\kappa(\vec{L}) +f^{s}(\vec{l}, \vec{L}-\vec{l}) f(\vec{L})
\end{equation}

for a statistical average with the CMB lensing convergence field and foreground fixed at some mode $\vec{L}$.

Therefore, the estimator for the CMB lensing convergence field, say with $T_A,T_B$ temperature maps

\begin{equation}
    \hat{\kappa}(\vec{L}) = \int_{\vec{\ell}} T_A(\vec{\ell})T_B(\vec{L}-\vec{\ell})g_{AB}(\vec{\ell}, \vec{L}-\vec{\ell}),
\end{equation}

picks up some foreground contribution \cite{Namikawa_2013}.

The idea behind bias hardening is to write an unbiased estimator for the foreground source

\begin{equation}
    \hat{s}(\vec{L}) = \int_{\vec{\ell}} T_A(\vec{\ell})T_A(\vec{L}-\vec{\ell})g_{s}(\vec{\ell}, \vec{L}-\vec{\ell}) \ .
\end{equation}

One can then write (we omit the argument $\vec{L}$)

\begin{equation}
\begin{pmatrix} \langle \hat{\kappa} \rangle \\ \langle \hat{s} \rangle \end{pmatrix} = \begin{pmatrix}
1 & R^{s \kappa}\\
R^{\kappa s} & 1
\end{pmatrix}\begin{pmatrix} \kappa \\ s \end{pmatrix}
\end{equation}

where $R^{ab}$ is the response of the estimator $a$ on the field $b$, $R^{ab}=\int_{\vec{l}} g_a(\vec{l}, \vec{L}-\vec{l})f_b(\vec{l}, \vec{L}-\vec{l})$. Inverting this system it then possible to write a weighting function for the bias hardened estimator\footnote{Another way to write the estimator, to easier reconnect to the CMB lensing literature, is the following. Set $g_{AB}=A_{\kappa}F_{\kappa}$, with some normalisation $A_{\kappa}$, and  $g_{s}=A_{s}F_{s}$, with some normalisation $A_s$. Then, $F_{BH} = F_{\kappa}-F_{s}A_{s}/A_{\kappa f}$, where $A_{\kappa s}=\left[\int_{\vec{l}} F_{\kappa} f_s\right]^{-1}$. The normalization is $A_{BH} = \left[\int_{\vec{l}}F_{BH} f_{\kappa}\right]^{-1}$.}

\begin{equation}
    g_{BH,AB} = \frac{1}{1-R^{s\kappa}R^{\kappa s}}(g_{AB}-R^{s\kappa}g_s)
\end{equation}

For composing bias hardening with the symmetric estimator, we just then set $g_{AB} \equiv g_{\mathrm{symm}}$, applying it on $T_A, T_B = T, T_{\mathrm{depr}}$, a map with the foreground, and another without it respectively.\footnote{Note that $g_s$ was built to act on $T_A, T_A$ only, and not on the deprojected map too. Therefore, when $g_s$ is applied on $T_A, T_B\ T_A \neq T_B$, it is a slighlty suboptimal estimator in terms of variance for the foreground source, although it is still capable of detecting it for removal.}

\section{Optimization explorations}

In this section we present a few details about the optimization results presented in the main text.

\subsubsection{Per mode results}
\begin{figure*}
    \centering
    \includegraphics[width=0.7\columnwidth]{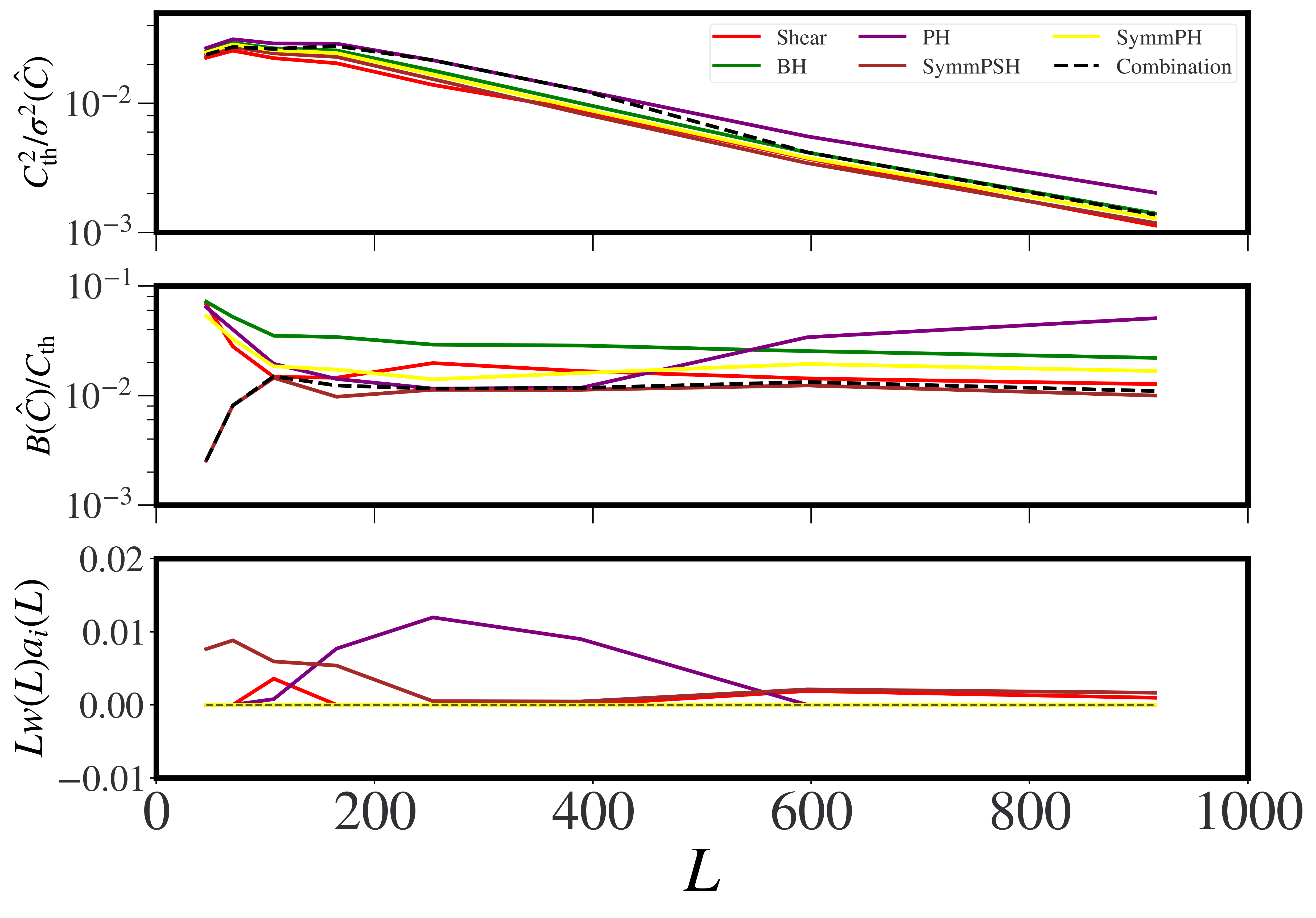}%
  \caption{Per mode plot for the case of $f_b=1$ for the cross power TT only data optimization. On the top plot we have the signal over noise squared per mode from $TT$, then a filtered version, with a Gaussian filter, of the absolute value of the total foreground bias for each estimator. On the bottom, the weights per mode for each estimator.} \label{fig:crossTTresults_per_L}
\end{figure*}

\begin{figure*}
    \centering
    \includegraphics[width=0.6\columnwidth]{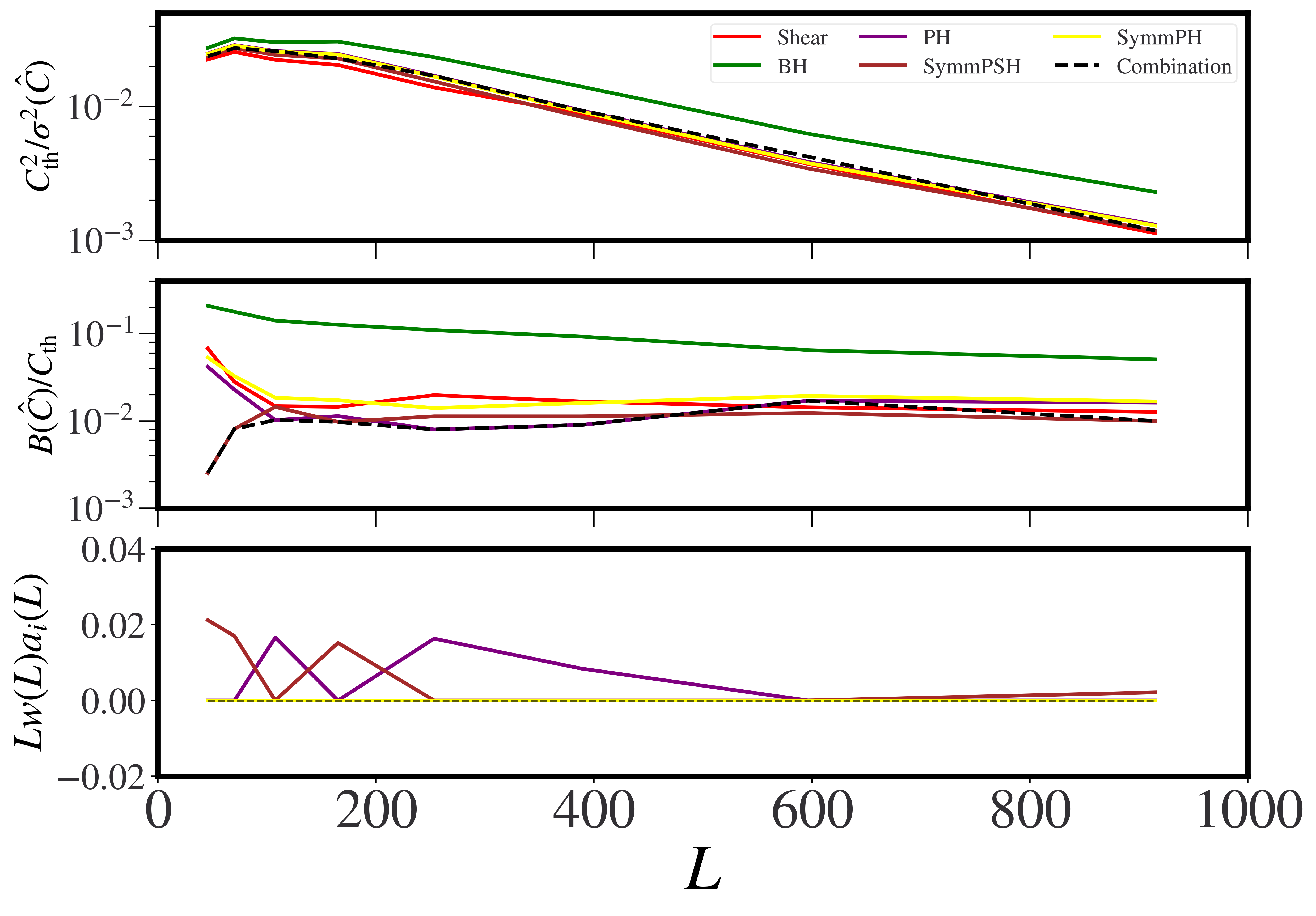}%
  \caption{Per mode plot for the case of $f_b=4$ for the cross power TT only data with an LSST like sample optimization. On the top plot we have the signal over noise squared per mode from $TT$, then a filtered version, with a Gaussian filter, of the absolute value of the total foreground bias for each estimator. On the bottom, the weights per mode for each estimator.} \label{fig:crossTTresults_per_L_fb4}
\end{figure*}

\begin{figure}
    \centering
    \includegraphics[width=0.6\columnwidth]{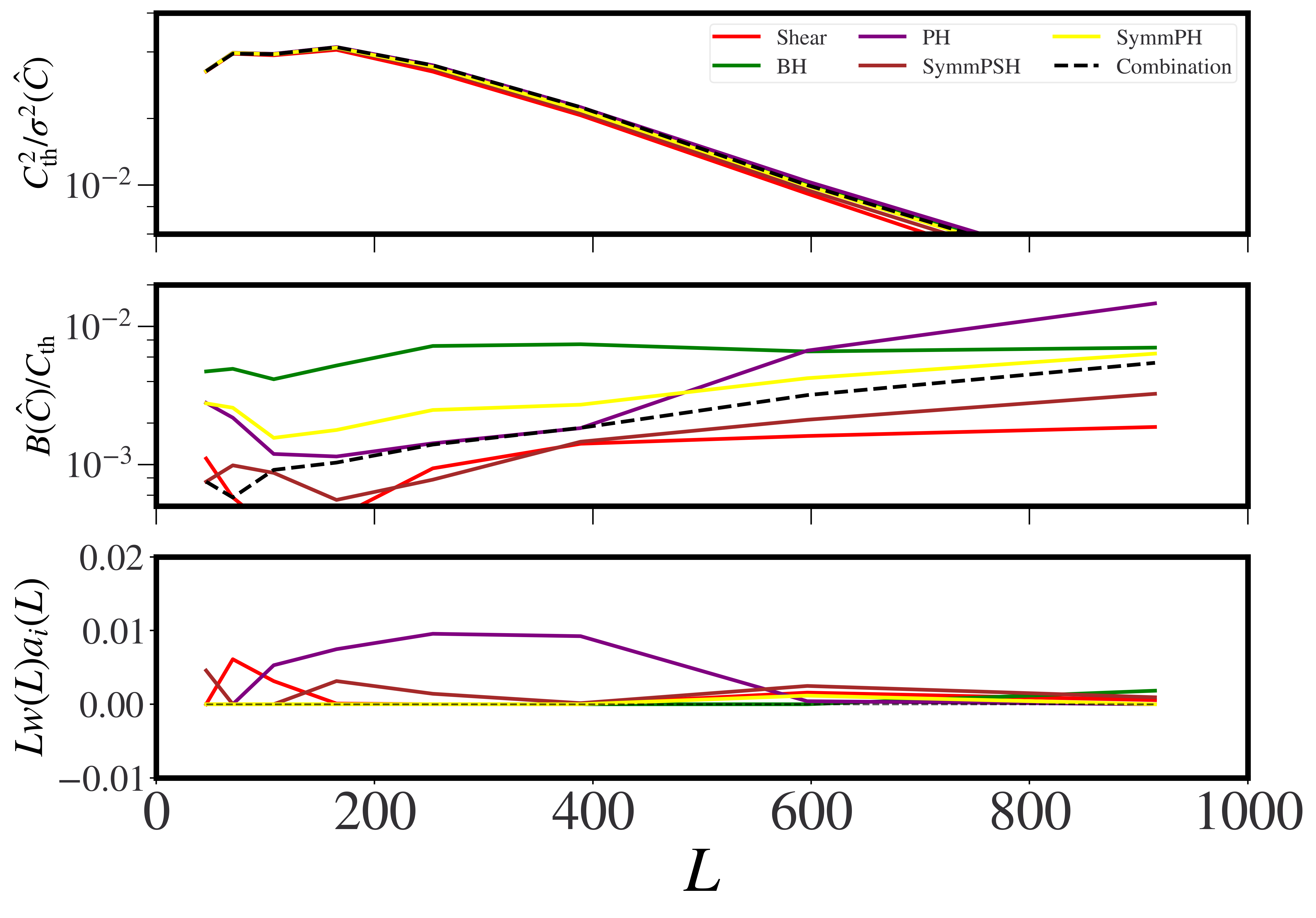}%
  \caption{As for Figure \ref{fig:autopolresults_per_L_fb1}, but now showing the case of $f_b=1$ for the cross power $TT$ plus polarization data optimization. It  can  be  seen  that,  for $fb = 1$  where bias and variance are assigned equal importance, the optimization  tries  to  compromise  between  bias  and  noise per mode.} \label{fig:crosspolresults_per_L_fb1}
\end{figure}

\begin{figure}
    \centering
    \includegraphics[width=0.6\columnwidth]{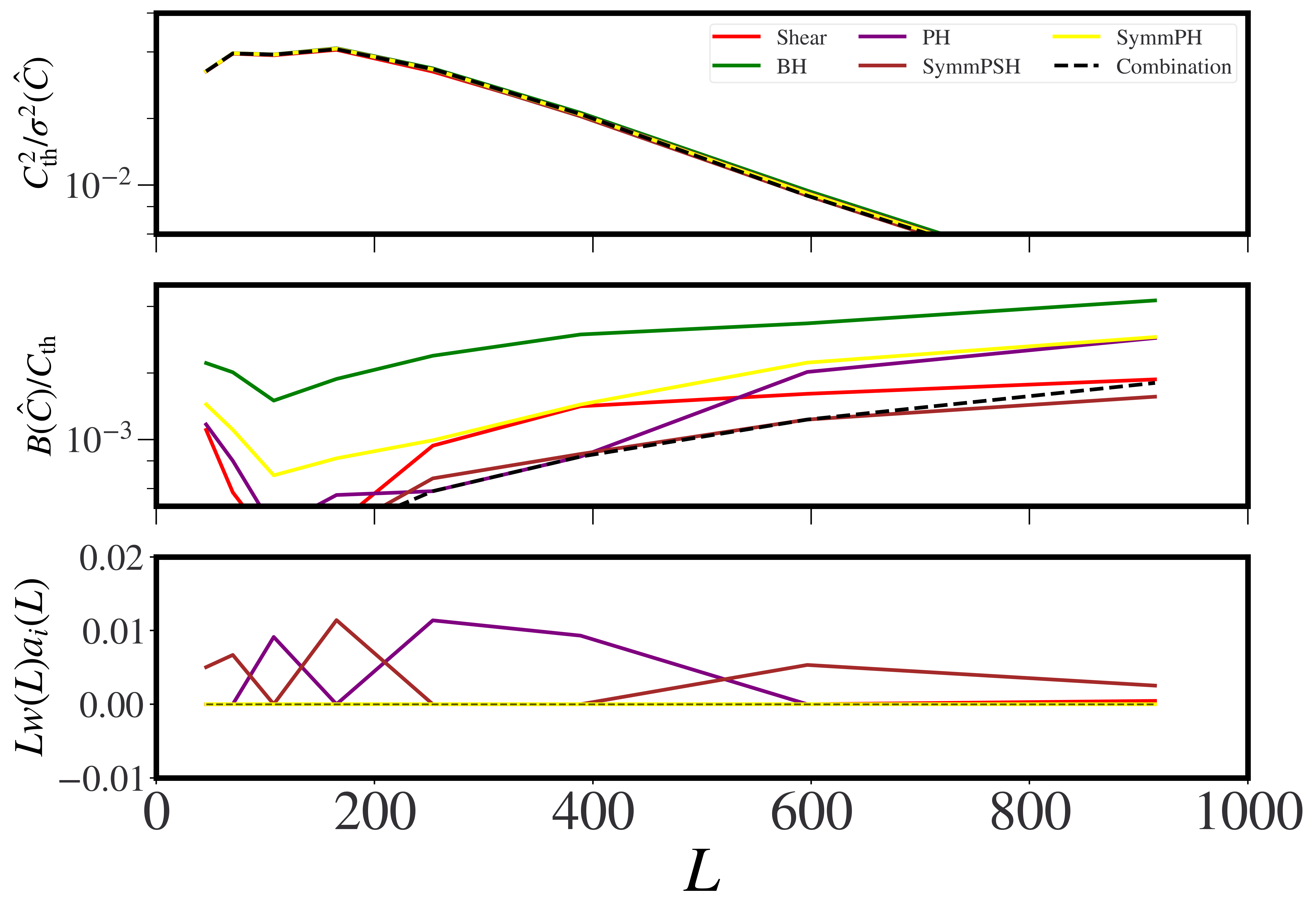}%
  \caption{As for Figure \ref{fig:autopolresults_per_L_fb1}, but now showing the case of $f_b=4$ for the cross power $TT$ plus polarization data optimization. It  can  be  seen  that,  for $fb = 4$   generally,  the  estimator  with  lowest  bias gets selected in the combination.} \label{fig:crosspolresults_per_L_fb4}
\end{figure}

\subsubsection{Dependence of biases and optimization on the point source mask}

\begin{figure}
  \centering
  \begin{minipage}{0.45\textwidth}
    \includegraphics[width=\textwidth]{Figures/configILCall_piechart_fbs_abs.png}
    \caption{Optimization results for the auto correlation for TT data. The pie charts represent the contribution from each estimator, calculated as $\int_{\vec{L}} w(L) a_{i}(L)$. We can see that for high deprojection, on the extreme left, we need a combination of geometric and multi-frequency deprojection methods.}
    \label{fig:standard}
  \end{minipage}
  \hfill
  \begin{minipage}{0.45\textwidth}
    \includegraphics[width=\textwidth]{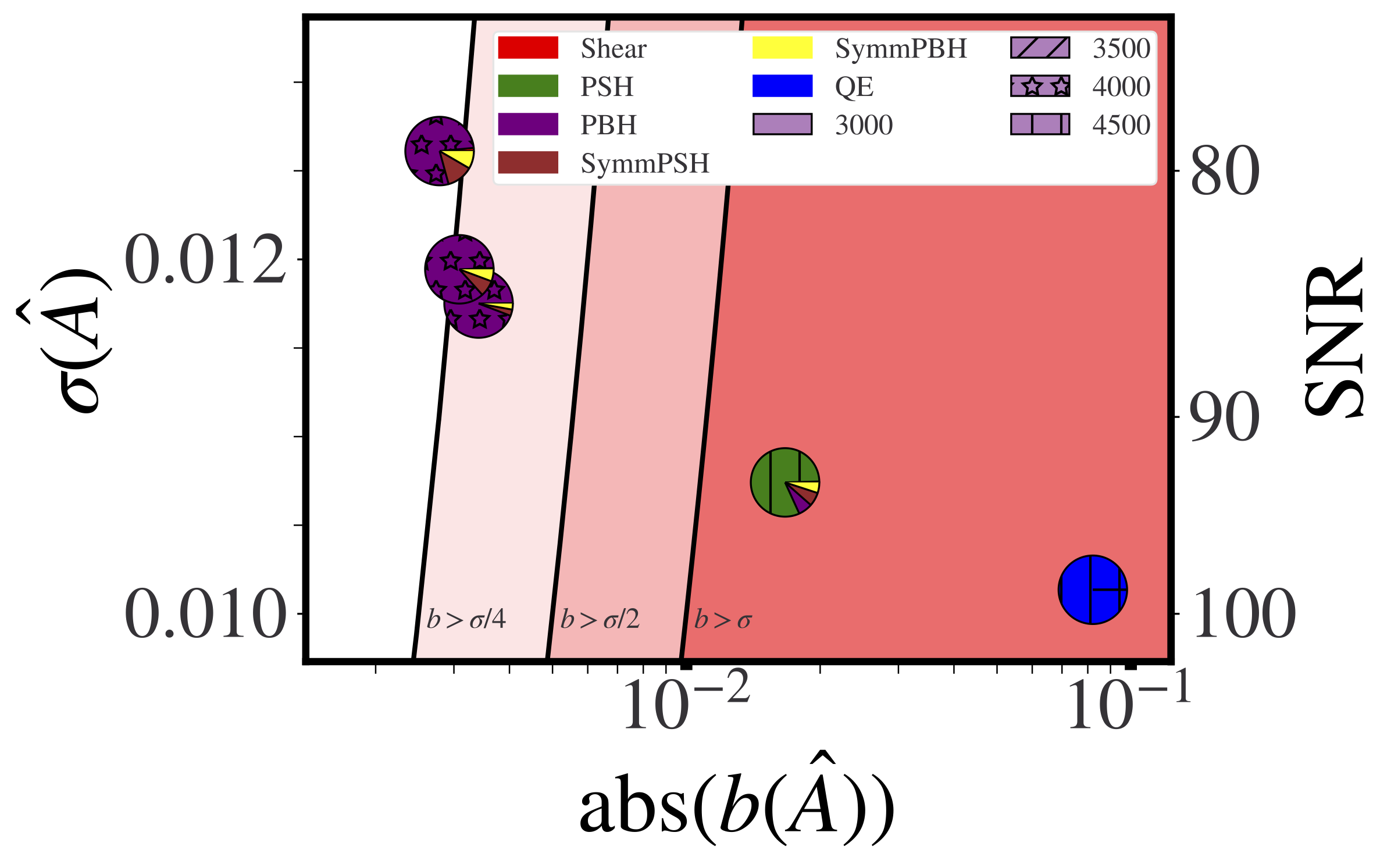}
    \caption{Optimization results for the auto correlation for TT data with aggressive mask. The pie charts represent the contribution from each estimator, calculated as $\int_{\vec{L}} w(L) a_{i}(L)$. When using the aggressive mask, we fix the ILC weights to the ones of the baseline mask, and maintain the same fraction of sky. We can see that for high deprojection, the results change compared to the baseline mask. The SymmPH has no trispectrum or primary term, just secondary. While the PH has all of them. The aggressive mask impacts more the trispectrum term, reducing it. The optimization process then prefers to use more PH, compared to the baseline mask, for high deprojection.}
    \label{fig:agressive}
  \end{minipage}
\end{figure}

We may run our pipeline with a more aggressive masking approach by running the matched filter for point sources at all the SO frequencies, and not just a single one as in the main text and keeping as our source catalog all the objects found at any frequency. The results of the aggressive masking optimization are in Figure \ref{fig:agressive}, to be compared to the standard ones, that for convenience are in Figure \ref{fig:standard}. During the optimization of the more aggressive masking, that covers $8\%$  of the observed sky, we do not consider the change in the available sky fraction, with respect to the standard mask, with
$3\%$ masking are. The sky fraction is fixed to the SO one, to $f_{sky}=0.4$. We can see that the noise penalty for high $f_b$ is lower in the aggressive mask case, compared to the standard one. This can be explained by a change in the biases for each estimator, where now the $PH$ has smaller biases, but has a good signal over noise compared to other estimators.

\section{Lensing validation}

\begin{figure}
     \centering
     \begin{subfigure}{0.45\textwidth}
         \centering
         \includegraphics[width=\textwidth]{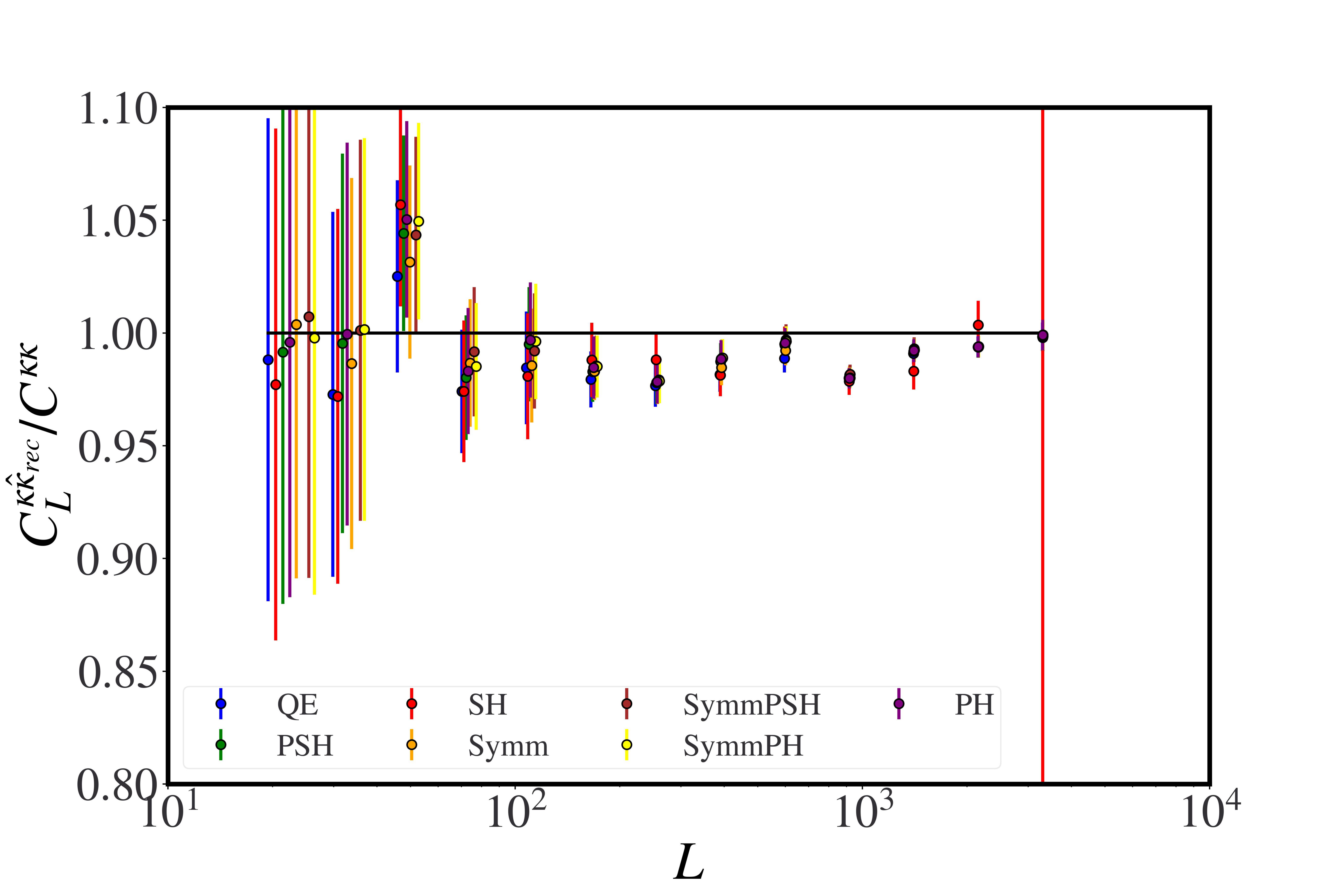}
         \caption{All the estimators have practically unit response. The large red bar on the right comes from the shear estimator, that becomes sub-optimal at small scales.}
         \label{fig:lensing_validation}
     \end{subfigure}
     \hfill
     \begin{subfigure}{0.45\textwidth}
         \centering
         \includegraphics[width=\textwidth]{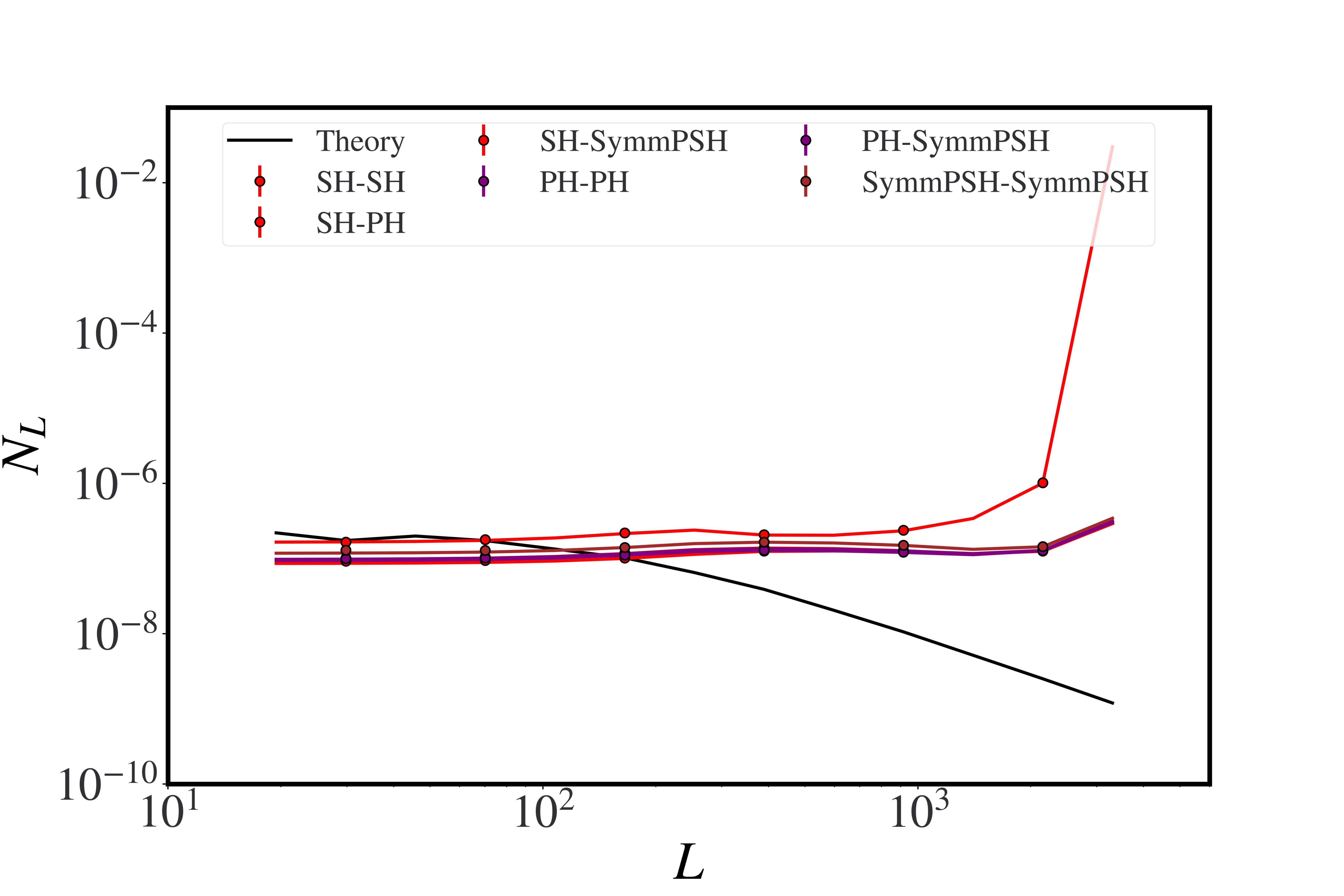}
         \caption{In solid, the Gaussian reconstruction noise from the theory expression, vs the measured power spectrum in dot of some lensing estimators when applied to a sim map with no lensing (Gaussian random field with power spectrum equal to the
total power spectrum). We can see that the dotted points agree with the theory curves.}
         \label{fig:noisecurves}
     \end{subfigure}
     \caption{ Validation of the pipeline for generating maps, lensing maps, and reconstructing lensing for QE, Shear, SymmPSH, SymmPH, PSH, Symm, evaluated on $80$ mock maps, $20 \times 20  \ \mathrm{deg}^2$ each.}
\end{figure}

We verified our pipeline with two checks. In Figure \ref{fig:lensing_validation} we show the lensing validation results, from the mean of the cross-spectrum of $80$ Gaussian lensed simulations with the corresponding $\kappa$ map (note that this is on a small area  $20 \times 20  \ \mathrm{deg}^2$ of cut sky so exact agreement with 1 is not expected). While in Figure \ref{fig:noisecurves} we show the analytical Gaussian noise of the CMB lensing power spectrum vs the measured power spectrum of a few estimators when applied to mock Gaussian CMB maps with a total power spectrum equal to the total power of the filter used in the estimators (lensed CMB + detector noise + foregrounds).

\section{Optimization details \label{app:optdetails}}

\subsection{Choosing among different loss functions}

When performing the optimization in practice the variance part will include only the total variance for power spectra, namely cosmic variance and (reconstruction) noise, without the bias contribution. The bias enters separately explicitly in another term, as seen in equation \ref{eq:totalfunction}.

We do not know the real true foreground biases in nature. The biases that we use to calculate the optimal combination in general will arise: from some theory model (e.g. halo model), or from simulations, as in this paper. In both cases, care is required when deciding what to include as a bias in the loss function. Here we list a few possibilities.

\begin{itemize}
    \item[1] Taking as bias the sum over foregrounds of the absolute values of the different parts of the bias, i.e. the trispectrum, primary and secondary contributions $\sum_{f}\left(|T|+|P|+|S|\right)$.
    \item[2] Taking as the bias the sum over foregrounds $\sum_{f}\left(T+P+S\right)$.
    \item[3] Taking as the bias the absolute value of the sum of biases over foregrounds. $|\sum_{f}\left(T+P+S\right)|$.
    \item[4] Taking as the bias the total bias arising from the sum of the foreground maps $\left(T+P+S\right)_{\mathrm{total}}$, where $\mathrm{total}=\mathrm{tSZ}+\mathrm{kSZ}+\mathrm{CIB}+\mathrm{radio}$.
    \item[5] Taking as the bias the absolute value of the sum of the foreground maps $|\left(T+P+S\right)_{\mathrm{total}}|$.
    \item[6] Taking as the bias $\left(|T|+|P|+|S|\right)_{\mathrm{total}}$.
\end{itemize}

Among these choices, 2, 3, 4, and 5 allow for cancellations among the trispectrum, primary and secondary contribution that lead to zero crossing. In particular, 2 and 4 allow for cancellations among estimator biases at the same $\vec{L}$ or among different $\vec{L}$'s in the integral of \ref{eq:totalfunction}. The zero crossing effect and any type of cancellation can lead to very misleading conclusions, as these are highly simulation/model dependent, and depend on the experimental configuration and the $l_{\mathrm{max}}$'s of reconstruction. Therefore, using 2 or 4 in equation \ref{eq:totalfunction} is not a wise choice.\footnote{Another important point to be made is that with zero crossing sometimes, an estimator that is practically not used, will enter the combination only in one mode, just because its bias in that mode is much smaller than the one of other estimators because of zero crossing. One then gets spikes in the weighting combination, that do not carry big physical meaning, as the zero crossing might depend on the simulation fidelity, on the experimental configuration and analysis choices. This might not give general results. This effect can be mitigated with a regularizer.}

To overcome this, one might want to use one of the other  choices, where there are absolute values that do not allow proper zero crossing. In 1 and 6, zero crossings are by construction not allowed, although in 3 and 5 they might be allowed, as there could be cancellations among $T,P,S$ inside the absolute value. On the other hand, we checked that 1 and 6 will make the optimization process prefer configurations with low $l_{\mathrm{max}}$-es (of around 2500-3000) as the bias contribution becomes too dominant, and the statistical significance of the measurement decreases greatly. Therefore, 3 and 5 are the only choices that remains. Between these the more realistic one is 5, as biases are calculated from the map of the sum of the foregrounds: this is what we usually have in a data analysis.

Now, how do we solve for the possible zero crossing problem of 5? One way is to use a regularizer $g_r(\vec{a}, B^{ij}, ....)$: indeed the idea is that we want to not mathematically achieve the best possible solution for a fixed configuration. Ways to regularize are:
\begin{enumerate}
    \item Smooth the input bias with some kernel, so that a bias at some bin is weighted with neighbouring biases.
    \item Ignore the zero crossing. Optimise regardless, and choose by hand configurations that do not have the smallest total function equation \ref{eq:totalfunction} above, i.e. noise squared plus bias squared.
    \item Introduce priors on the weights, such that the cost function increases if weight is given to estimators with zero crossing. It should not change the optimal solution much, but allows for better behaved solutions.
\end{enumerate}
In this work we opt for the simple choice of option 1 for regularization. To summarize, we take $|\left(T+P+S\right)_{\mathrm{total}}|$ and we smooth it with some function $K$, such that the input bias is\footnote{We take $K$ to be a Gaussian, $K(L)=\frac{1}{Z} \exp\left(-\frac{(L-L_0)^2}{2\sigma^2}\right)$, with $Z$ a normalization constant, and we take $\sigma= 1.5\Delta / \sqrt{8 \log{2}}$, with $\Delta$ the width of the bin edges whose center is $L_0$. Then the smoothed version of the bias at $L_0$ is just $\sum_{\mathrm{bins}} B(L_{\mathrm{bin}})K(L_{\mathrm{bin}})$.}
\begin{equation}
    B_{\mathrm{input}} = K(|\left(T+P+S\right)_{\mathrm{total}}|)
\end{equation}
In practice we take $K$ to be a Gaussian with $\sigma=1.5$ in band-powers. This choice of the input bias should be a realistic non-optimistic one.\footnote{Note, that another way to include uncertainty in foreground biases is by producing a range of simulations, or even better, using some theory foreground model in function of some parameters (e.g. given by halo model) to calculate the variation in the optimization in the weights, and check for the robustness of the optimization in function of varying bias deprojection request.}

\subsection{Optimization algorithm}

For the optimization algorithm we use Differential Evolution (DE), a gradient free global optimization algorithm. DE is based on \emph{mutation} and  \emph{crossover} steps that look for solutions, and a \emph{selection} step that drives in the right direction of global optimization.\footnote{We check the stability of the optimization procedure by optimizing the same configuration for several $\mathcal{O}(10)$ times, and we find the final total functional is the same, although some times the optimizer finds different solution with sub-percent differences in the total functional.} We use the implementation of the \rm{mystic} library \cite{mckerns2012building, mystic}.\footnote{\url{https://github.com/uqfoundation/mystic}.}

\subsection{Estimators over which to optimize }

Given $N_e$ lensing estimators, each calculated for $ N_{l_{max,i}}$ lmaxes, supposing a fixed $l_{min}$, we have a total of $\prod_i N_{l_{max,i}}$ configurations over which to optimize to choose the best ones for a given $f_b$. In particular, for a fixed $N_{l_{max}}$ number of lmaxes for each estimator, the number of configurations is $N_e\times N_{l_{max}}$. For a fixed configuration of estimators, we then have $n_{bins}\times (N_e+1)$ parameters. As we show in Appendix \ref{app:weightstheory} this can be reduced to $n_{bins}\times N_e$.

For a simple case of seven estimators and eight bins this is $8\times7 = 56$ total parameters over to optimize.

\begin{figure}
    \centering
    \includegraphics[width=0.8\columnwidth]{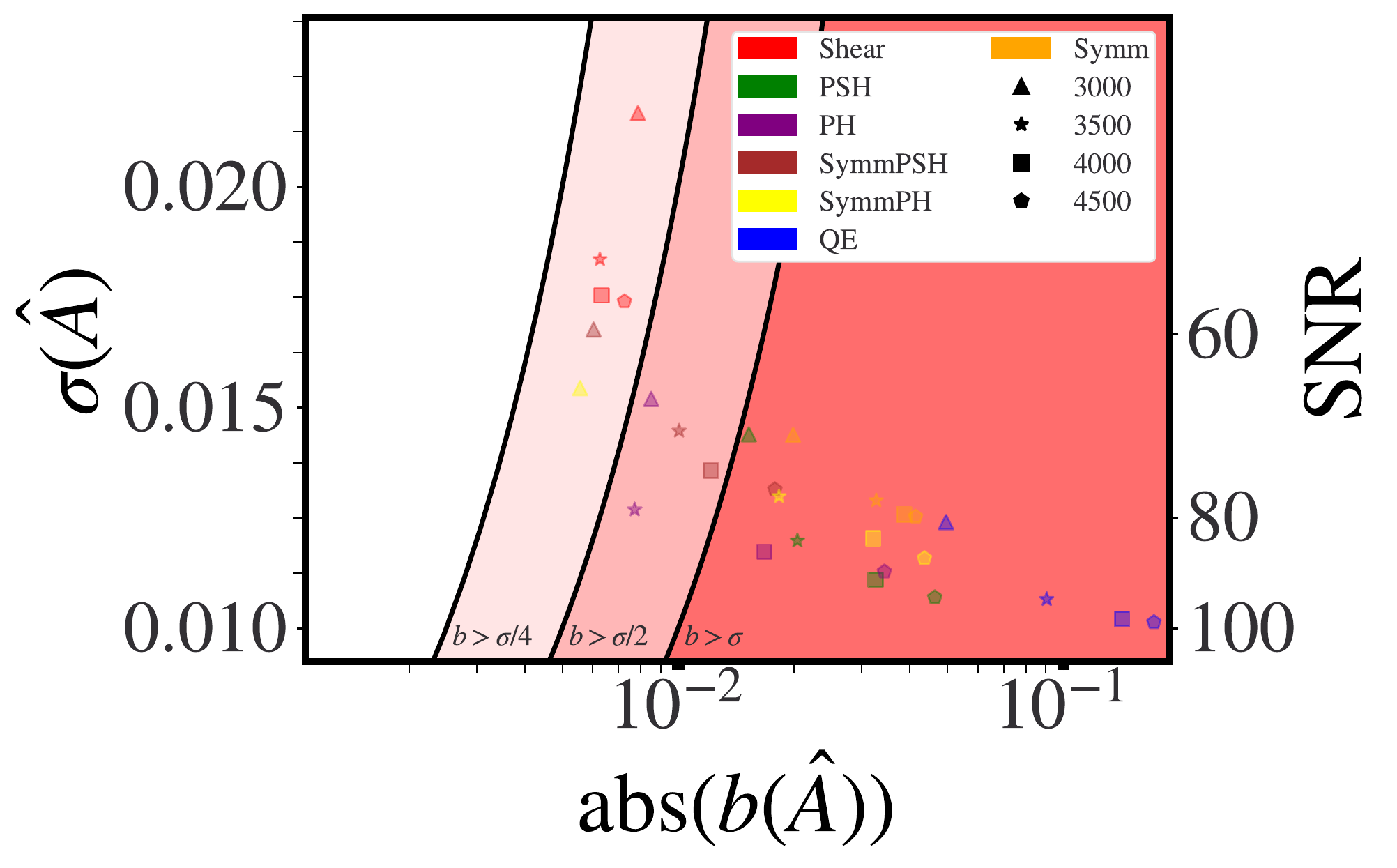}%
    \caption{Biases and noises on the CMB lensing amplitude $A$  for the auto spectrum, with $TT$ only data at different lmaxes of reconstruction. The bands bands represent regimes where the bias is greater than a fraction of the noise. We can see in action the bias-noise trade off: as we increase the noise, the biases get reduced, and vice versa. } \label{fig:alensperestimator}
\end{figure}

To optimize in a reasonable time (for a fixed configuration a couple of hours), we employ a heuristic rule: we optimize for several $l_{\mathrm{max}}$'s for each estimator, as show in Figure \ref{fig:alensperestimator} for the $TT$ only case. Then we choose the estimators with configurations nearer the origin in order to conduct further optimization. The idea is that an estimator very far from the origin will have a higher noise or higher bias, or both. Therefore, if for example we intend to minimize the bias, then we will discard the estimators with very high bias and far from the origin, as they will not be useful for the combined optimal estimator.
Hence, unless $f_b=0$, we choose to optimize over [Shear, PSH, PH, SymmPSH, SymmPH], with possible lmaxes for each estimator of $3000+i500, i=0, 1, 2, 3$: this gives 1024 total configurations per $f_b$, which is numerically tractable.

\begin{figure}
    \centering
    \includegraphics[width=0.8\columnwidth]{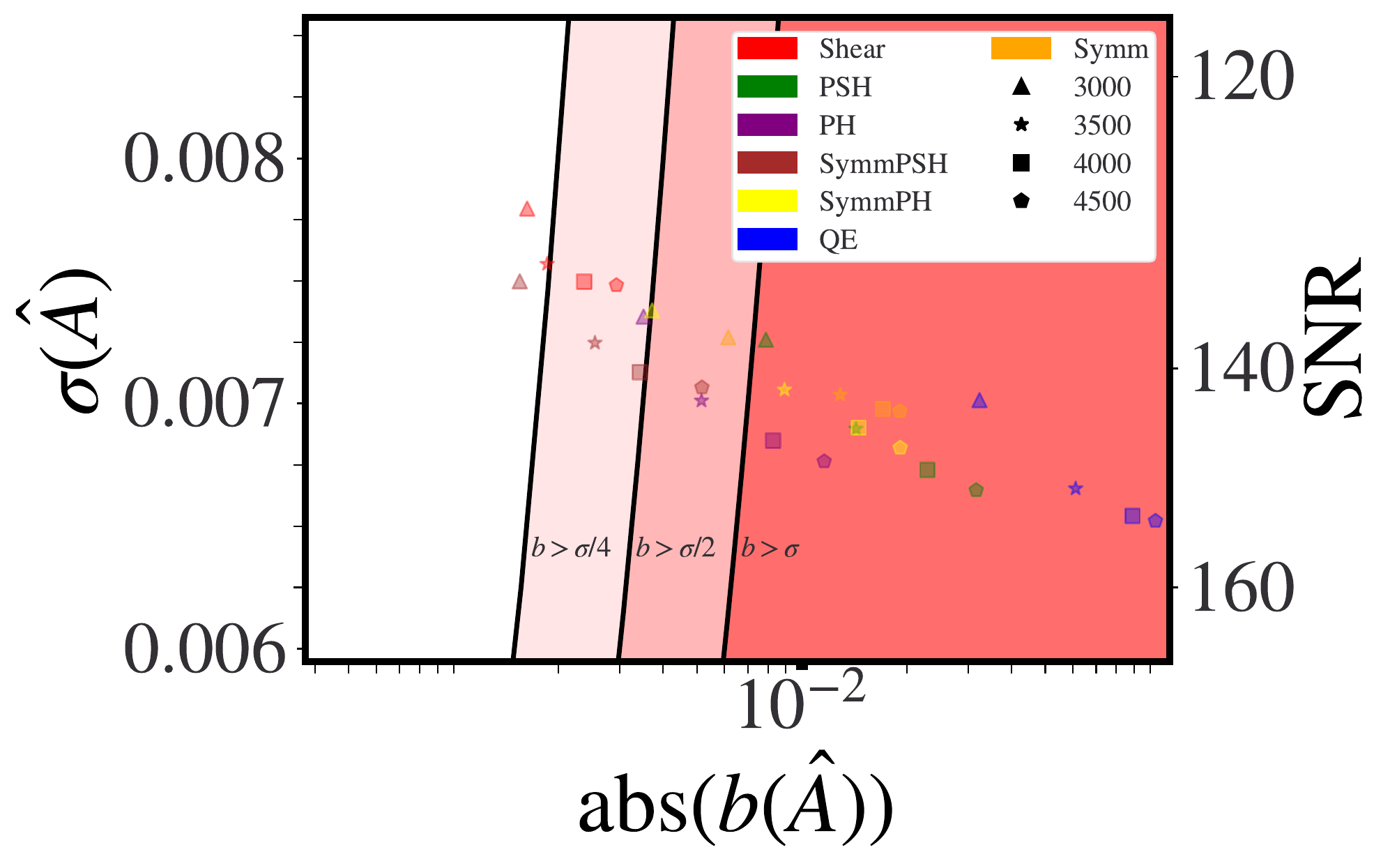}%
    \caption{Biases and noises on the CMB lensing amplitude $A$  for the auto spectrum, with $TT$ plus (foreground free) polarization data, for the estimators presented in the main text, at different lmaxes of reconstruction. Thanks to how clean the polarization is, we are easily able to reduce the bias without a high cost in SNR.} \label{fig:alensperestimatorpol}
\end{figure}

\begin{figure}
    \centering
    \includegraphics[width=0.8\columnwidth]{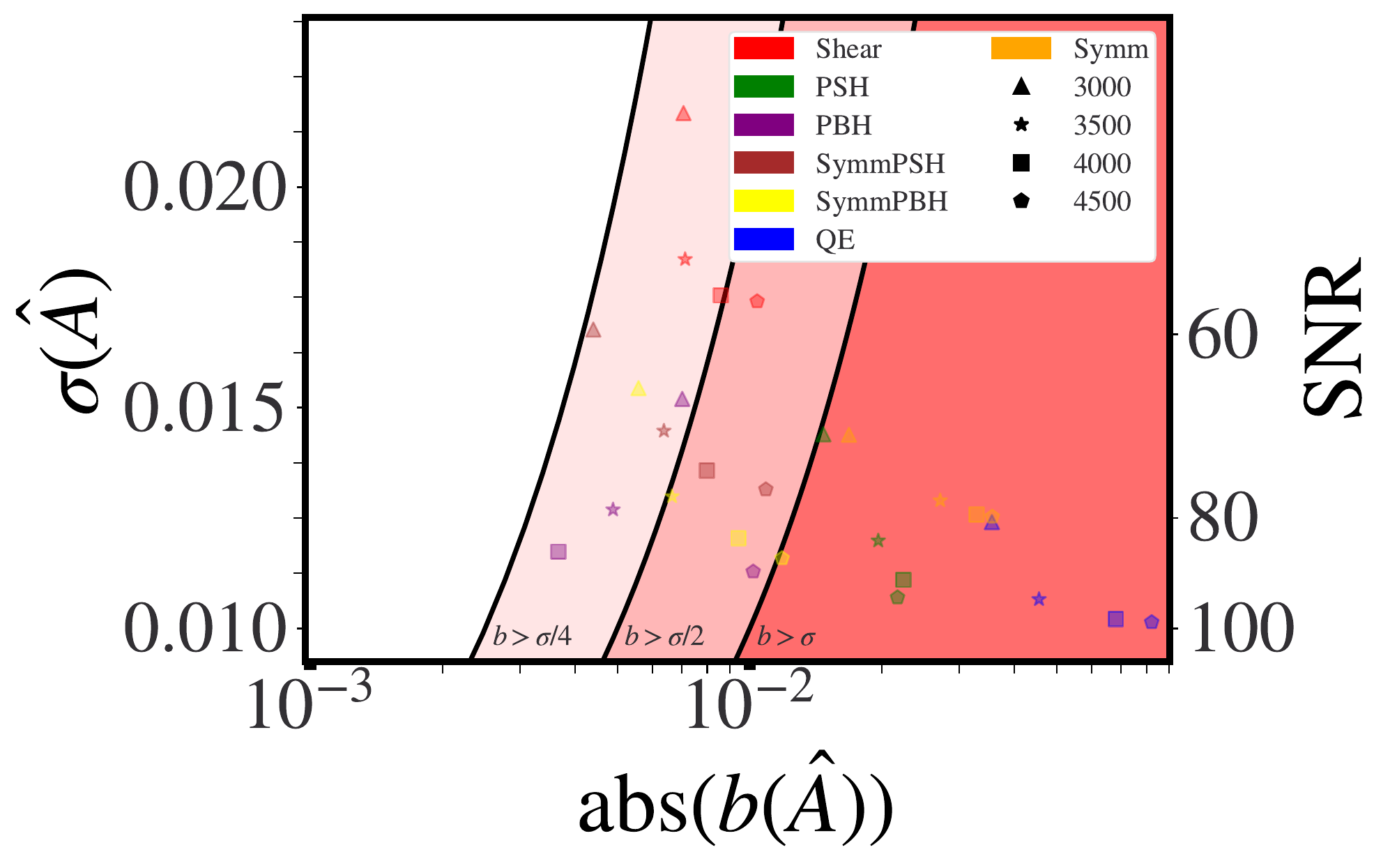}%
    \caption{Same as Figure \ref{fig:alensperestimator} for $TT$ with a more aggressive masking. In general, the aggressive masking will have more impact on the trispectrum term of the foreground induced CMB lensing bias. This will lead to a change in hierarchy among the biases at different $l_{\mathrm{max,TT}}$ for the same estimator, e.g. for PH going to $l_{\mathrm{max,TT}}=4000$ leads to a lower bias with respect to $l_{\mathrm{max,TT}}=3000$, }  \label{fig:alensperestimatoragressive}
\end{figure}

\begin{figure}
    \centering
    \includegraphics[width=0.8\columnwidth]{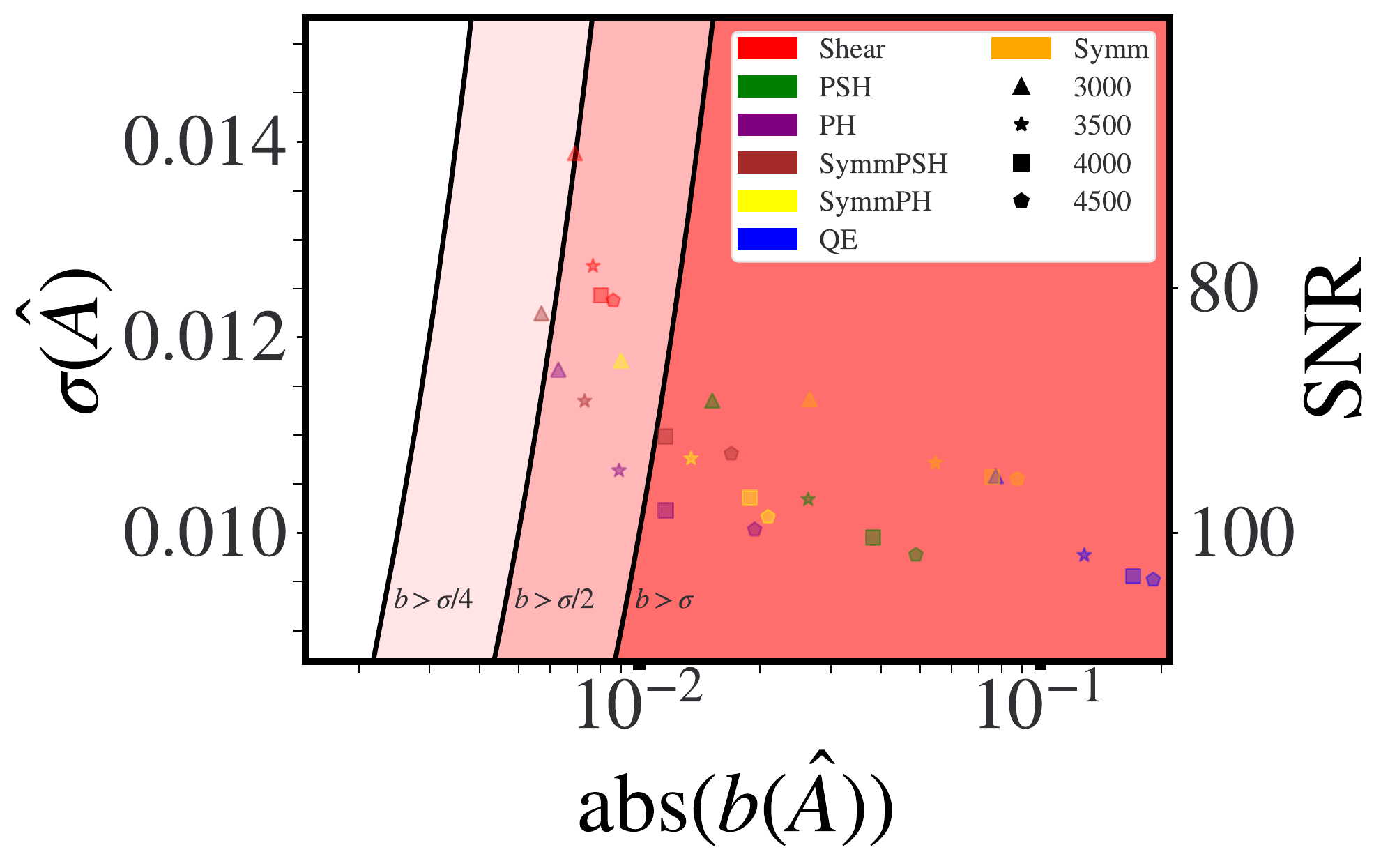}%
    \caption{Biases and noises on the CMB lensing amplitude $A$  for the cross spectrum with an LSST-like sample, with $TT$ only data, for the estimators presented in the main text, at different lmaxes of reconstruction. Here, the bias barely reaches the regime of $b<\sigma/2$. Therefore, for this case an optimization is needed to reach a subdominant bias with respect to the noise.} \label{fig:crossalensperestimator}
\end{figure}

\begin{figure}
    \centering
    \includegraphics[width=0.8\columnwidth]{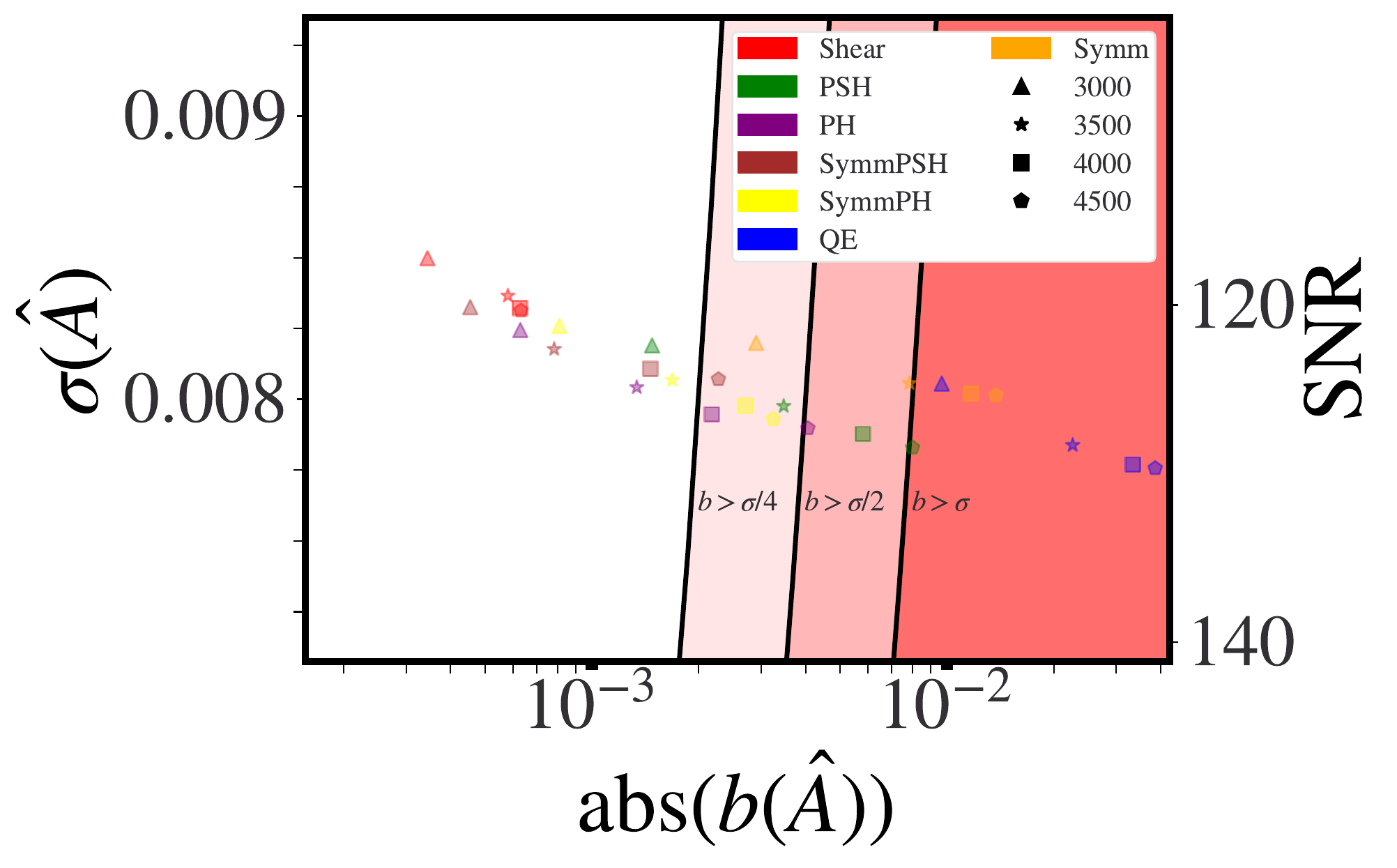}%
    \caption{Same as Figure \ref{fig:crossalensperestimator} for the cross spectrum with an LSST-like sample, for $TT$ plus (foreground free) polarization. Thanks to how clean the polarization is, we are easily able to reach a regime where the bias is subdominant compared to the noise for several estimators at several $l_{\mathrm{max},TT}-$es, as the cross-primary term goes as $\alpha_TT^2$.} \label{fig:crossalensperestimatorpol}
\end{figure}

\section{Understanding the weights}{\label{app:weightstheory}}

In this appendix we want to understand the optimal weights analytically for some specific case. Let's write again
\begin{multline}
    \mathcal{L}= \int_{\vec{L}} w^2(\vec{L})\frac{1}{4\pi f_{sky}}\sum_{ijmn}  \frac{\Theta^{ijmn}(\vec{L})}{(C_L^{\kappa\kappa})^2} a_i(\vec{L})a_j(\vec{L})a_m(\vec{L})a_n(\vec{L}) \\ + f^2_{b}  w(\vec{L})a_i(\vec{L})a_j(\vec{L}) \frac{B^{ij}(\vec{L})}{C_L^{\kappa\kappa}} \int_{\vec{L'}} w(\vec{L'}) a_n(\vec{L'})a_m(\vec{L'}) \frac{B^{mn}(\vec{L'})}{C_{L'}^{\kappa\kappa}}
\end{multline}

Solving this analytically is a very difficult problem.\footnote{For the case $f_b=0$ it is possible to write the minimum variance problem in a matrix form, where the tensor is written as a matrix (e.g. \cite{Voigt}). Or, it is possible to just notice that everything is positive. When $f_b\neq0$ the same trick becomes difficult to reapply, and one in general might want to introduce extra constraints.} In the case of positive only integrands, e.g. only positive input biases in the optimizer, to minimize $f$ we need to minimize the integrand at each point, as we do not have cancellations among different $L$s. Therefore, if we have the optimal configuration for each bin, then we could imagine to minimize over $\vec{a}$, then over $w$. Let's then fix the weights combination per mode, and just vary $w$. We will use the Lagrange multipliers method. The constraints are $\int_{\vec{L}}w(\vec{L})=1$, and $w(\vec{L})>0, \forall \vec{L}$.
\begin{equation}
    \mathcal{L} = \int_{\vec{L}} w^2(\vec{L})\Theta(\vec{L}) + f_b^2 \left(\int_{\vec{L}} w(\vec{L}) B(\vec{L}) \right)^2
    +\lambda (1-\int_{\vec{L}}w(\vec{L}))+ \mu(\vec{L}) (s^2(\vec{L})-w(\vec{L}))
\end{equation}
where $\Theta$ is the combined variance per mode divided by the theory squared, $B$ is the combined bias per mode divided by the theory, $\lambda$ is a Lagrange multiplier, and $\mu$ is a Lagrange multiplier function, and $s$ is a slack variable for the inequality part of the problem.  For simplicity, from now on we will call $w(\vec{L})=w(\vec{L})$, and omit the arguments of $\Theta$ and $B$. We note that $4\pi f_{sky}$ can be absorbed into the definition of $f_b^2$, so it can be ignored when obtaining the solution, and substituted again as $f_b^2 \rightarrow 4\pi f_{sky}f_b^2$. Basically, a lower $f_{sky}$ gives larger noise as there are less modes, effectively making the deprojection harder.

When the inequality constraint is not active anywhere

\begin{equation}
w(\vec{L}) = \frac{C_L^2}{\sigma^2_L}\frac{1}{\int_{\vec{L}}\frac{C_L^2}{\sigma^2_L}} +
 4\pi f_{sky} f_b^2 \left ( \frac{\frac{\int_{\vec{L}}\frac{B_LC_L}{\sigma^2_L}}{\int_{\vec{L}}\frac{C_L^2}{\sigma^2_L}}}{1+4\pi f_{sky}f_b^2\int_{\vec{L}}\frac{B^2_L}{\sigma^2_L}-4\pi f_{sky}f_b^2\frac{\int_{\vec{L}}\frac{B_LC_L}{\sigma^2_L}}{\int_{\vec{L}}\frac{C_L^2}{\sigma^2_L}} \int_{\vec{L}}\frac{B_LC_L}{\sigma^2_L}} \right)\times
  \left( \frac{C_L^2}{\sigma^2_L}\frac{1}{\int_{\vec{L}}\frac{C_L^2}{\sigma^2_L}}  \int_{\vec{L}}  \frac{B_LC_L}{\sigma^2_L} - \frac{B_L C_L}{\sigma^2_L} \right)
\end{equation}

where the quantity in the first big round brackets is the total integrated bias

\begin{equation}
    b =  \left ( \frac{\frac{\int_{\vec{L}}\frac{B_LC_L}{\sigma^2_L}}{\int_{\vec{L}}\frac{C_L^2}{\sigma^2_L}}}{1+4\pi f_{sky}f_b^2\int_{\vec{L}}\frac{B^2_L}{\sigma^2_L}-4\pi f_{sky}f_b^2\frac{\int_{\vec{L}}\frac{B_LC_L}{\sigma^2_L}}{\int_{\vec{L}}\frac{C_L^2}{\sigma^2_L}} \int_{\vec{L}}\frac{B_LC_L}{\sigma^2_L}} \right)
\end{equation}

We can see that this is just the minimum variance solution, with a correction depending on the bias:
\begin{equation}
    w(\vec{L}) = w_L^{MV}(1+4\pi f_{sky}f_b^2b\int_{\vec{L}} \frac{B_LC_L}{\sigma^2_L})-4\pi f_{sky}f_b^2b\frac{B_LC_L}{\sigma^2_L}
\end{equation}
When the constraint is active somewhere, then we have to substitute the expression for $w(\vec{L})$ there, and basically $\mu$ enforces that $w(\vec{L})=0$, for the specific $\vec{L}$. We verify that using this expression leads to the same results as not leaving $w(\vec{L})$ free for some particular cases of $f_b$.

\section{CMB Lensing biases from  quadratic estimators\label{app:biaseslensing}}

\subsection{Biases from simulations}

\begin{figure}
    \centering
    \includegraphics[width=0.6\columnwidth]{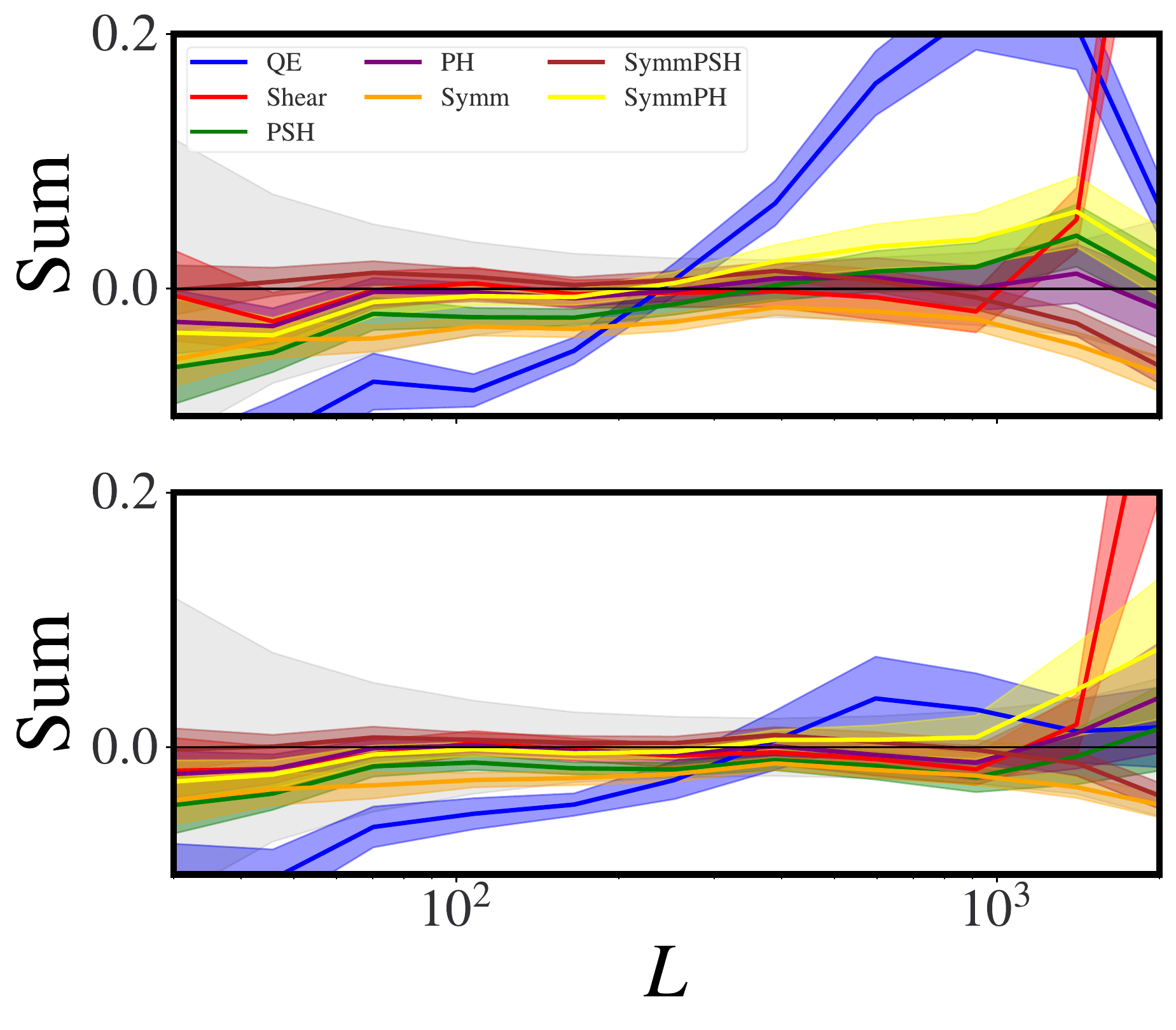}%
    \caption{Total relative foreground-induced auto-spectrum CMB lensing bias per mode for each estimator considered in this work, for $l_{\mathrm{max}}=3500$. In the top panel is shown the contribution arising when using the point source mask constructed on $148$ GHz data; in the bottom panel, when using the point source mask constructed from the product of point source masks on the SO frequencies. When calculating the ILC combinations we fix the weights to theory ones, without considering the change in mask.  We can see how the aggressive masking reduces the biases, with particular emphasis on the QE, meaning that the trispectrum-foreground-induced bias term is the most affected by the masking operation.} \label{fig:biases_auto_per_L}
\end{figure}

\begin{figure}
    \centering
    \includegraphics[width=0.6\columnwidth]{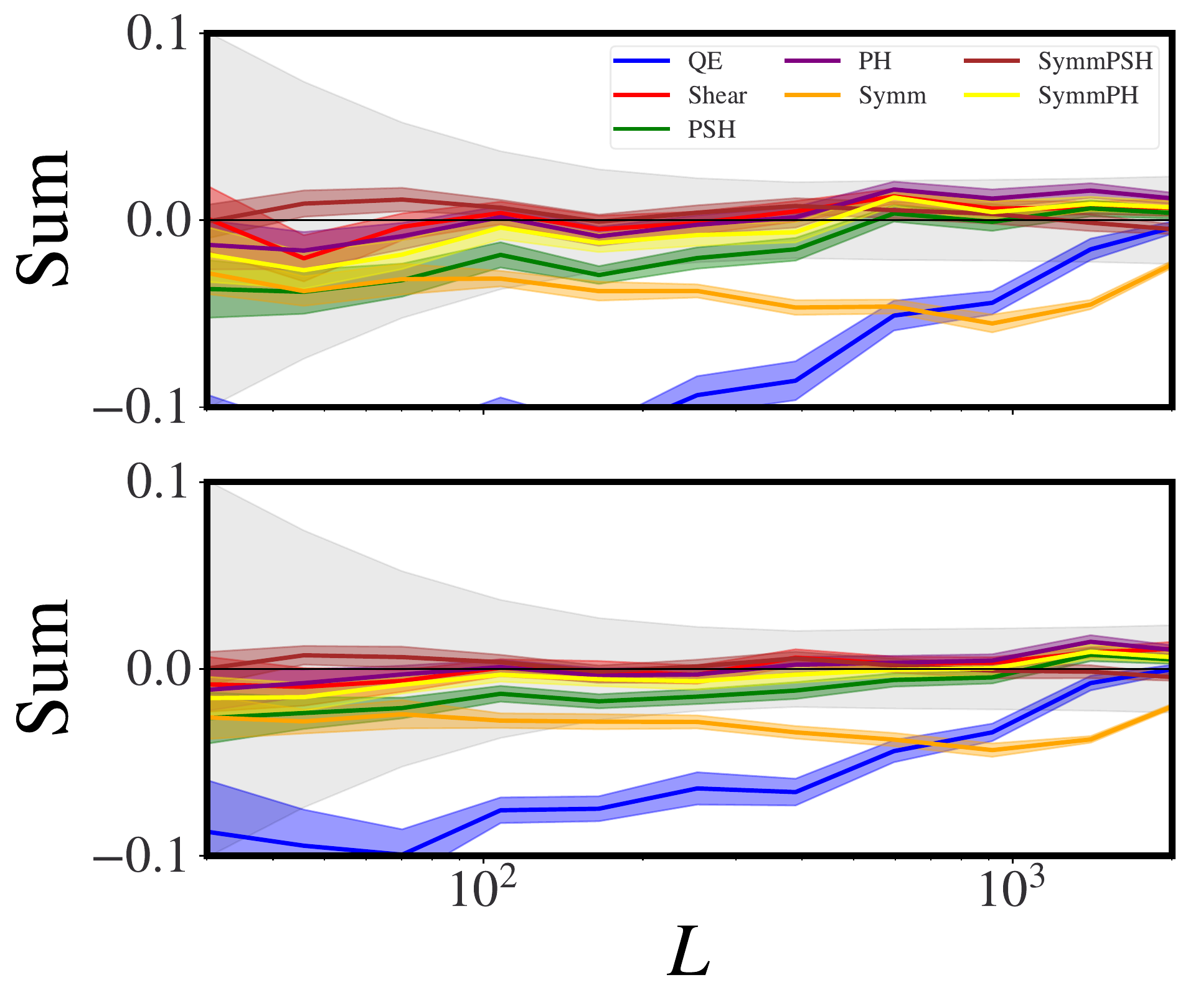}%
    \caption{Same as Figure \ref{fig:biases_auto_per_L} but for the cross-spectrum of CMB lensing with an LSST-like sample. We can see how the aggressive masking reduces the biases, with particular emphasis on the QE.} \label{fig:biases_cross_per_L}
\end{figure}

In Figures \ref{fig:biases_auto_per_L}, \ref{fig:biases_cross_per_L} we show foreground biases at ILC for a configuration where CMB modes used in lensing reconstruction come from $l_{\mathrm{min}}=30$ and $l_{\mathrm{max}}=3500$, and the amplitude is calculated from CMB lensing scales between $L_{\mathrm{min}}=30$ and $L_{\mathrm{max}}=1200$.

\clearpage
\bibliographystyle{prsty.bst}
\bibliography{main}

\end{document}